\documentclass[12pt]{iopart}
\usepackage{epsfig,graphicx}

\newcommand{\bra}{\langle}
\newcommand{\ket}{\rangle}
\newcommand{\bigbra}{\left\langle}
\newcommand{\bigket}{\right\rangle}
\newcommand{\order}{{\mathcal O}}

\newcommand{\erf}{\textrm{erf}}

\newcommand{\one}{{\rm 1\!\!I}}

\newcommand{\bnull}{{\mbox{\boldmath $0$}}}
\newcommand{\be}{\begin{equation}}
\newcommand{\ee}{\end{equation}}
\newcommand{\bd}{\begin{displaymath}}
\newcommand{\ed}{\end{displaymath}}
\newcommand{\vsp}{\vspace*{3mm}}
\newcommand{\room}{\rule[-0.1cm]{0cm}{0.6cm}}

\newcommand{\R}{{\rm I\!R}}

\newcommand{\bv}{\ensuremath{\mathbf{v}}}

\newcommand{\bx}{\ensuremath{\mathbf{x}}}

\newcommand{\bz}{\ensuremath{\mathbf{z}}}

\newcommand{\bR}{\ensuremath{\mathbf{R}}}

\newcommand{\bpsi}{{\mbox{\boldmath $\psi$}}}
\newcommand{\bphi}{{\mbox{\boldmath $\phi$}}}

\newcommand{\hC}{\hat{C}}

\newcommand{\hK}{\hat{K}}
\newcommand{\hL}{\hat{L}}

\newcommand{\hbv}{\widehat{\mathbf{v}}}

\newcommand{\here}{\makebox(0,0)}

\newcommand{\chiR}{\chi_{\rm\!_R}}

\newcommand{\tp}{t^\prime}
\newcommand{\nt}{m(\bv(t),\bz(t))}
\newcommand{\ntp}{m(\bv(\tp),\bz(\tp))}
\newcommand{\hv}{\hat{v}}

\begin{document}

\title[Batch minority games with arbitrary strategy numbers]{Generating functional analysis of batch minority games with
arbitrary strategy numbers}
\author{N Shayeghi and A C C Coolen}
\address{
Department of Mathematics, King's College London\\ The Strand,
London WC2R 2LS, UK }

\begin{abstract}
Both the phenomenology and the theory of minority games (MG) with
more than two strategies per agent are different from those of the
conventional and extensively studied case $S=2$. MGs with $S>2$
exhibit nontrivial statistics of the frequencies with which the
agents select from their available decision making strategies,
with far-reaching implications. In the few theoretical MG studies
with $S>2$ published so far, these statistics could not be
calculated analytically. This prevented solution even in ergodic
stationary states; equations for order parameters could only be
closed approximately, using simulation data. Here we carry out a
generating functional analysis of fake history batch MGs with
arbitary values of $S$, and give an analytical solution of the
strategy frequency problem. This leads to closed equations for
order parameters in the ergodic regime, exact expressions for
strategy selection statistics, and phase diagrams. Our results
find perfect confirmation in numerical simulations.
\end{abstract}

\ead{nima@mth.kcl.ac.uk,ton.coolen@kcl.ac.uk}

\section{Introduction}

Minority games \cite{Arth94,ChalZhan97} were proposed as simple
models with which to increase our understanding of the origin of
the observed nontrivial dynamics of markets; these dynamics are
believed to result from an interplay of cooperation, competition
and adaptation of interacting agents. Minority games were indeed
found to show intriguing nontrivial relations between observables
such as the market volatility, overall bid correlations  and
sensitivity to market perturbations, and the information available
for agents to act upon. Nevertheless, over the years we have found
that they can to a large extent be solved analytically using
equilibrium and non-equilibrium statistical mechanical techniques
(see e.g. the recent textbooks \cite{MGbook1,MGbook2} and
references therein). This desirable combination of complexity and
solvability is their beauty and appeal.

There have been many advances in the mathematical study of MGs,
but nearly all calculations have so far been carried out for
$S=2$, where $S$ denotes the number of trading strategies
available to each agent. One typically finds phase transitions
separating an ergodic from a non-ergodic regime, which for $S=2$
can be located exactly. For $S>2$, in contrast, no such exact
results are available. Although mathematical theory has been
developed along the lines of the $S=2$ case
\cite{ChallMarsZha00,MarsChalZecc00,MarsChall01,Bianconi_etal06},
the authors of the latter studies ultimately ran into the
nontrivial problem of calculating the statistics of agents'
strategy selections, a problem which resisted analytical solution.
In this paper we generalize the theorists' toolbox: we develop a
generating functional analysis \cite{DeDominicis} for MGs with
arbitrary finite values of $S$, and show how to derive exact
closed equations for observables and predict phase transition
lines. We will do so for the simplest so-called batch version of
the MG \cite{HeimelCoolen01,CoolHeimSher01}, with fake (i.e.
purely random) external information \cite{Cavagna99}.

Our paper is organized as follows. We start with model definitions
and review the phenomenology of MGs with $S>2$ as observed in
simulations, focusing on the crucial differences with the more
familiar $S=2$ case and on the core problem of determining the
statistics of strategy selection frequencies. The next step is a
formal generating functional analysis (of which most details are
relegated to an appendix), leading as usual to an effective single
agent process, here describing the evolution of $S$ strategy
valuations. We then state the strategy selection frequency problem
in the language of the effective agent process, and solve it. The
result is an exact (but nontrivial) set of closed equations for
the relevant observables in MGs with arbitrary $S$ in the ergodic
phase, and predictions for the location of the phase transition
(where ergodicity breaks down). $S=3$ is the simplest nontrivial
situation to which our new theory applies, so we present extensive
applications to $S=3$ and test each prediction against numerical
simulations. This is followed by some further (more limited)
applications to $S=4$ and $S=5$ MGs, again verified via simulation
experiments, and predictions regarding market volatility and
predictability. Although the theory is initially set up for
arbitrary types and levels of decision noise, we will concentrate
in this paper mostly on MGs with either additive or absent
decision noise. We end our paper with a discussion of our results
and their implications.

\section{Definitions}

The MG describes $N$ agents in a market, labeled by $i=1\ldots N$.
Each agent $i$ is required to submit a bid $b_i(\ell)\in\{-1,1\}$
(e.g. `sell' or `buy') to this market, at each round
$t\in\{0,1,2,\ldots\}$ of the game, in response to public
information which is distributed to all agents.
 Those who subsequently find themselves in the minority group, i.e. those $i$ for which
$b_i(t)[\sum_j b_j(t)]<0$ (who sell when most wish to buy, or buy
when most wish to sell), make profit. In order to be successful in
the game, agents must therefore anticipate how their competitors
are likely to respond to the public information.

The mathematical implementation of the game is as follows. The
public information at time $t$ (e.g. the state of the market) is
represented by an integer number $\mu(t)\in\{1,\ldots,p\}$. Each
agent $i$ has $S$ private strategies $\bR^{ia}\in\{-1,1\}^{p}$ at
his disposal, labeled by $a=1\ldots S$, with which to convert the
observed information into a binary bid. A strategy is a look-up
table giving prescribed bids for each of the $p$ possible states
of the market: upon observing $\mu(t)=\mu$ at time $t$, strategy
$a$ of agent $i$ would prescribe submitting the bid
$b_i(t)=R^{ia}_\mu$. All $pNS$ entries $R_\mu^{ia}\in\{-1,1\}$ are
drawn randomly and independently at the start of the game, with
equal probabilities. The strategy used by agent $i$ at time $t$ is
called his `active strategy', and denoted by
$a_i(t)\in\{1,\ldots,S\}$. Given the active strategies of all
agents, their collective responses are fully deterministic; the
dynamics of the MG evolves around the evolution of the $N$ active
strategies $a_i(t)$.

Agents in the MG select their active strategies on the basis of
strategy valuations $v_{ia}(t)$, which indicate how often each
strategy would have been profitable if it had been played from the
start of the game. In the so-called batch version of the MG these
valuations are continually updated following
\be
v_{ia}(t+1)=v_{ia}(t)+\theta_{ia}(t)-\frac{\tilde{\eta}}{\sqrt{N}}\sum_{\mu=1}^p
A_\mu(t)R_{\mu}^{ia} \ee
 Here $A_\mu(t)=N^{-1/2}\sum_j
R_\mu^{j a_j(t)}$ is the re-scaled overall market bid at time $t$
that would be observed upon presentation of external information
$\mu$, $\tilde{\eta}$ (the `learning rate') is a parameter that
controls the characteristics time scales of the process, and
$\theta_{ia}(t)$ denotes a (small) perturbation that enables us to
define response functions later. We abbreviate
$\bv^i(t)=(v_{i1}(t),\ldots,v_{iS}(t))\in\R^S$. The active
strategies are now determined at each time $t$ and for each agent
$i$ by a function $m:\R^S\to \{1,\ldots,S\}$ that promotes
strategies with large valuations, but may allow for a degree of
randomness. Typical choices are
\begin{eqnarray}
{\rm deterministic:}&~~~& m(\bv)={\rm argmax}_a [v_a]
\label{eq:deterministic}
\\ {\rm additive~noise:}&~~~&
m(\bv)={\rm argmax}_a [v_a+Tz_a(\ell)] \label{eq:additive}
\\
{\rm multiplicative~noise:}&~~~& m(\bv)=r(\ell)~{\rm argmax}_a
[v_a]+[1-r(\ell)]a(\ell) \label{eq:multiplicative}
\end{eqnarray}
Here $z_a(t)$ is a zero-average and unit-variance random variable,
drawn independently for each $(a,t)$, $T\geq 0$ is a parameter to
control the randomness in (\ref{eq:additive}), $r(t)\in\{0,1\}$ is
drawn randomly and independently for each $\ell$ from some
distribution $P(r)$, and $a(t)$ is drawn randomly and
independently for each $t$ from $\{1,\ldots,S\}$ (with equal
probabilities). We recover the deterministic case
(\ref{eq:deterministic}) by taking the limit $T\to 0$ in
 (\ref{eq:additive}), or $P(r)\to \delta(r-1)$ in
 (\ref{eq:multiplicative}).
In order to compactify subsequent equations we will write all
random variables at time $t$ in
(\ref{eq:additive},\ref{eq:multiplicative}), which will be drawn
separately and independently for each of the $N$ agents, as
$\bz(t)$. This allows us to write generally
$a_i(t)=m(\bv^i(t),\bz^i(t))$. The (stochastic) equations
describing the MG with arbitrary $S$ can therefore be written as
\begin{eqnarray}
v_{ia}(t+1)&=&v_{ia}(t)+\theta_{ia}(t)-\frac{\tilde{\eta}}{N}\sum_{\mu=1}^p\sum_{j=1}^N
R_{\mu}^{ia}  R_\mu^{j m(\bv^j(t),\bz^j(t))} \label{eq:defs}
\end{eqnarray}
Alternatively we can write the process (\ref{eq:defs}) in
probabilistic form, i.e. as an evolution equation for the
probability density $P(\bv^1,\ldots,\bv^N)$ of the $N$ strategy
valuation vectors. Upon abbreviating
$\{\bz\}=(\bz^1,\ldots,\bz^N)$ and $\{\bv\}=(\bv^1,\ldots,\bv^N)$
this can be written as
\begin{eqnarray}
P_{t+1}(\{\bv\})&=&
\int\!d\{\bv^\prime\}~W_t(\{\bv\};\{\bv^\prime\})P_{t}(\{\bv^\prime\})
\label{eq:process1}
\\
W_t(\{\bv\};\{\bv^\prime\}) &=& \bigbra \prod_{i a}
\delta\Big[v_{ia}-v^\prime_{ia}-\theta_{ia}(t)+\frac{\tilde{\eta}}{N}\sum_{\mu
j} R_{\mu}^{ia}R_\mu^{j m(\bv^{j\prime}\!,\bz^j)}\Big]
\bigket_{\!\!\{\bz\}}
\label{eq:process2}
\end{eqnarray}
Averages over the process (\ref{eq:process1},\ref{eq:process2})
are written as $\bra \ldots\ket$. The market fluctuations in the
MG are characterized by the average $\bra
A(t)\ket=p^{-1}\sum_{\mu} \bra A_\mu(t)\ket$, which one expects to
be zero, and the covariance kernel $\Xi_{tt^\prime}=p^{-1}\sum_\mu
\bra [A_\mu(t)-\bra A(t)\ket][A_\mu(t^\prime)-\bra
A(t^\prime)\ket]\ket$. In particular, the volatility $\sigma$ is
defined by $\sigma^2=\lim_{\tau\to\infty}\tau^{-1}\sum_{t=1}^\tau
\Xi_{tt}$, and the direct predictability is measured by
$H=p^{-1}\sum_\mu [\lim_{\tau\to\infty}\tau^{-1}\sum_{t=1}^\tau
(A_\mu(t)-\bra
A(t)\ket)]^2=\lim_{\tau\to\infty}\tau^{-2}\sum_{tt^\prime=1}^\tau
\Xi_{tt^\prime}$. For detailed discussions of the relations
between the above and various alternative MG versions we refer to
\cite{MGbook1,MGbook2}.

\section{Phenomenology of MGs with arbitrary values of $S$}

\begin{figure}[t]
\vspace*{-3mm} \hspace*{40mm} \setlength{\unitlength}{0.30mm}
\begin{picture}(300,450)

   \put(20,300){\includegraphics[height=125\unitlength,width=135\unitlength]{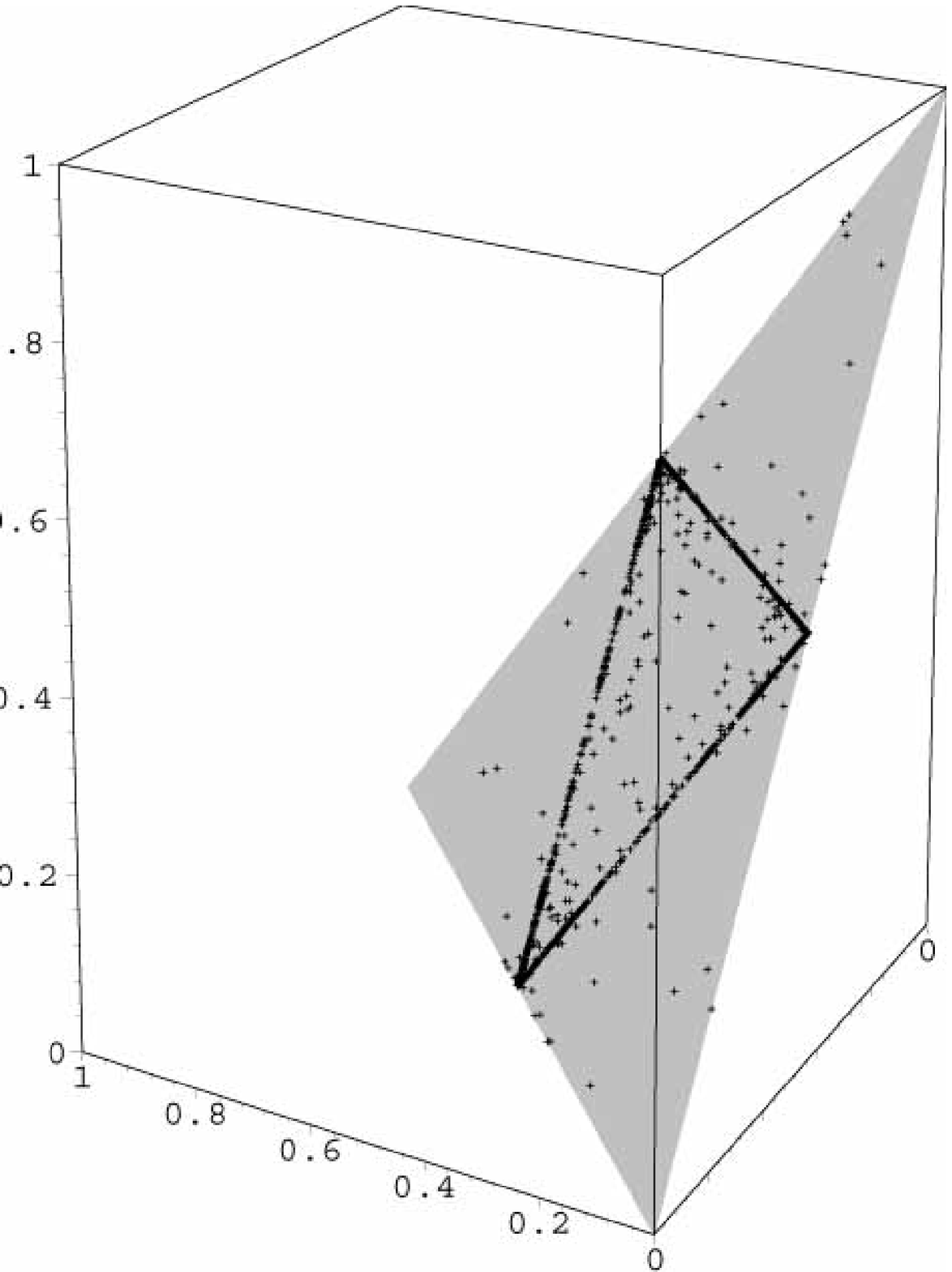}}
  \put(-40,350){\here{$\alpha=1/64$}}

 \put(20,150){\includegraphics[height=125\unitlength,width=135\unitlength]{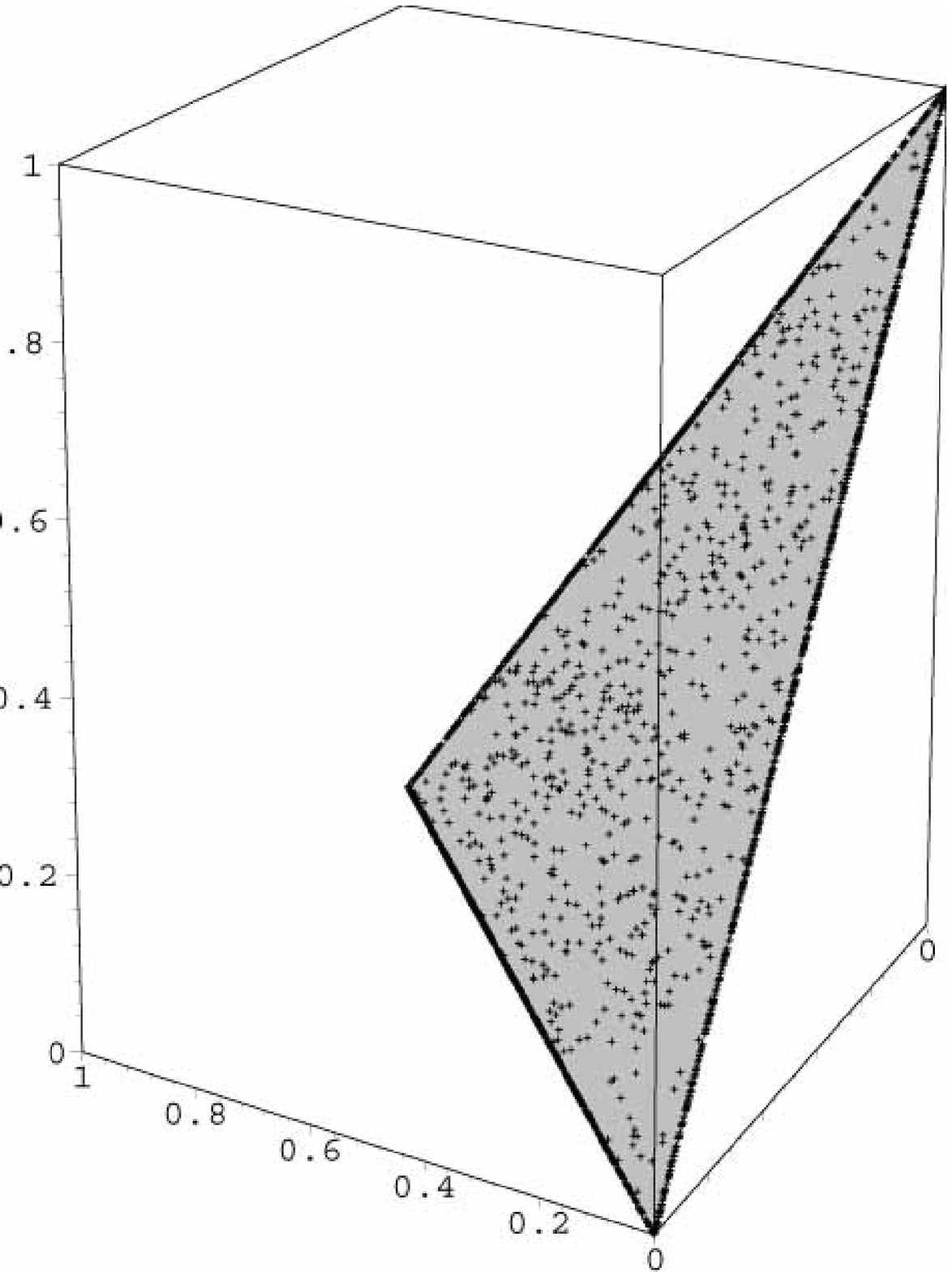}}
  \put(-40,200){\here{$\alpha=1$}}

 \put(20,0){\includegraphics[height=125\unitlength,width=135\unitlength]{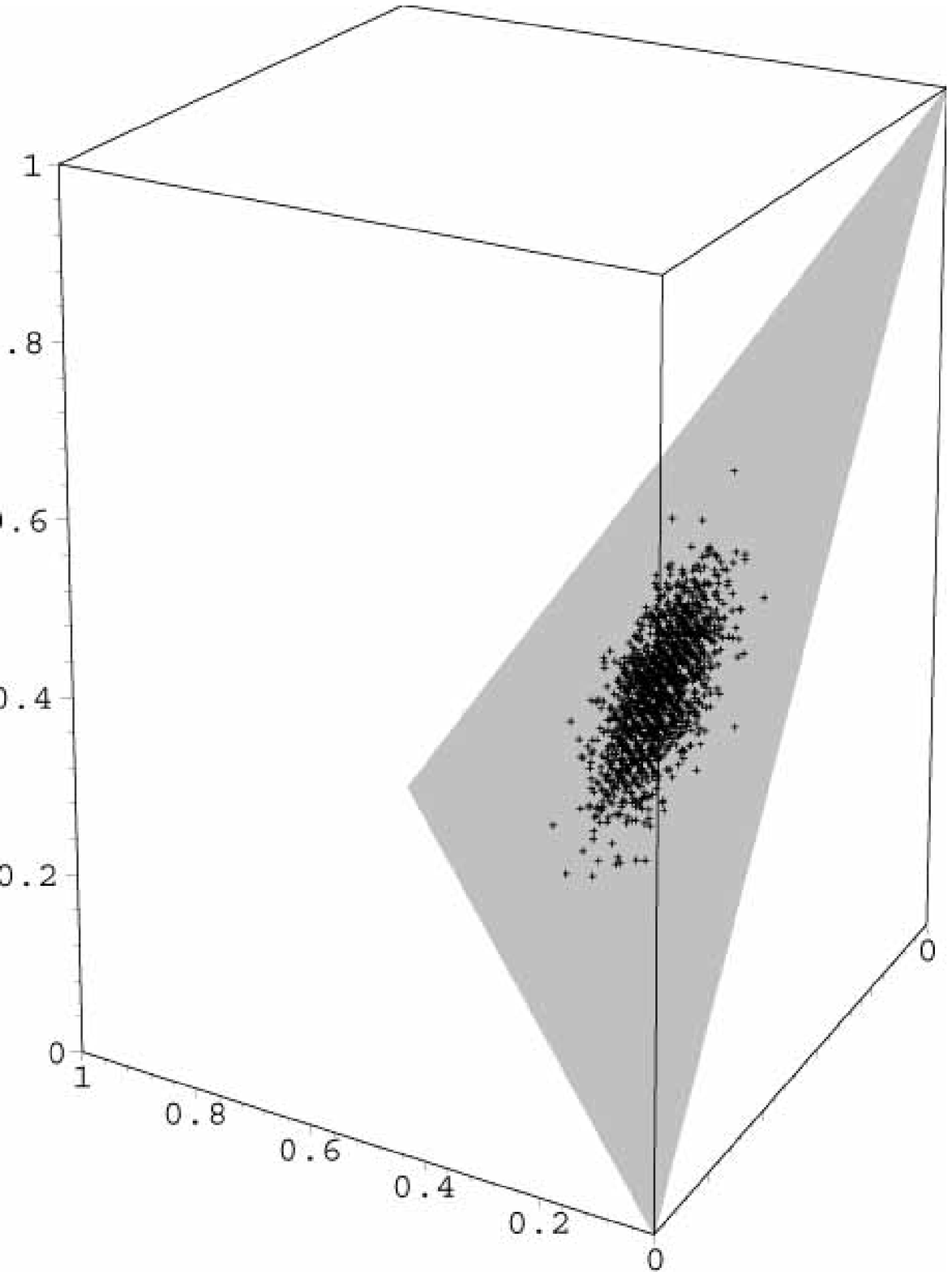}}
   \put(-40,50){\here{$\alpha=64$}}

   \put(190,300){\includegraphics[height=125\unitlength,width=135\unitlength]{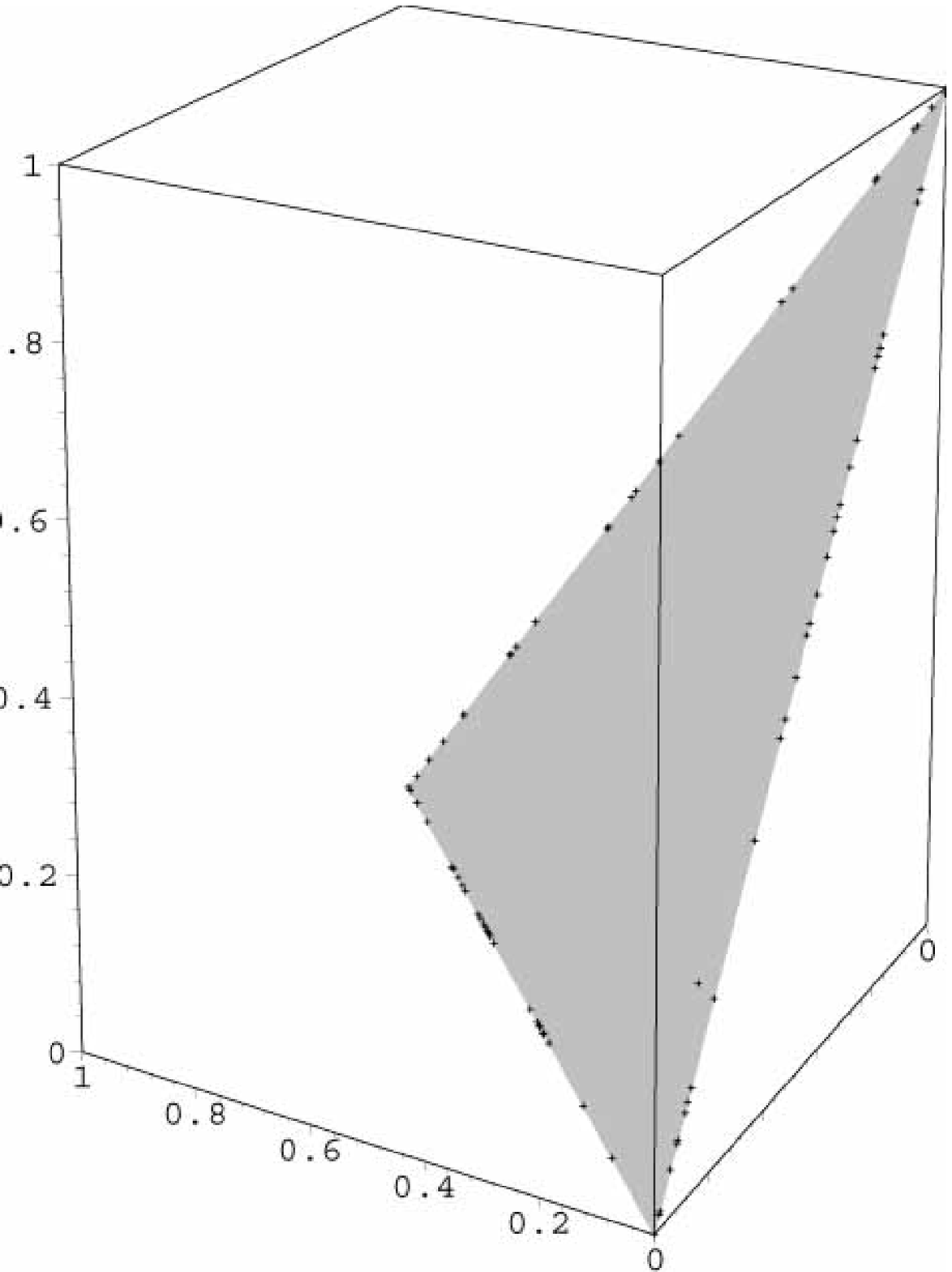}}

 \put(190,150){\includegraphics[height=125\unitlength,width=135\unitlength]{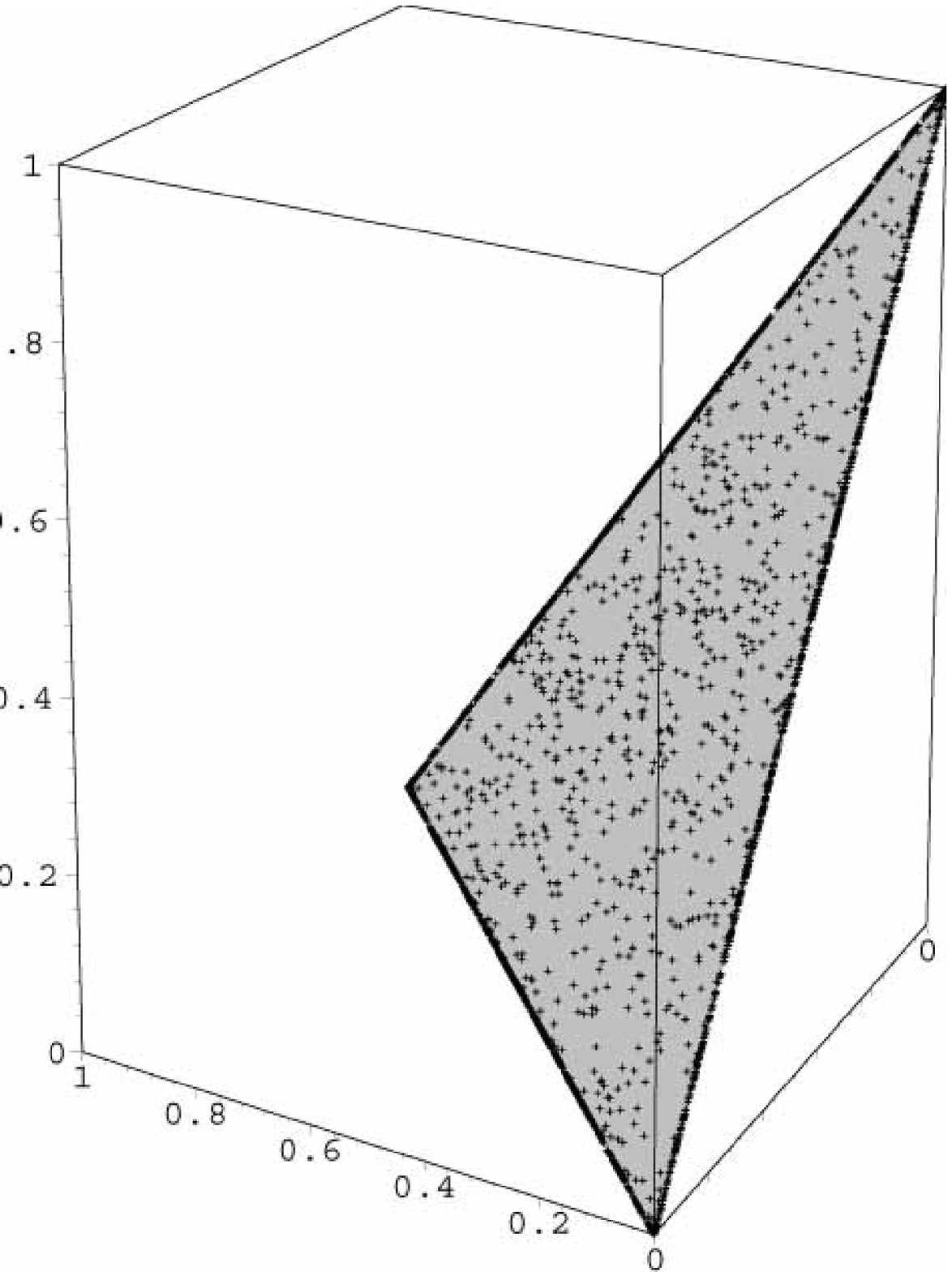}}

 \put(190,0){\includegraphics[height=125\unitlength,width=135\unitlength]{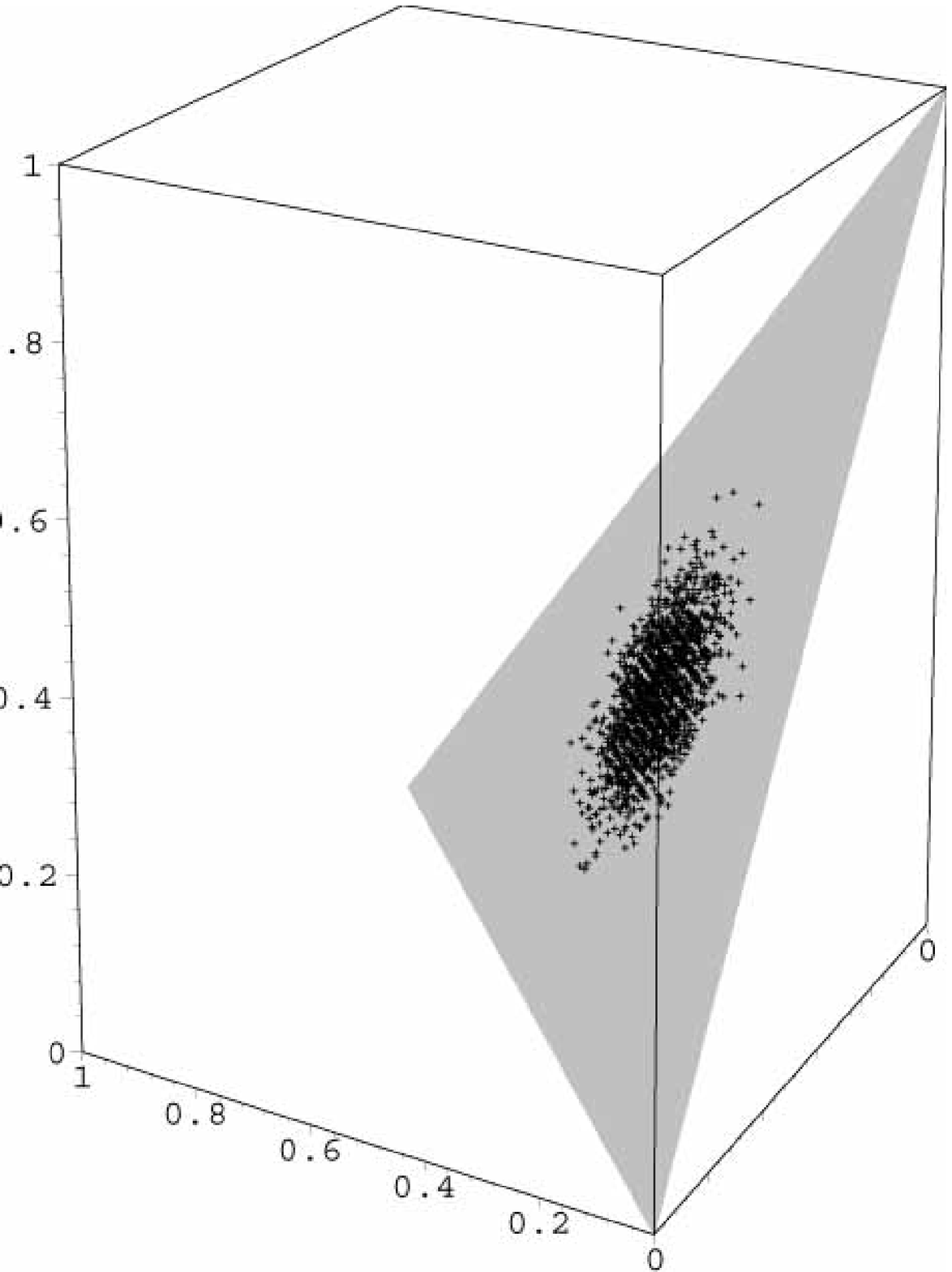}}

\end{picture}
 \vspace*{5mm}
\caption{Each plot shows the frequency vectors ${\bf
f}_i=(f_1^i,f_2^i,f_3^i)$ with which the agents use their three
strategies, drawn for each agent $i$ as a point in $[0,1]^3$
(giving $N$ points per plot), as obtained from numerical
simulations of an MG without decision noise with $N=4097$ and
$S=3$. The constraint $\sum_a f_a^i=1$ for all $i$ implies that
all points are in the plane that goes through the three corners
$\{(1,0,0),(0,1,0),(0,0,1)\}$ (each corner represents agents
`frozen' into using a single strategy). Left plots: unbiased
initial conditions (random initial strategy valuations drawn from
$[-10^{-4},10^{-4}]$). Right plots: biased initial conditions
(random initial strategy valuations drawn from $[-10,10]$). }
\label{fig:freq_3d_S3} \vspace*{-2mm}
\end{figure}

Before diving into theory it is perhaps helpful to describe first
the phenomenology of MGs with $S>2$, as observed in numerical
simulations, with emphasis on those aspects in which they are
distinct from $S=2$ MGs. All simulations of which results are
shown in the remainder of this paper involved MGs with $N=4097$
agents and no decision noise, and stationary state measurements
were taken either over the time interval $500\leq t\leq 3000$ (for
$\alpha<32$) or over $100\leq t\leq 1100$ (for $\alpha\geq 32$).

\begin{figure}[t]
\vspace*{-6mm} \hspace*{15mm} \setlength{\unitlength}{0.40mm}
\begin{picture}(290,200)
  \put(0,0){\epsfxsize=160\unitlength\epsfbox{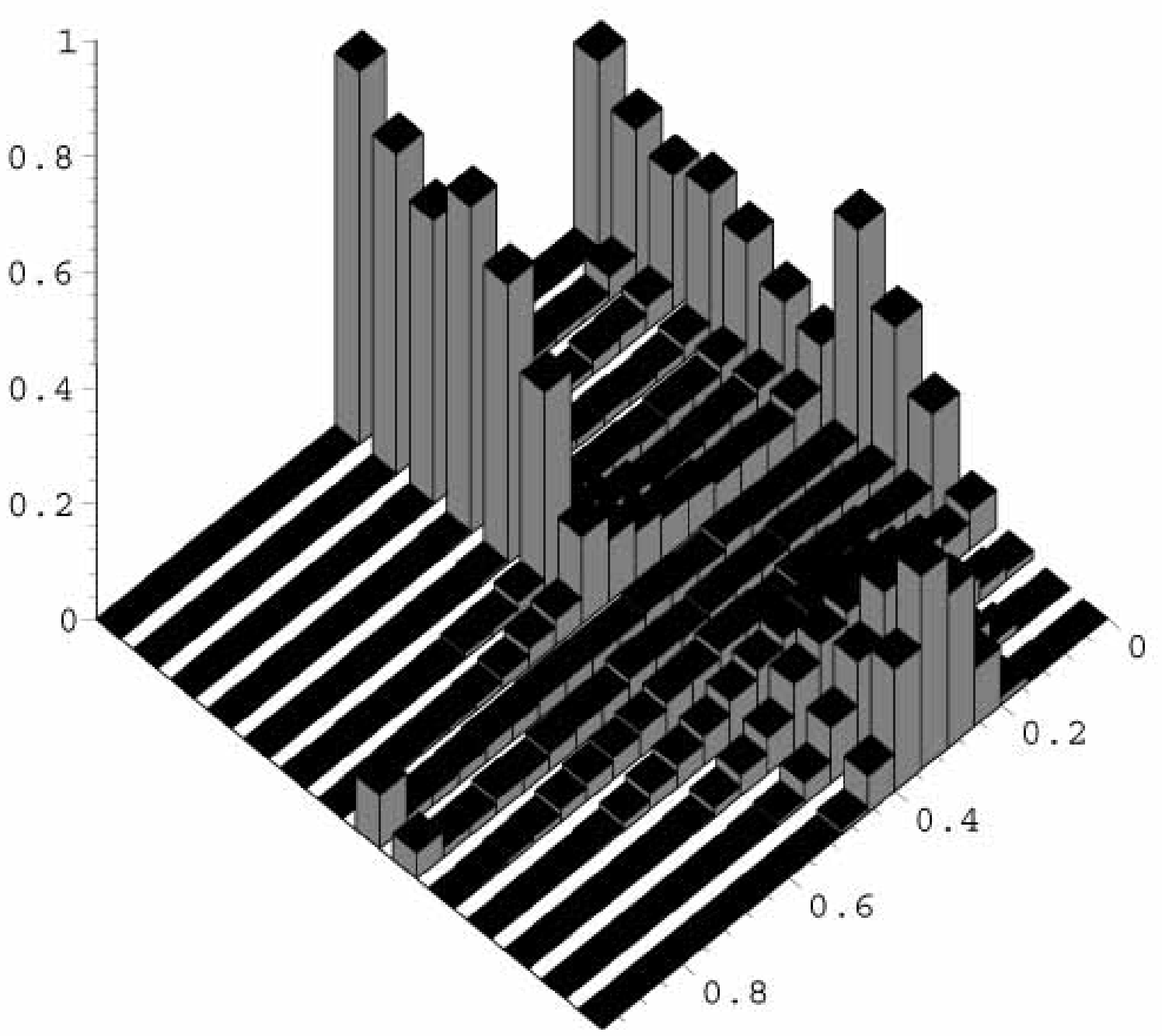}}
  \put(125,35){\here{$f_1$}}\put(36,36){\here{$\alpha$}}
  \put(-20,110){$\varrho(f_1)$}
    \put(200,0){\epsfxsize=160\unitlength\epsfbox{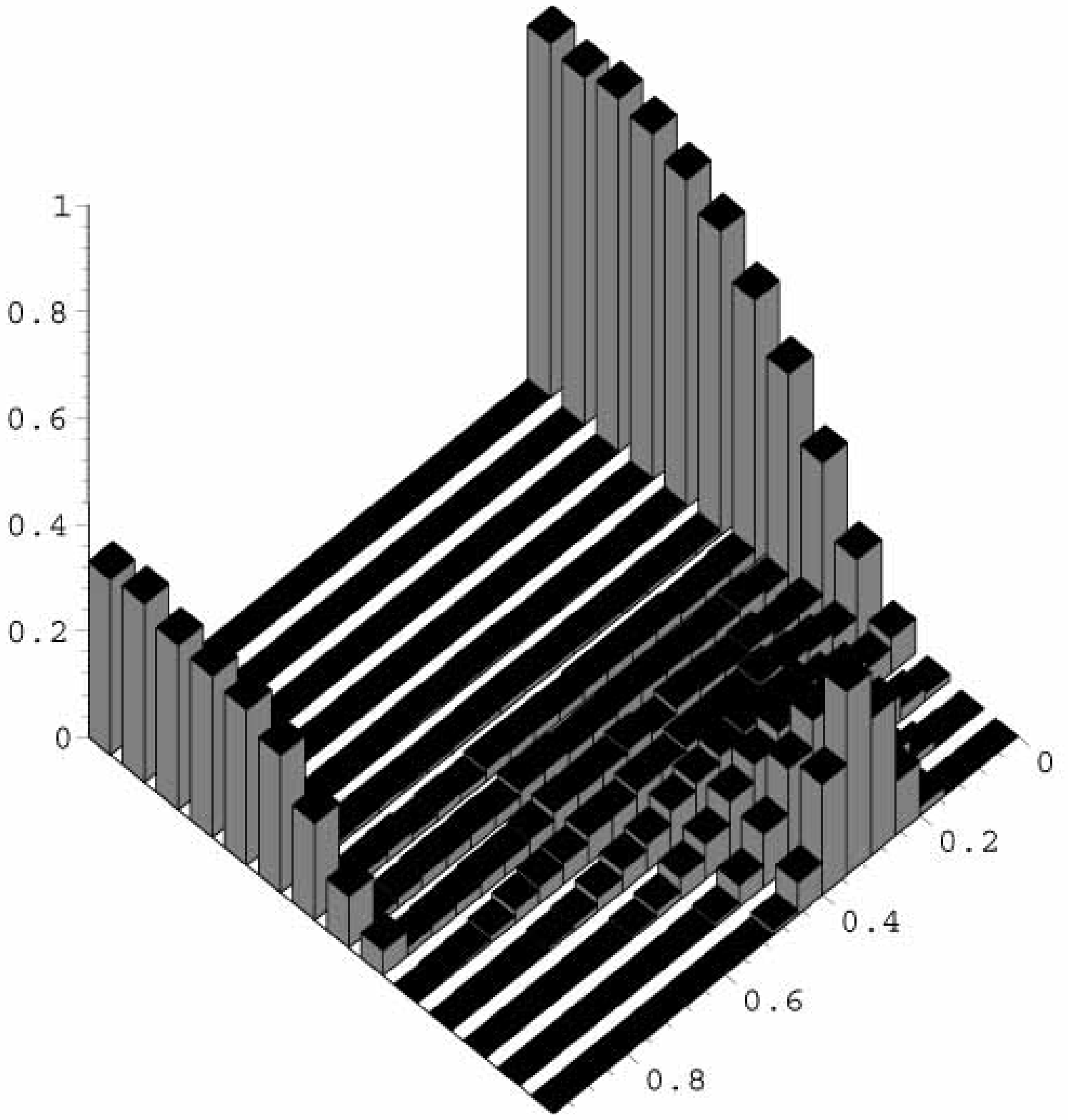}}
  \put(325,35){\here{$f_1$}}\put(236,36){\here{$\alpha$}}
  \put(180,110){$\varrho(f_1)$}
\end{picture}
\vspace*{-3mm} \caption{Histograms of the fraction $\varrho(f_1)$
of agents that play strategy 1 with frequency $f_1$ in the
stationary state  (observed in simulations with $N=4097$ and
$S=3$), for $\alpha\in\{1/128,1/64,\ldots,32,64\}$ (increasing by
a factor 2 at each step).
 Left: unbiased
initial conditions (random initial strategy valuations drawn from
$[-10^{-4},10^{-4}]$). Right: biased initial conditions (random
initial strategy valuations drawn from $[-10,10]$). }
 \label{fig:S3freqhisto}
\end{figure}

In early (numerical and theoretical) studies on MGs with $S>2$ the
emphasis was often on the behaviour of the volatility. Its
dependence on $\alpha$ was found to be very similar for different
values of $S$, but with the phase transition point $\alpha_c(S)$
(that separates a nonergodic regime at small values of $\alpha$
from an ergodic regime for large $\alpha$) increasing with
increasing values of $S$. This latter dependence was conjectured
in \cite{MarsChalZecc00} to be roughly $\alpha(S)\approx
\alpha_c(2)+\frac{1}{2}(S-2)$. The fundamental differences between
$S=2$ MGs and $S>2$ MGs become clear as soon as one tries to go
beyond measuring observables derived from the overall market bid
(like the volatility),
 but turns to quantities such as the fraction of `frozen' agents or the
long-time correlations. For $S>2$ it is no longer obvious how such
objects must be defined.  Some agents are found to play just one
strategy, some play two, some play three, etc.; for $S>2$ agents
can apparently be `frozen' to various extents, which cannot be
captured by a single number $\phi$ (the fraction of frozen agents
for $S=2$, see e.g. \cite{MGbook1,MGbook2}). Similarly, it is not
a priori clear which variables should be measured to define
correlation functions.

\begin{figure}[t]
\vspace*{-10mm} \hspace*{-5mm} \setlength{\unitlength}{0.42mm}
\begin{picture}(290,200)
  \put(0,0){\epsfxsize=215\unitlength\epsfbox{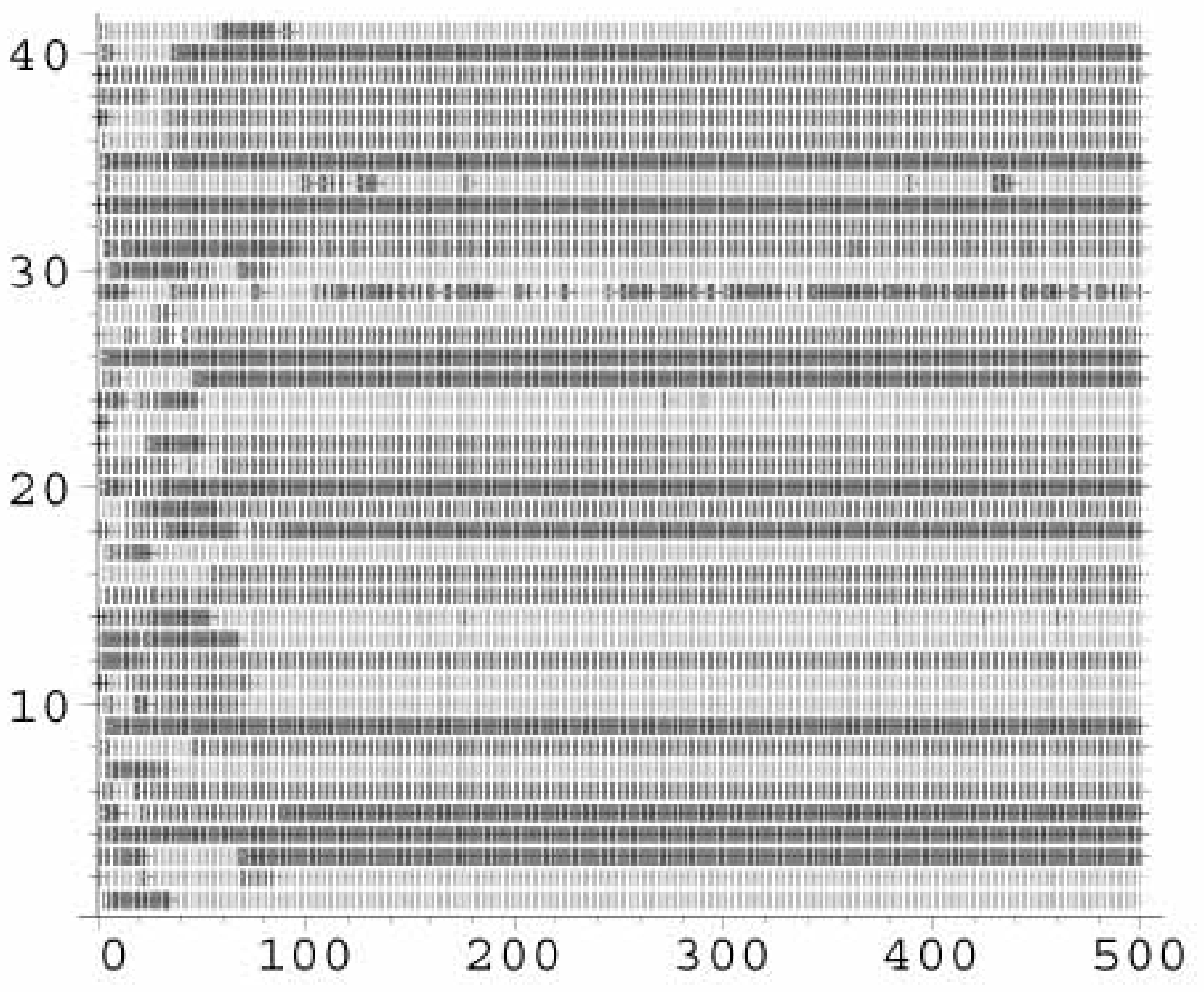}}
  \put(105,-5){\here{$t$}}
  \put(-2,90){$i$}
    \put(185,0){\epsfxsize=215\unitlength\epsfbox{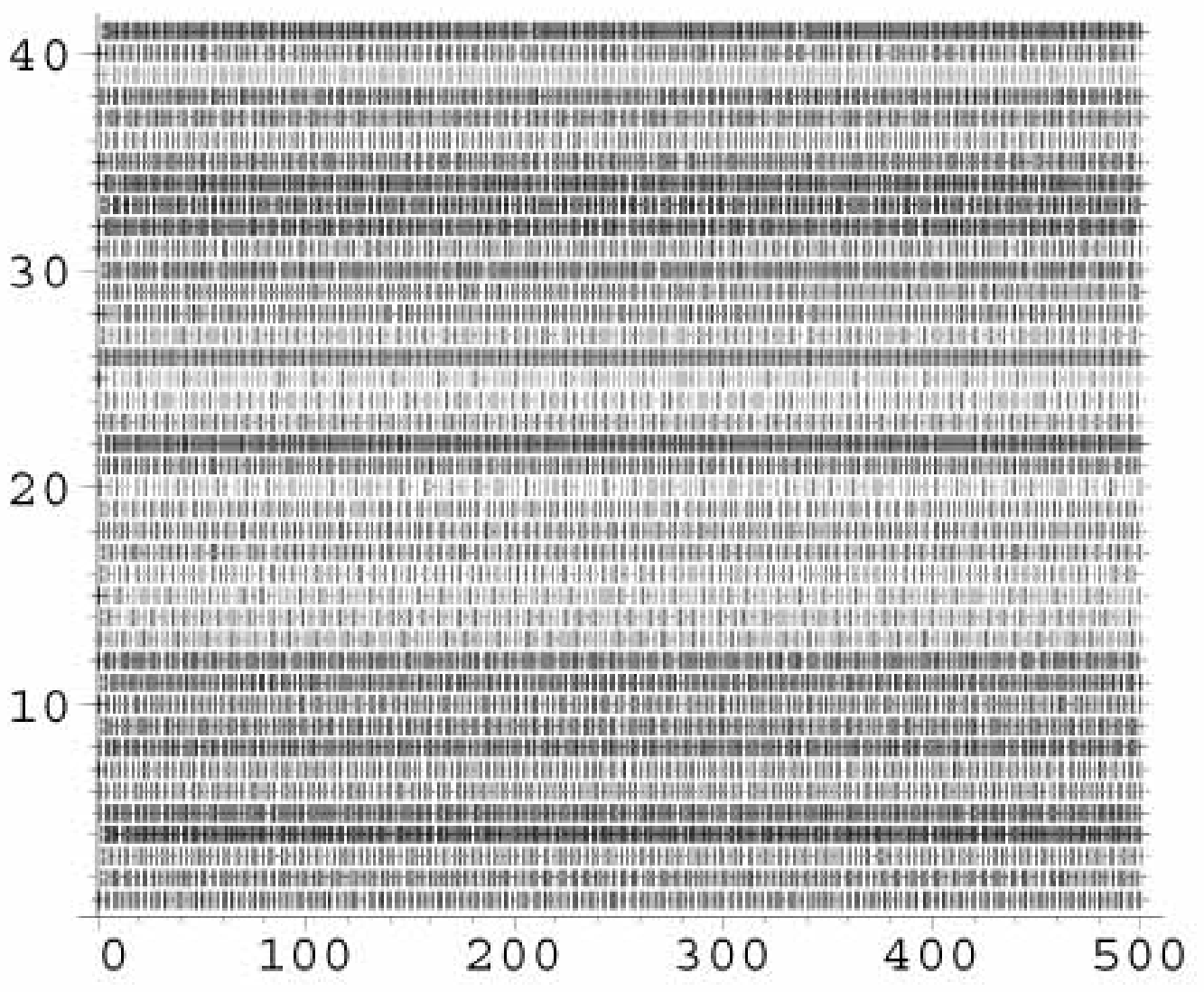}}
     \put(290,-5){\here{$t$}}
\end{picture}
\vspace*{5mm} \caption{Evolution in time of active strategy
selections for the first 41 out of the total number $N=4097$ of
agents in an MG with $S=3$, following unbiased initial conditions.
The chosen strategies are indicated by grey levels: white means
$a_i(t)=1$, grey means $a_i(t)=2$, and black means $a_i(t)=3$.
Left graph: $\alpha=1/128$ (in the nonergodic regime). Right
graph: $\alpha=8$ (in the ergodic regime). }
 \label{fig:S3evolution}
\end{figure}

\begin{figure}[t]
\vspace*{-3mm} \hspace*{40mm} \setlength{\unitlength}{0.30mm}
\begin{picture}(300,450)

   \put(20,300){\includegraphics[height=125\unitlength,width=135\unitlength]{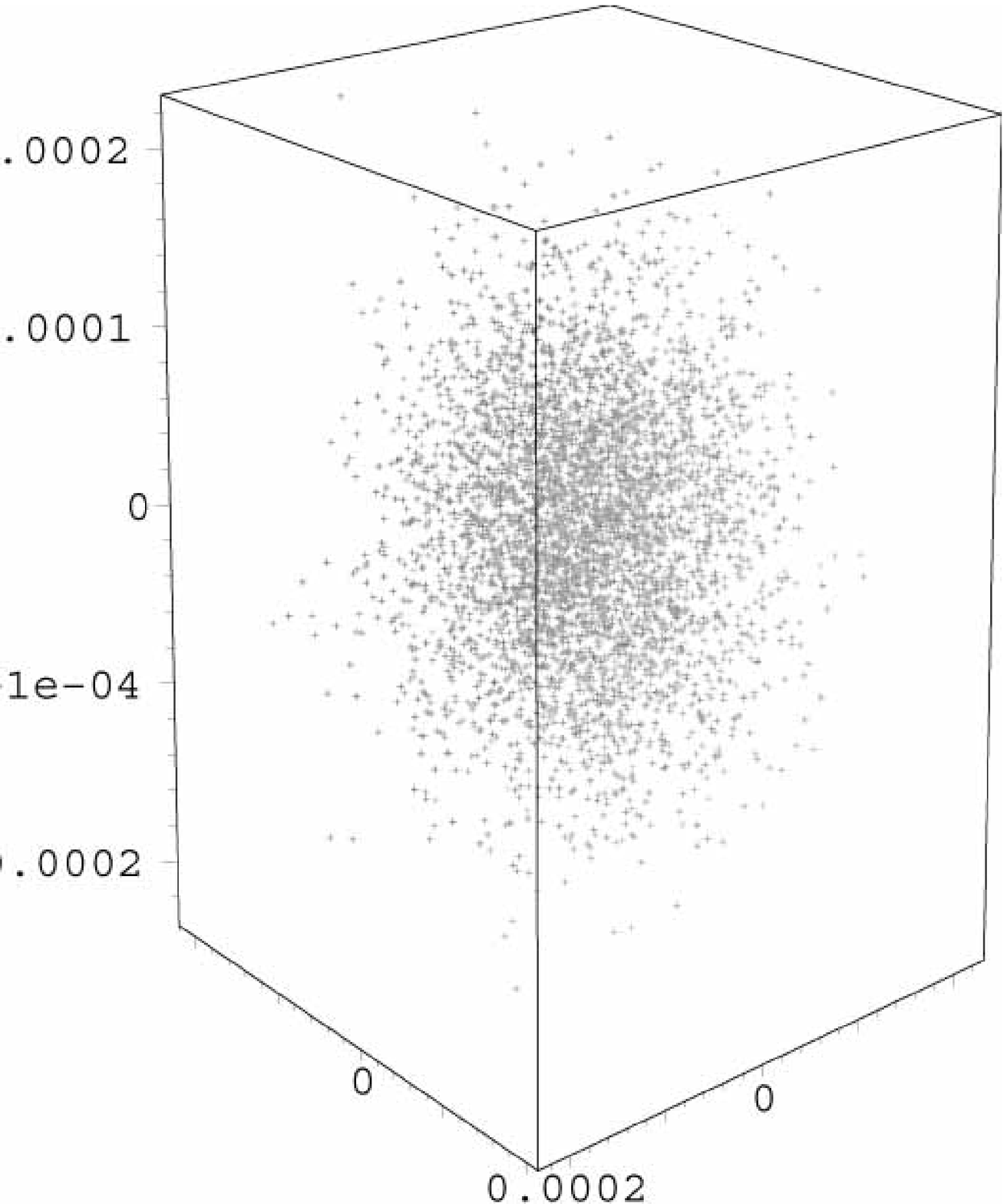}}
  \put(-40,350){\here{$\alpha=1/64$}}

 \put(20,150){\includegraphics[height=125\unitlength,width=135\unitlength]{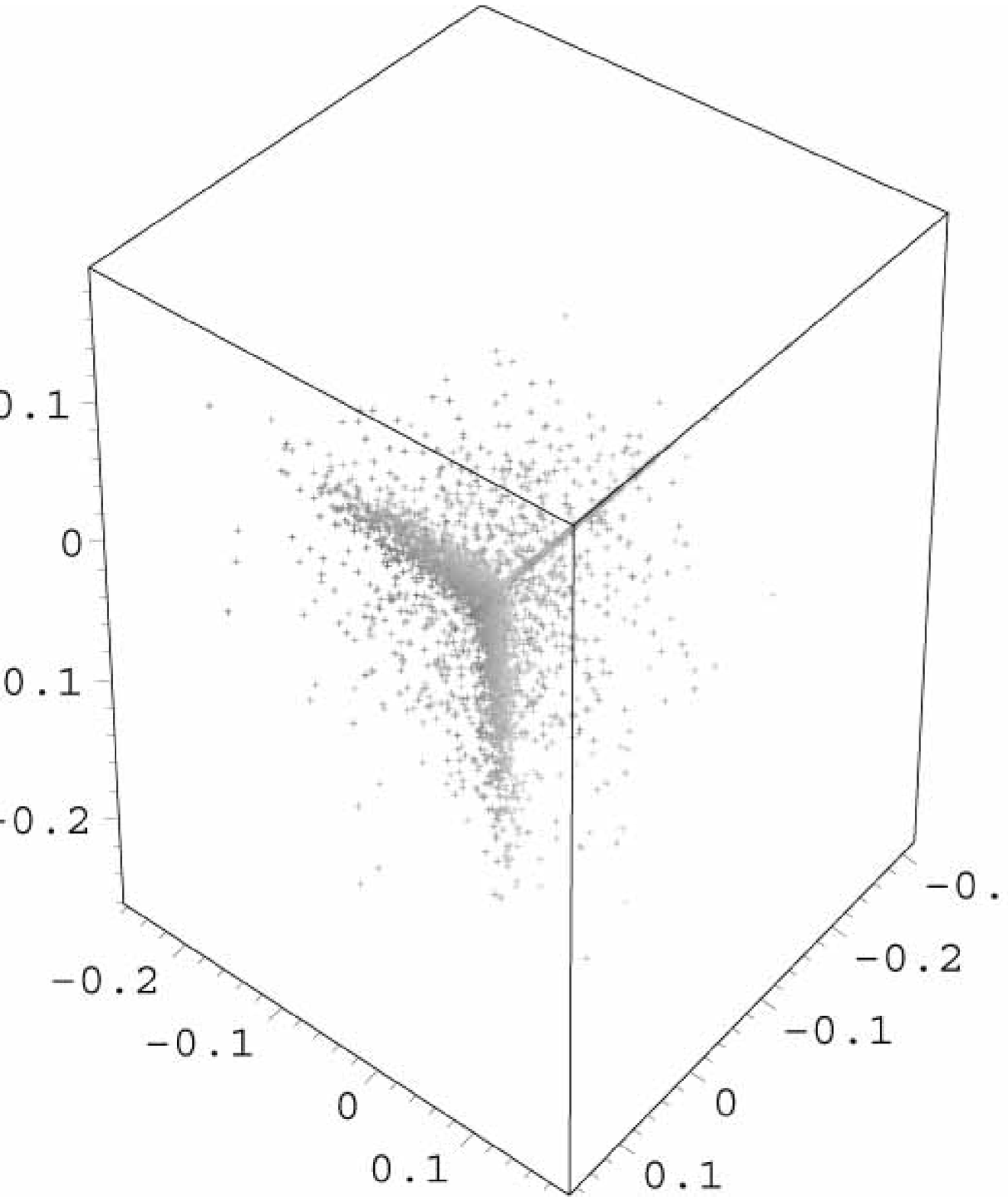}}
  \put(-40,200){\here{$\alpha=1$}}

 \put(20,0){\includegraphics[height=125\unitlength,width=135\unitlength]{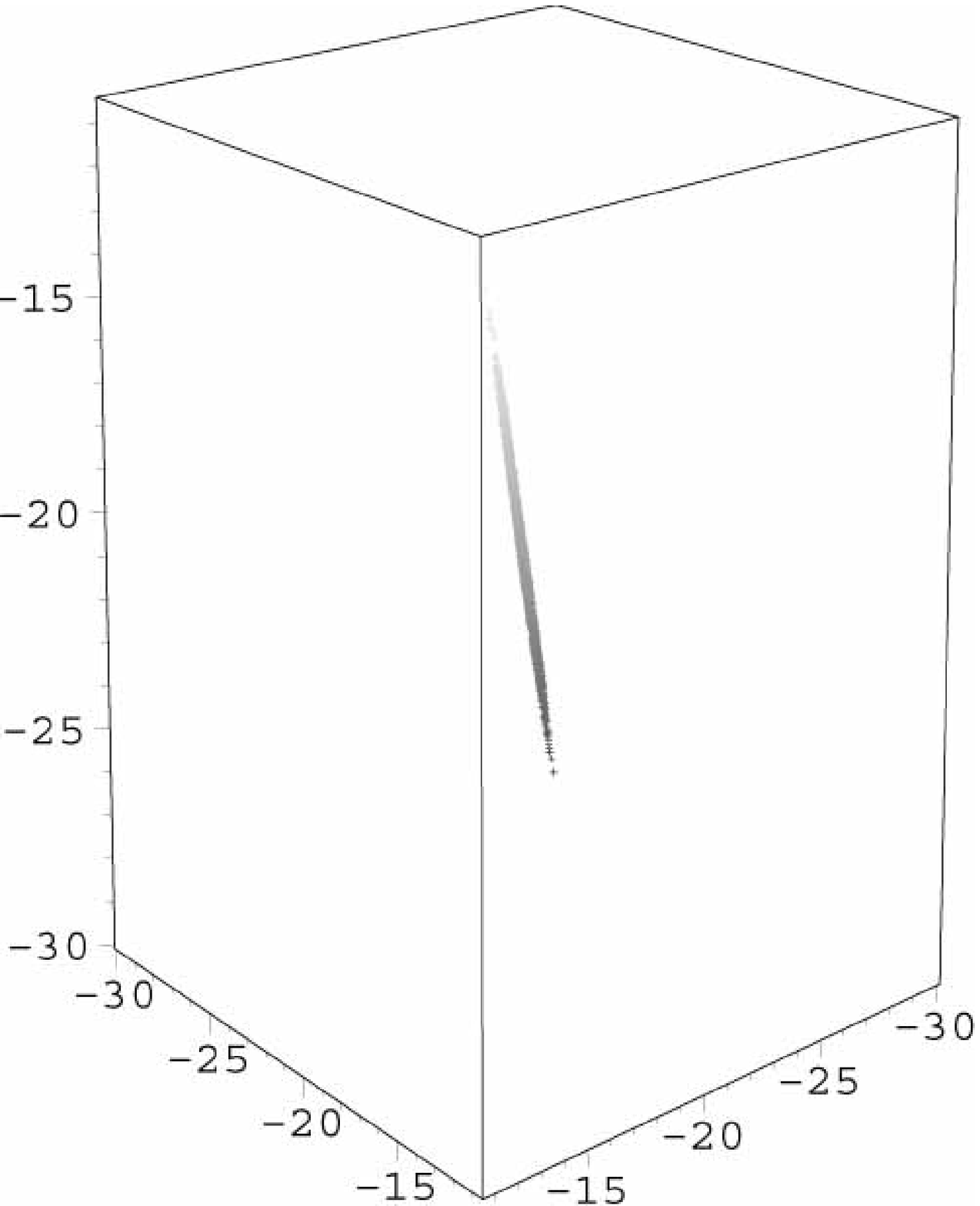}}
   \put(-40,50){\here{$\alpha=64$}}

   \put(190,300){\includegraphics[height=125\unitlength,width=135\unitlength]{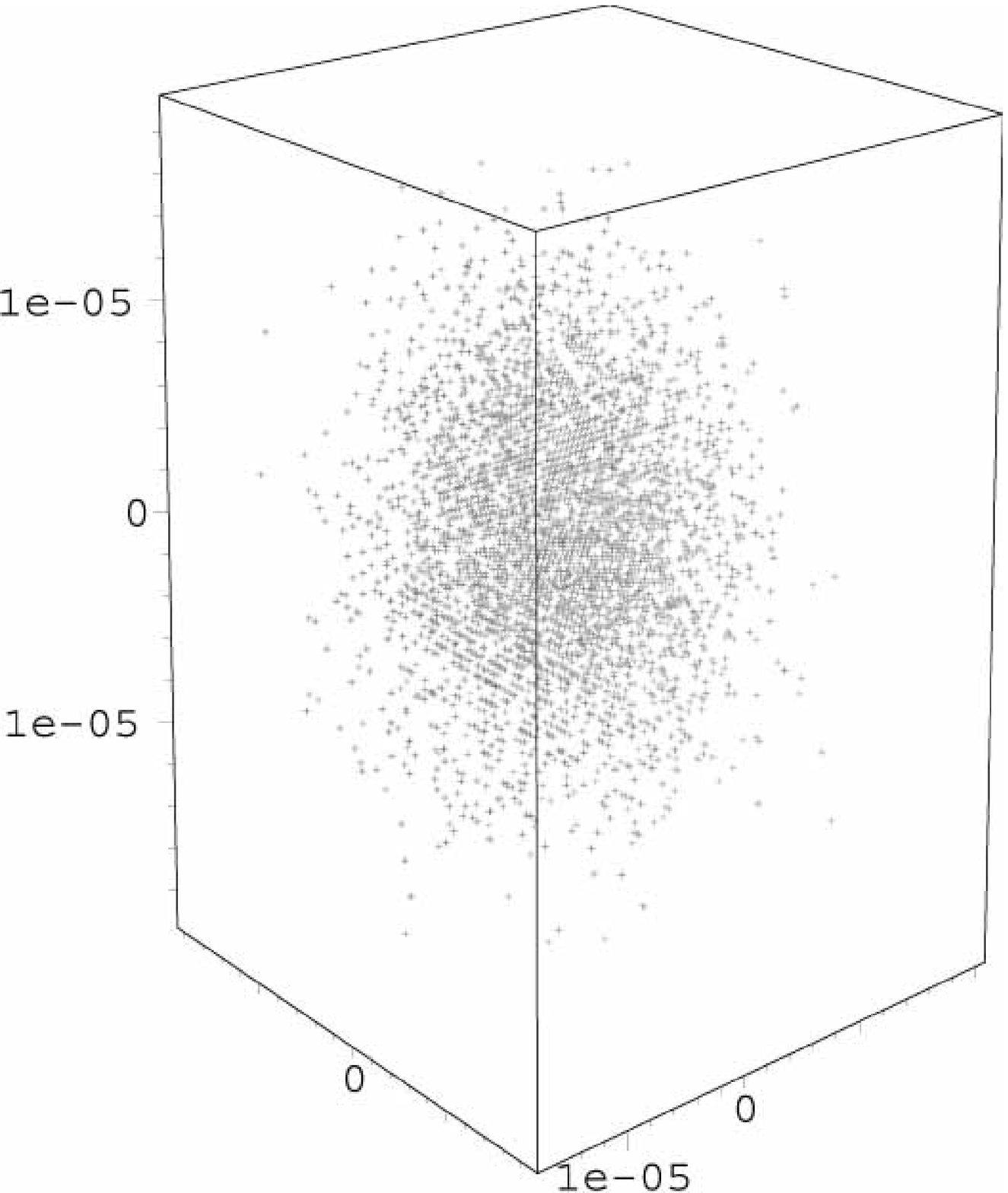}}

 \put(190,150){\includegraphics[height=125\unitlength,width=135\unitlength]{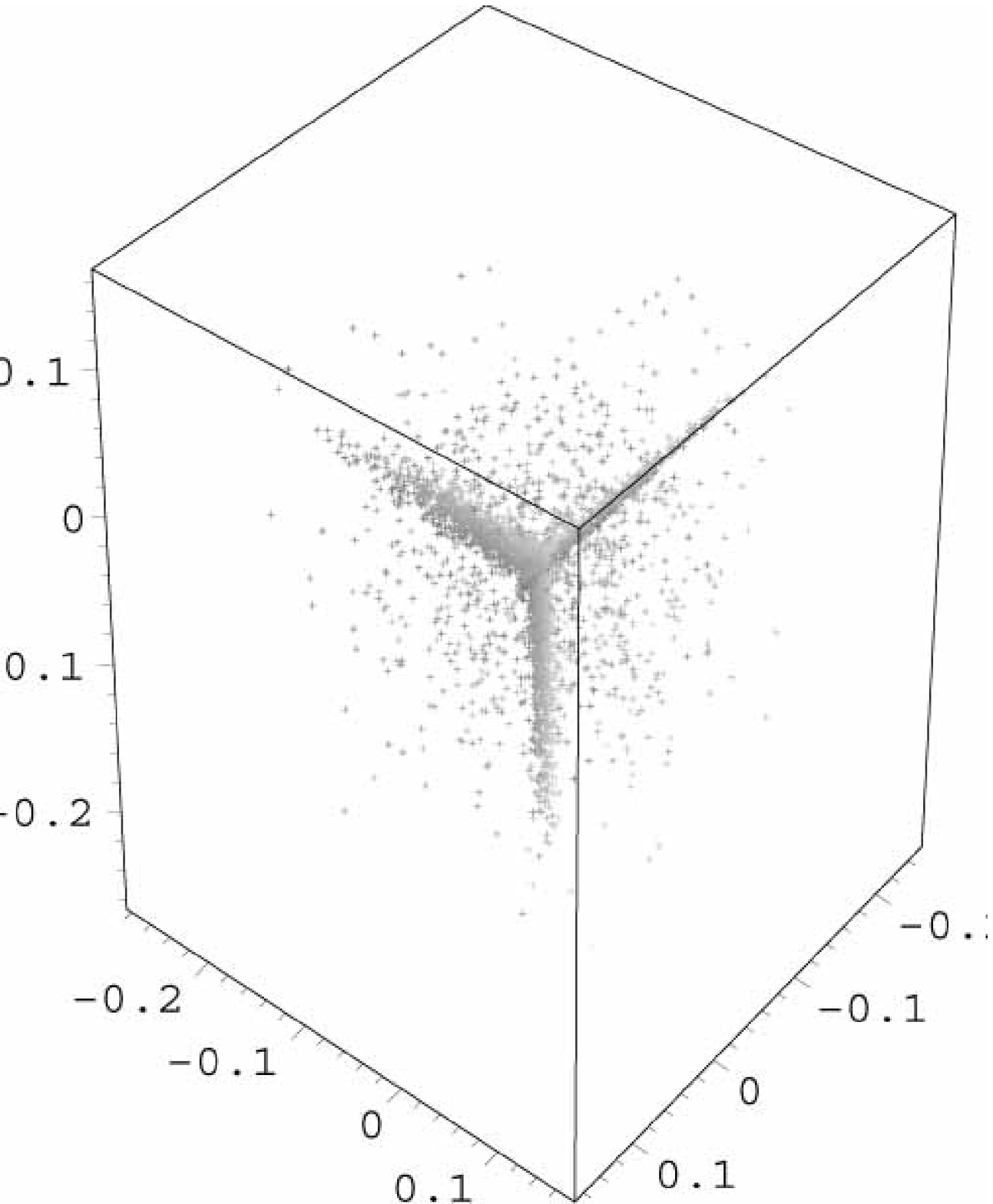}}

 \put(190,0){\includegraphics[height=125\unitlength,width=135\unitlength]{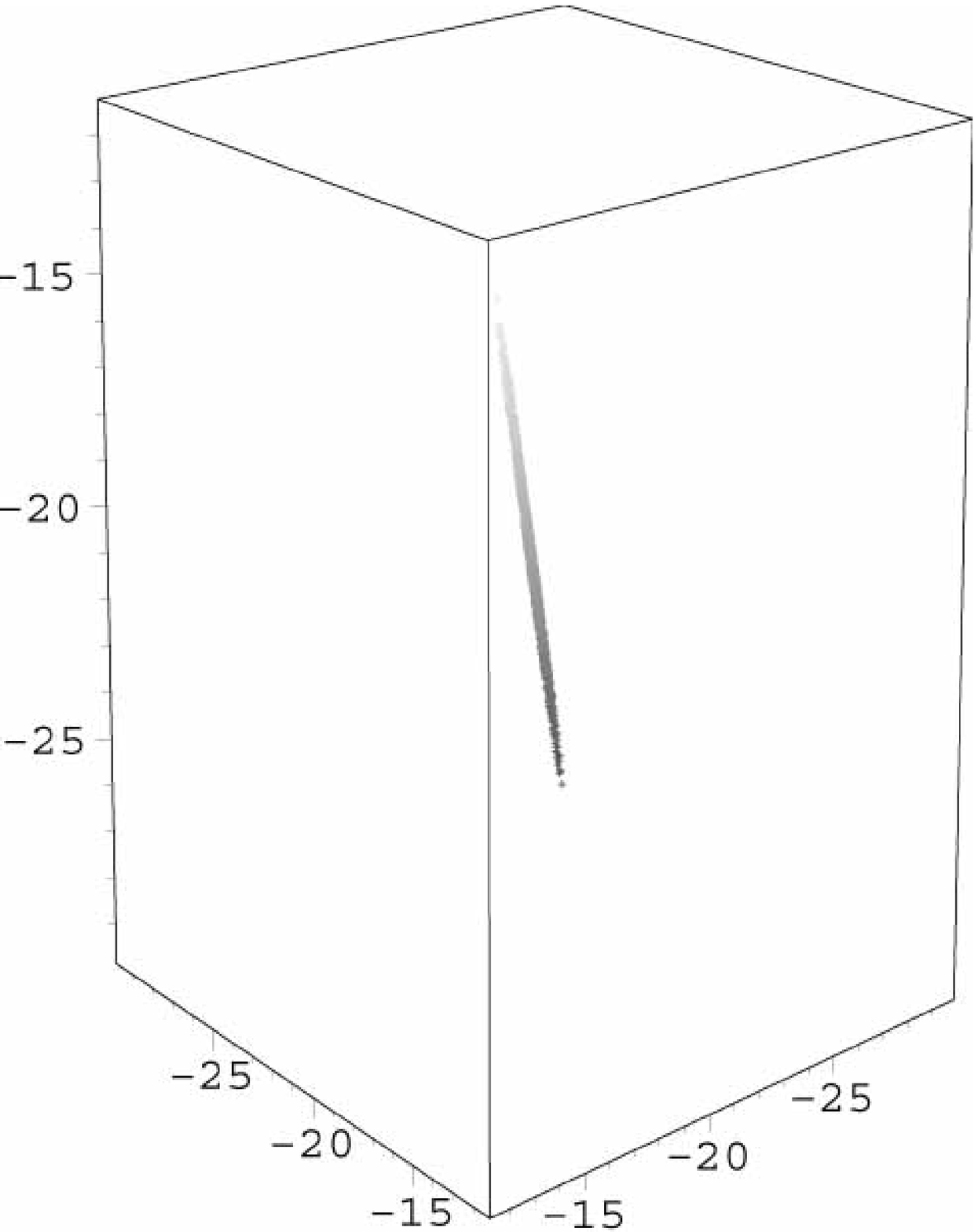}}

\end{picture}
 \vspace*{5mm}
\caption{Each plot shows the $N$ valuation velocity vectors
$(\overline{v}_1^i,\overline{v}_2^i,\overline{v}_3^i)$ as points
in $\R^3$, obtained from numerical simulations of an MG without
decision noise, with $N=4097$ and $S=3$. Left plots: unbiased
initial conditions (random initial strategy valuations drawn from
$[-10^{-4},10^{-4}]$). Right plots: biased initial conditions
(random initial strategy valuations drawn from $[-10,10]$). Note
the vastly different scales in the three graphs (increasing from
top to bottom). } \label{fig:velo_3d_S3}
\end{figure}

The dynamics of MGs is about the selection of active strategies
$a_i(t)\in\{1,\ldots,S\}$ by the agents, so let us observe in
simulations how agents select strategies for $S>2$ in the
stationary state. We define the frequencies
$f_a^i=\lim_{\tau\to\infty}\tau^{-1}\sum_{t\leq
\tau}\delta_{a,a_i(t)}$, where $f_a^i$ measures the fraction of
time during which agent $i$ played strategy $a$. Each vector ${\bf
f}_i=(f_1^i,\ldots,f_S^i)$ is a point in the $(S-1)$-dimensional
plane $\sum_{a=1}^S f_a^i=1$ in $[0,1]^S$. The collection of these
$N$ points gives a view on the collective stationary state of the
system. For $S=3$ the vectors ${\bf f}_i$ can be plotted directly
as points in $[0,1]^3$, giving rise to figures such as fig.
\ref{fig:freq_3d_S3}. We can extract relevant information from
these graphs. For large $\alpha$ the agents tend to involve all
three strategies, but not with identical frequencies (otherwise
one would have seen ${\bf
f}_i=(\frac{1}{3},\frac{1}{3},\frac{1}{3})$ for all $i$). As
$\alpha$ is reduced, the vectors ${\bf f}_i$ tend to concentrate
on the borders of the plane $\sum_{a=1}^S f_a^i=1$ in $[0,1]^S$,
which is where one of the $f_a^i$ equals zero, implying agents who
play only two of their strategies, but again not with identical
frequencies. Upon reducing $\alpha$ further we enter the
nonergodic regime (which explains the difference between the two
graphs for $\alpha=1/64$), with points either concentrating in the
corners $\{(1,0,0),(0,1,0),(0,0,1)\}$ (for biased initialization)
or on more constrained subsets where two strategies are played in
very specific combinations (for unbiased initialization). If we
measure the distribution $\varrho(f_a)=N^{-1}\sum_i
\delta(f_a-f_a^i)$ we also obtain quantitative information on the
{\em density} of points in various regions of $[0,1]^3$,
complementing the information in figure \ref{fig:freq_3d_S3}.
Typical examples are shown in  figure \ref{fig:S3freqhisto}, for
different values of $\alpha$; note that the symmetry of the
problem guarantees that for $N\to\infty$ all $S$ distributions
$\varrho(f_a)$ must be identical (this is confirmed in
simulations). Temporal information on how such macroscopic states
are realized can be obtained by showing the variables $a_i(t)$ as
functions of time. For $S=3$ and unbiased initial conditions this
has been done using grey-scale coding in figure
\ref{fig:S3evolution}, showing the difference between the small
$\alpha$ regime (left), where agents tend to alternate two of
their strategies equally, and the large $\alpha$ regime, where
agents tend to involve all three strategies, at non-uniform rates.

Strategy selections are made on the basis of strategy valuations,
which in MGs are known to grow potentially linearly with time.
Only those strategies with the largest growth rates will be
played.  If one measures the growth rates (or `strategy
velocities') of the valuations,
$\overline{v}_a^i=\tau^{-1}(v_a^i(t+\tau)-v_a^i(t))$, in the
stationary state (i.e. for large $\tau$ and $t$) for $S=3$, and
subsequently plots the $N$ velocity vectors
$(\overline{v}_1^i,\overline{v}_2^i,\overline{v}_3^i)$ as points
in $\R^3$, one obtains graphs as in figure \ref{fig:velo_3d_S3}.
For small $\alpha$ all velocities are concentrated very close to
the origin; as $\alpha$ is increased they become consistently more
negative, with concentration of points with two identical
components at intermediate $\alpha$ (consistent with agents
playing two strategies only) and concentration of points along the
diagonal where all three components are identical for large
$\alpha$ (consistent with agents playing all three strategies).
\vsp

\begin{figure}[t]
\vspace*{-3mm} \hspace*{40mm} \setlength{\unitlength}{0.30mm}
\begin{picture}(300,450)

   \put(20,300){\includegraphics[height=125\unitlength,width=135\unitlength]{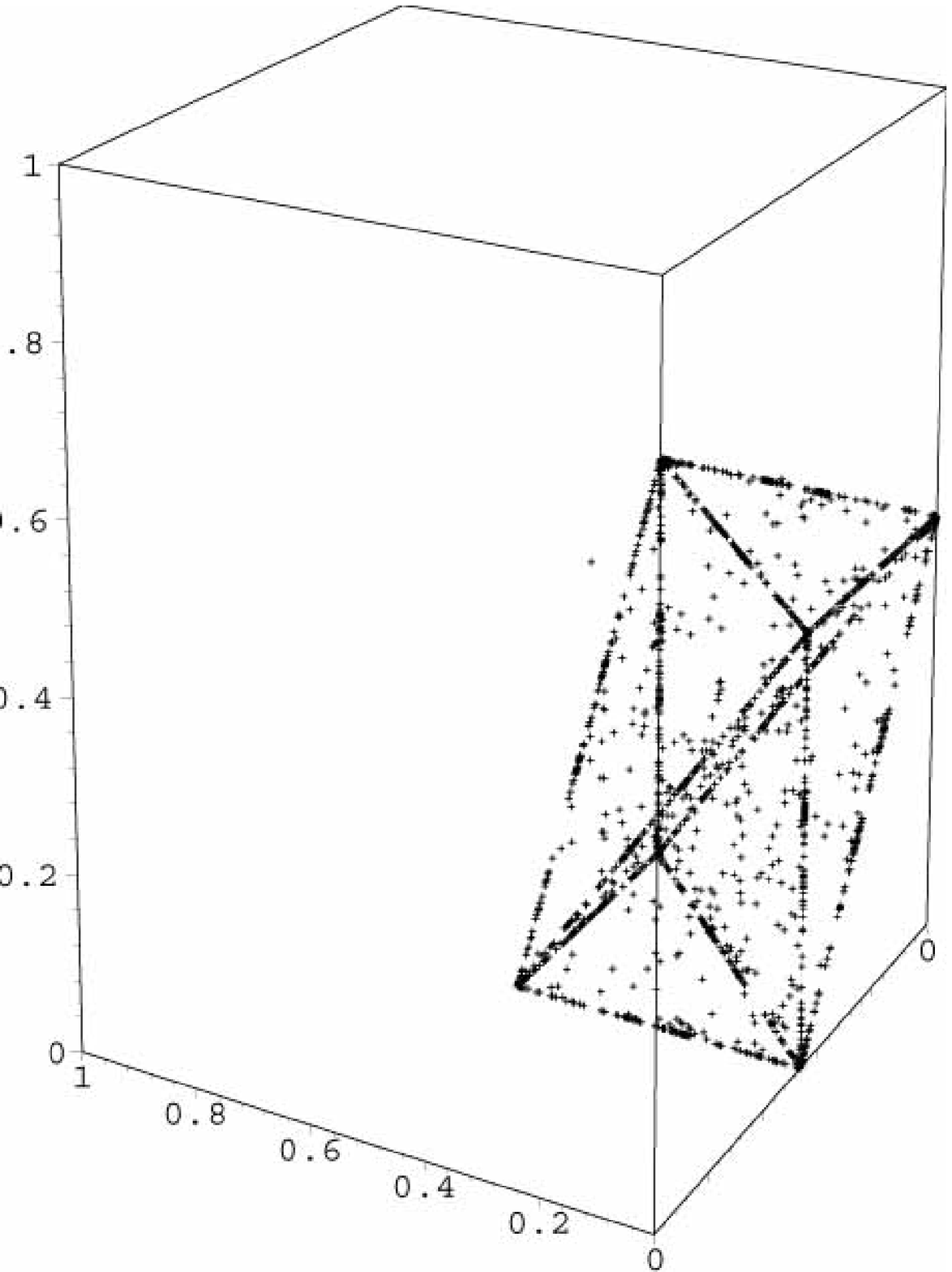}}
  \put(-40,350){\here{$\alpha=1/64$}}

 \put(20,150){\includegraphics[height=125\unitlength,width=135\unitlength]{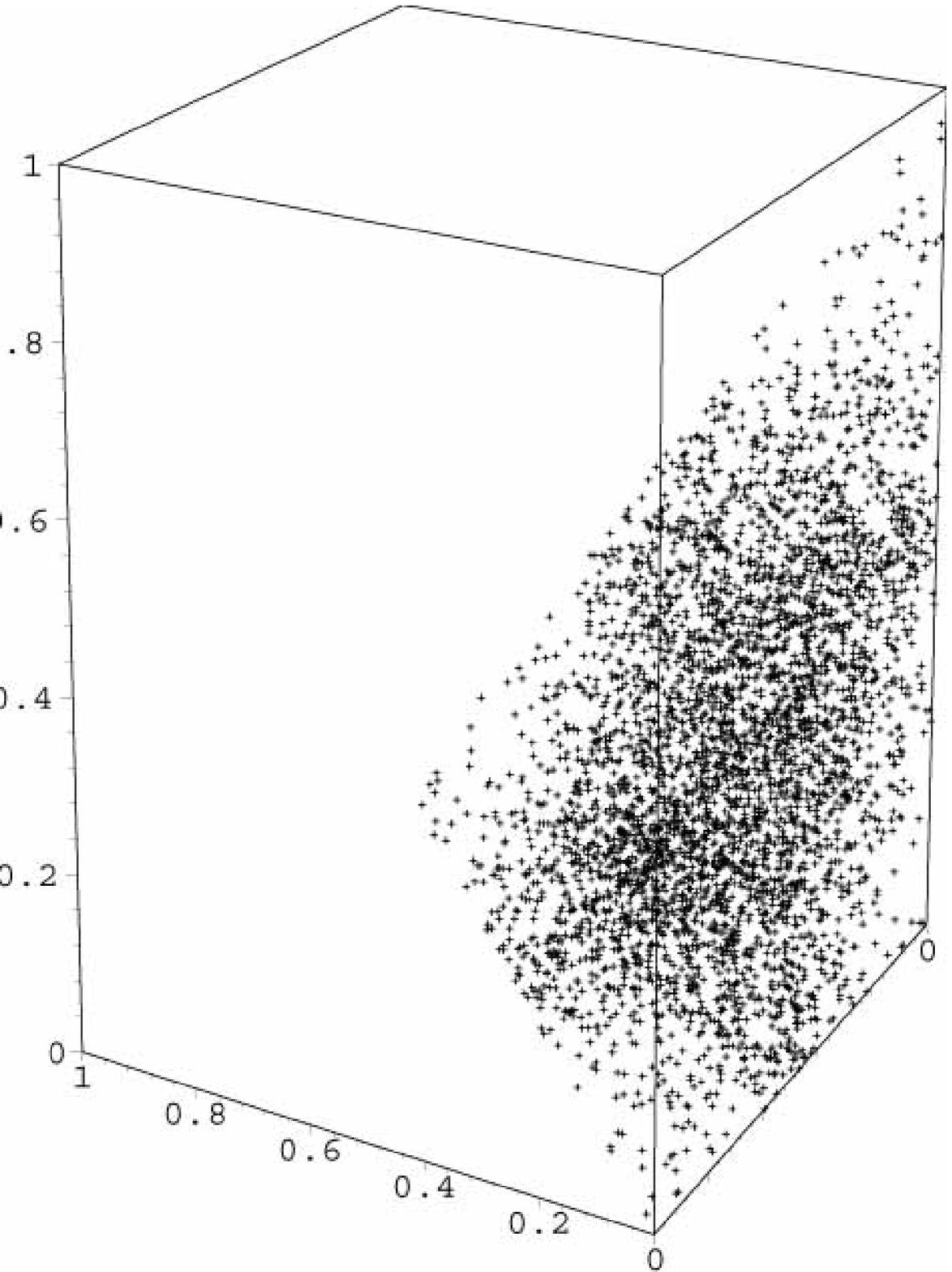}}
  \put(-40,200){\here{$\alpha=1$}}

 \put(20,0){\includegraphics[height=125\unitlength,width=135\unitlength]{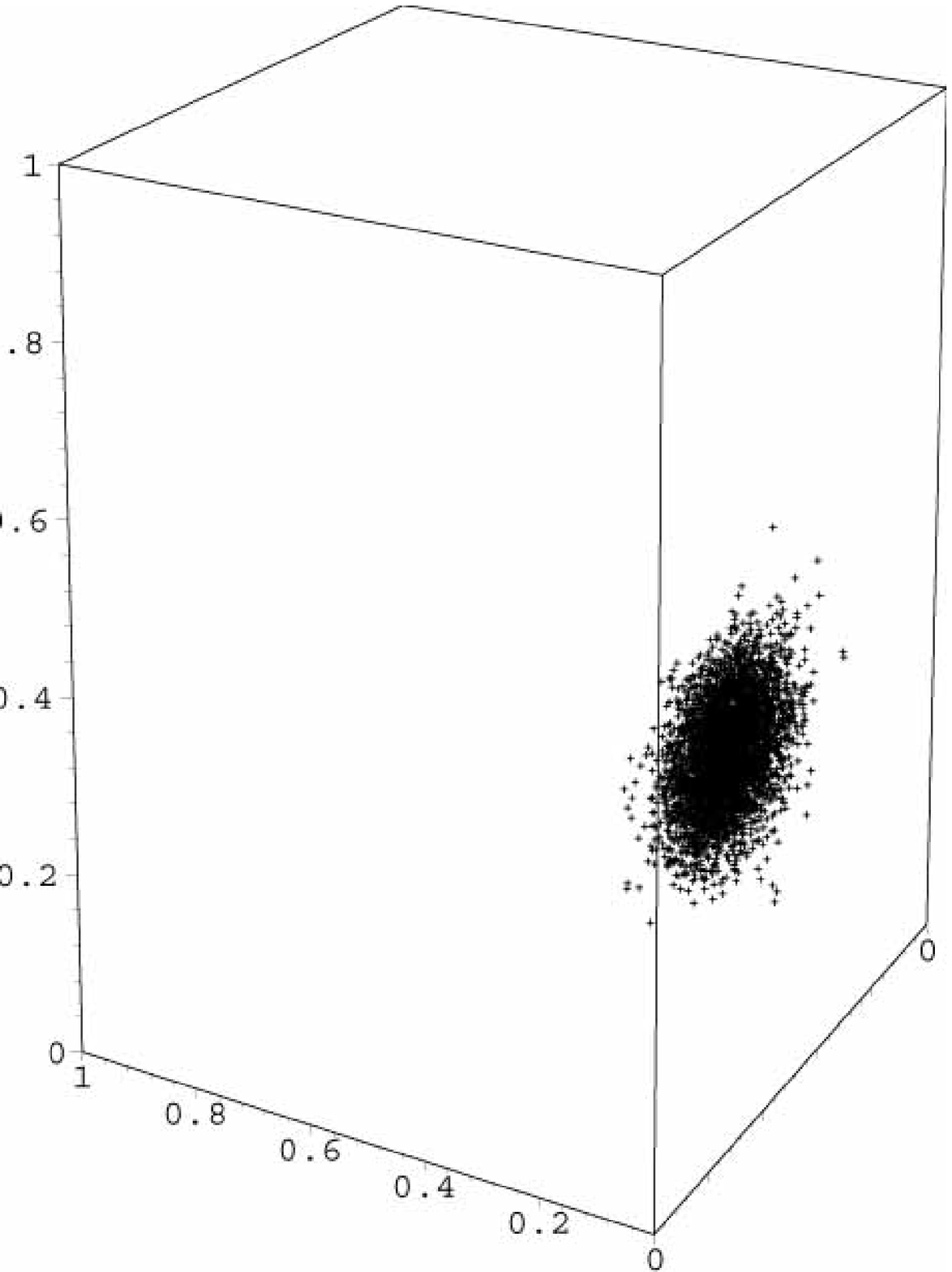}}
   \put(-40,50){\here{$\alpha=64$}}

   \put(190,300){\includegraphics[height=125\unitlength,width=135\unitlength]{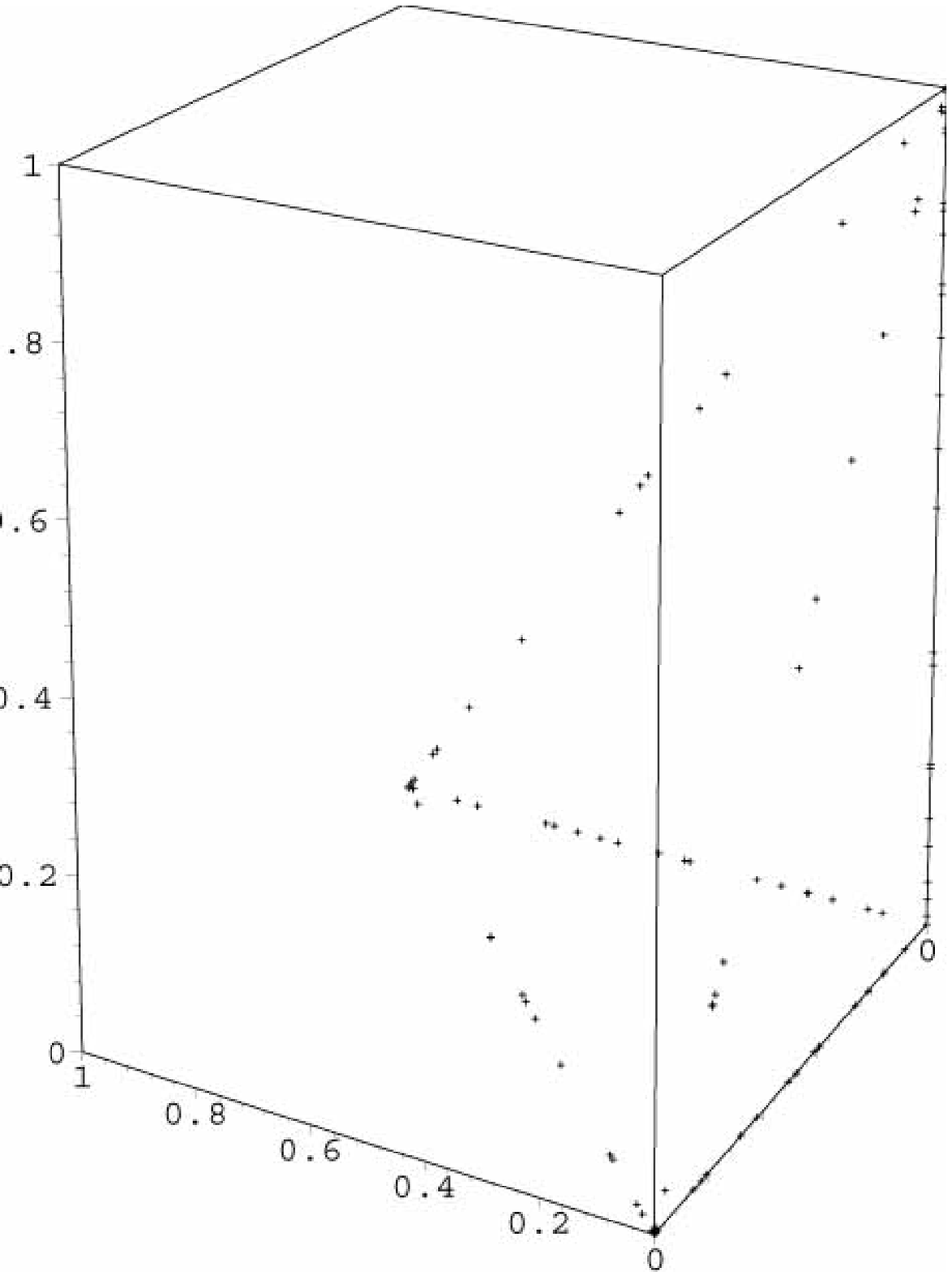}}

 \put(190,150){\includegraphics[height=125\unitlength,width=135\unitlength]{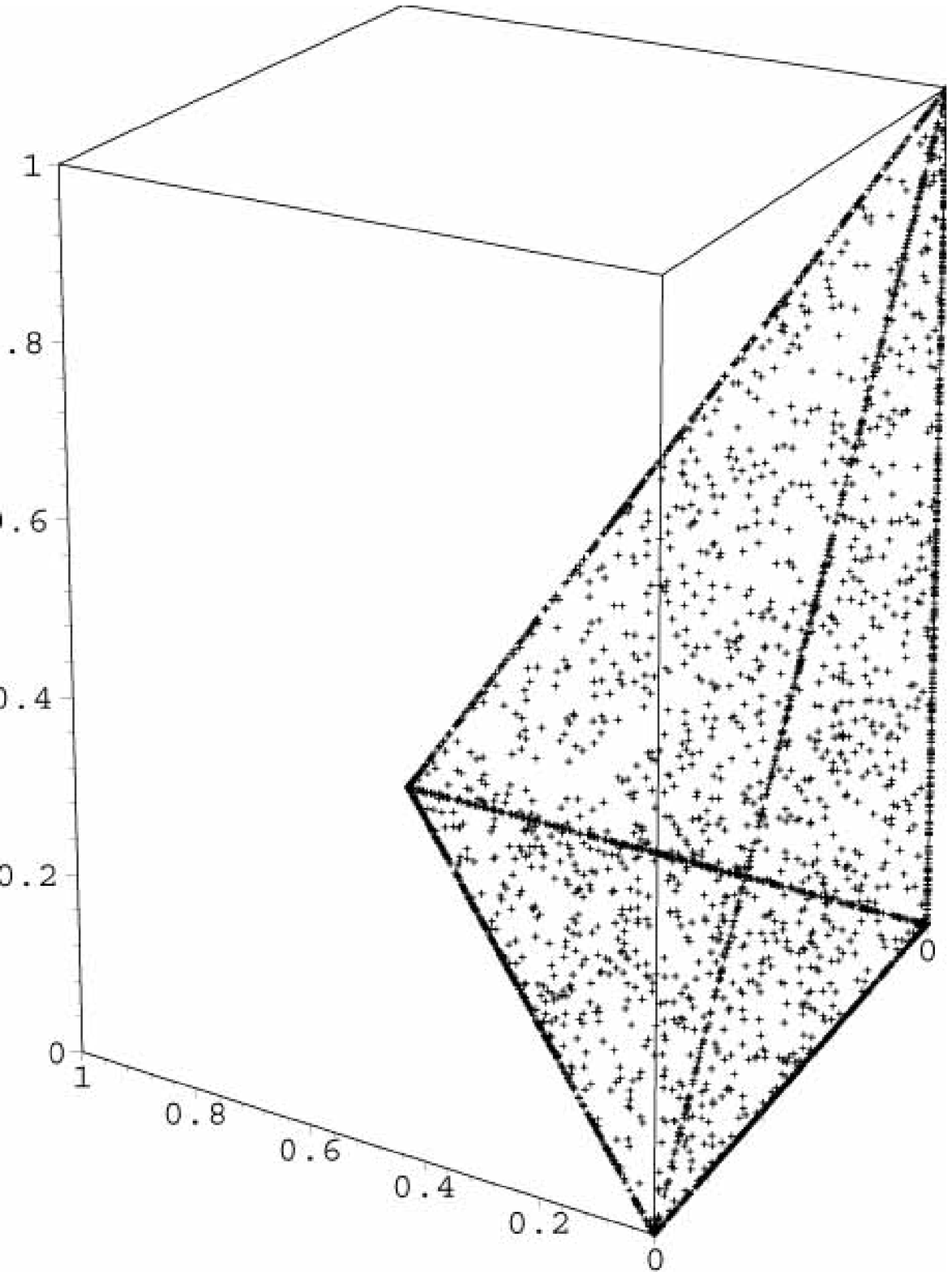}}

 \put(190,0){\includegraphics[height=125\unitlength,width=135\unitlength]{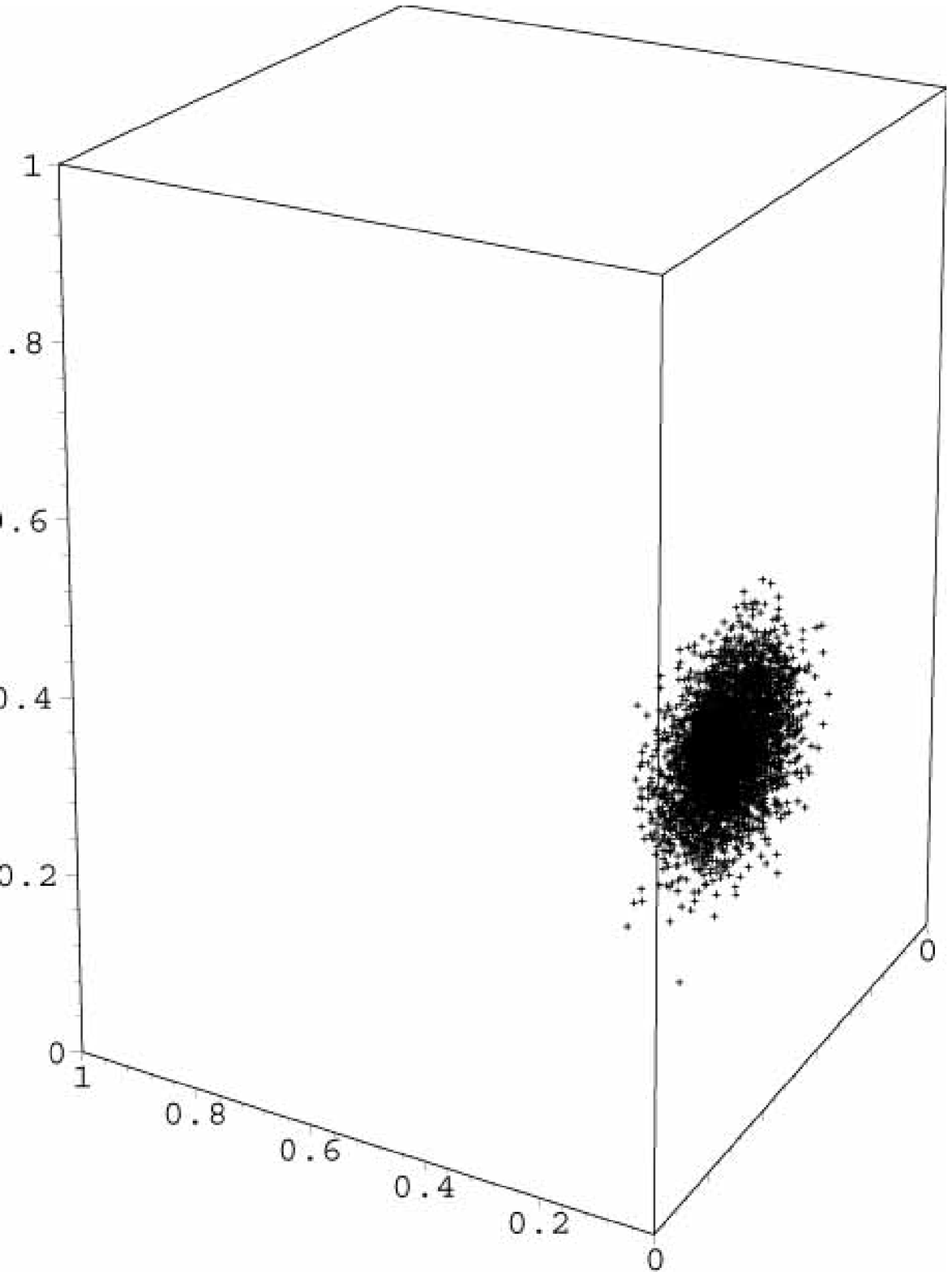}}

\end{picture}
 \vspace*{5mm}
\caption{The plots show the first three components of the
frequency vectors ${\bf f}_i=(f_1^i,f_2^i,f_3^i,f_4^i)$ with which
the agents use their three strategies, drawn for each agent $i$ as
a point in $[0,1]^3$ (giving $N$ points per plot), obtained from
numerical simulations of an MG without decision noise, with
$N=4097$ and $S=4$. The constraint $\sum_a f_a^i=1$ for all $i$
now implies that all points are in a the hyper-plane of which in
the present graph one sees a 3-dimensional projection. Left plots:
unbiased initial conditions (random initial strategy valuations
drawn from $[-10^{-4},10^{-4}]$). Right plots: biased initial
conditions (random initial strategy valuations drawn from
$[-10,10]$). } \label{fig:freq_3d_S4} \vspace*{-2mm}
\end{figure}

\begin{figure}[t]
\vspace*{-6mm} \hspace*{15mm} \setlength{\unitlength}{0.40mm}
\begin{picture}(290,200)
  \put(0,0){\epsfxsize=160\unitlength\epsfbox{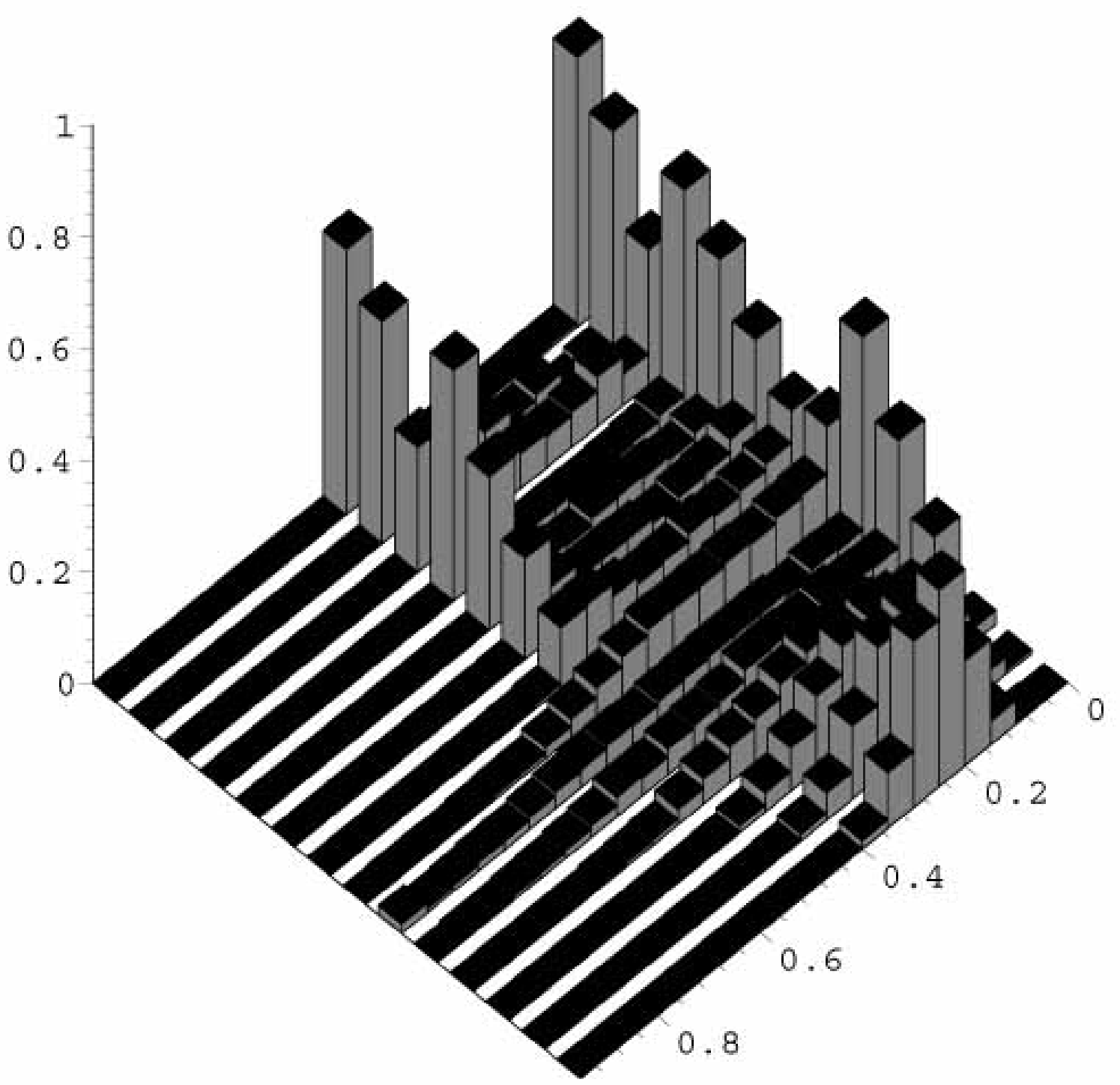}}
  \put(125,35){\here{$f_1$}}\put(36,36){\here{$\alpha$}}
  \put(-20,110){$\varrho(f_1)$}
    \put(200,0){\epsfxsize=160\unitlength\epsfbox{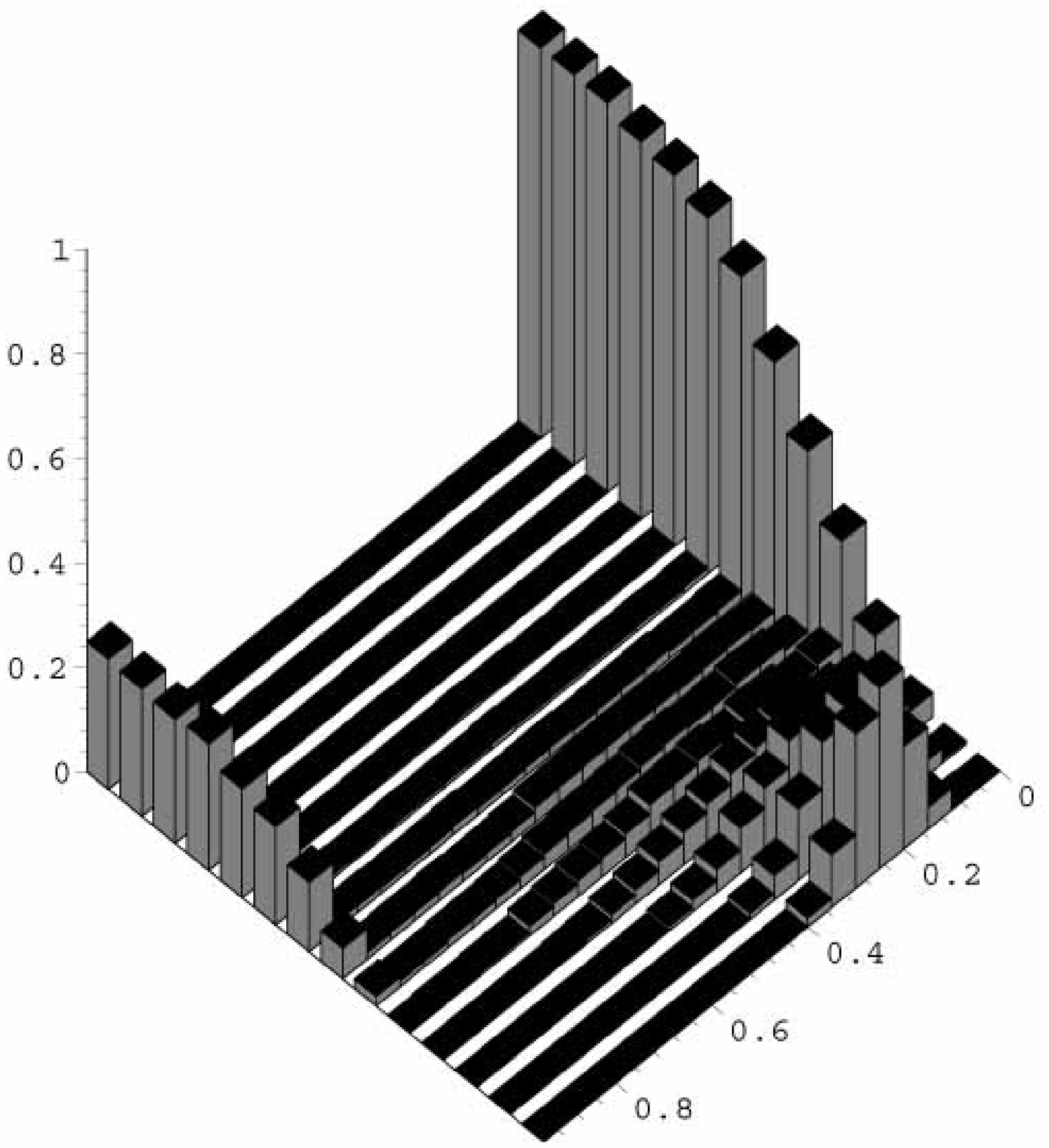}}
  \put(325,35){\here{$f_1$}}\put(236,36){\here{$\alpha$}}
  \put(180,110){$\varrho(f_1)$}
\end{picture}
\vspace*{-3mm} \caption{Histograms of the fraction $\varrho(f_1)$
of agents that play strategy 1 with frequency $f$ in the
stationary state  (observed in simulations with $N=4097$ and
$S=4$), for $\alpha\in\{1/128,1/64,\ldots,32,64\}$ (increasing by
a factor 2 at each step).
 Left: unbiased
initial conditions (random initial strategy valuations drawn from
$[-10^{-4},10^{-4}]$). Right: biased initial conditions (random
initial strategy valuations drawn from $[-10,10]$). }
 \label{fig:S4freqhisto}
\end{figure}

\begin{figure}[t]
\vspace*{-3mm} \hspace*{40mm} \setlength{\unitlength}{0.30mm}
\begin{picture}(300,450)

   \put(20,300){\includegraphics[height=125\unitlength,width=135\unitlength]{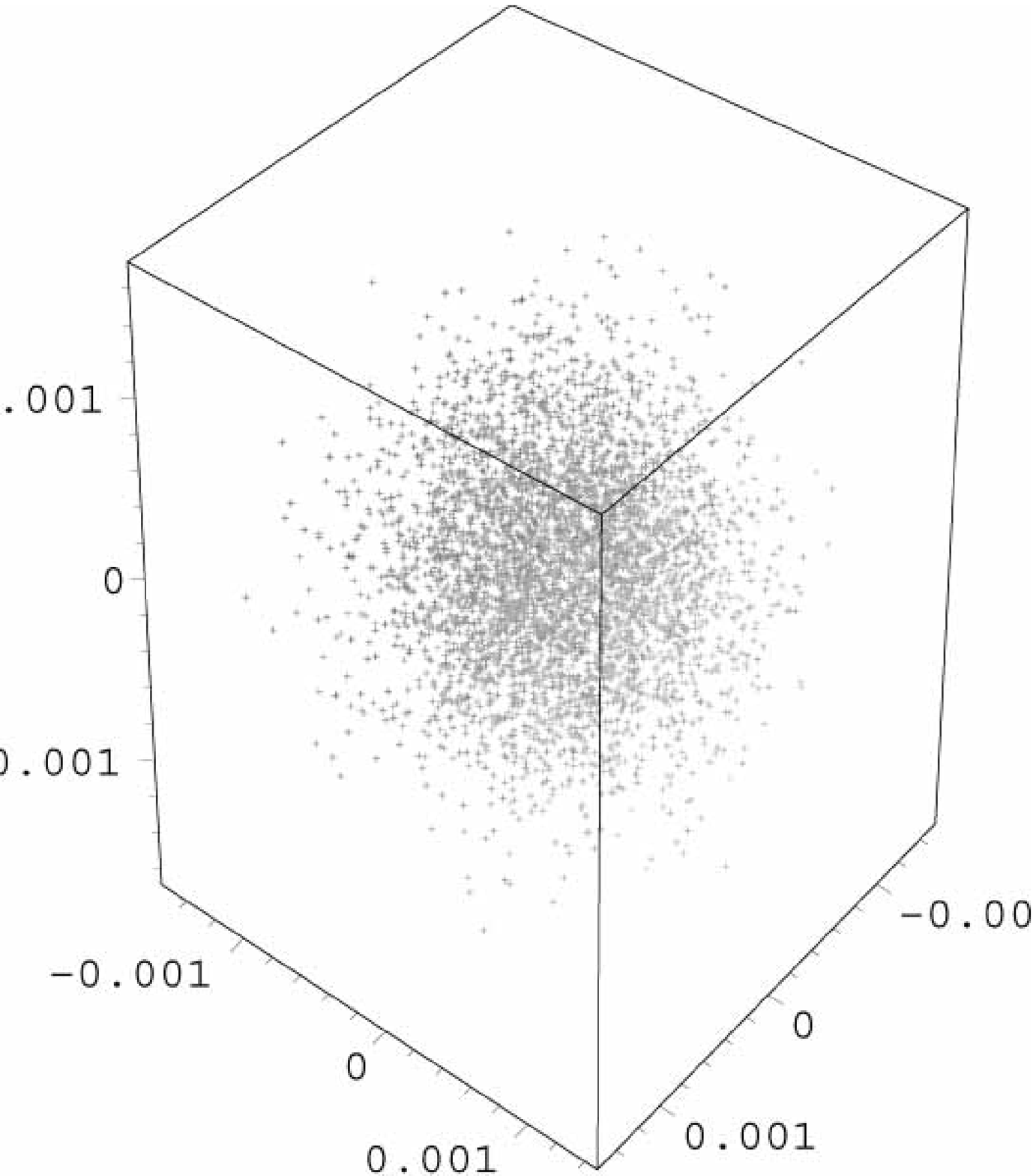}}
  \put(-40,350){\here{$\alpha=1/64$}}

 \put(20,150){\includegraphics[height=125\unitlength,width=135\unitlength]{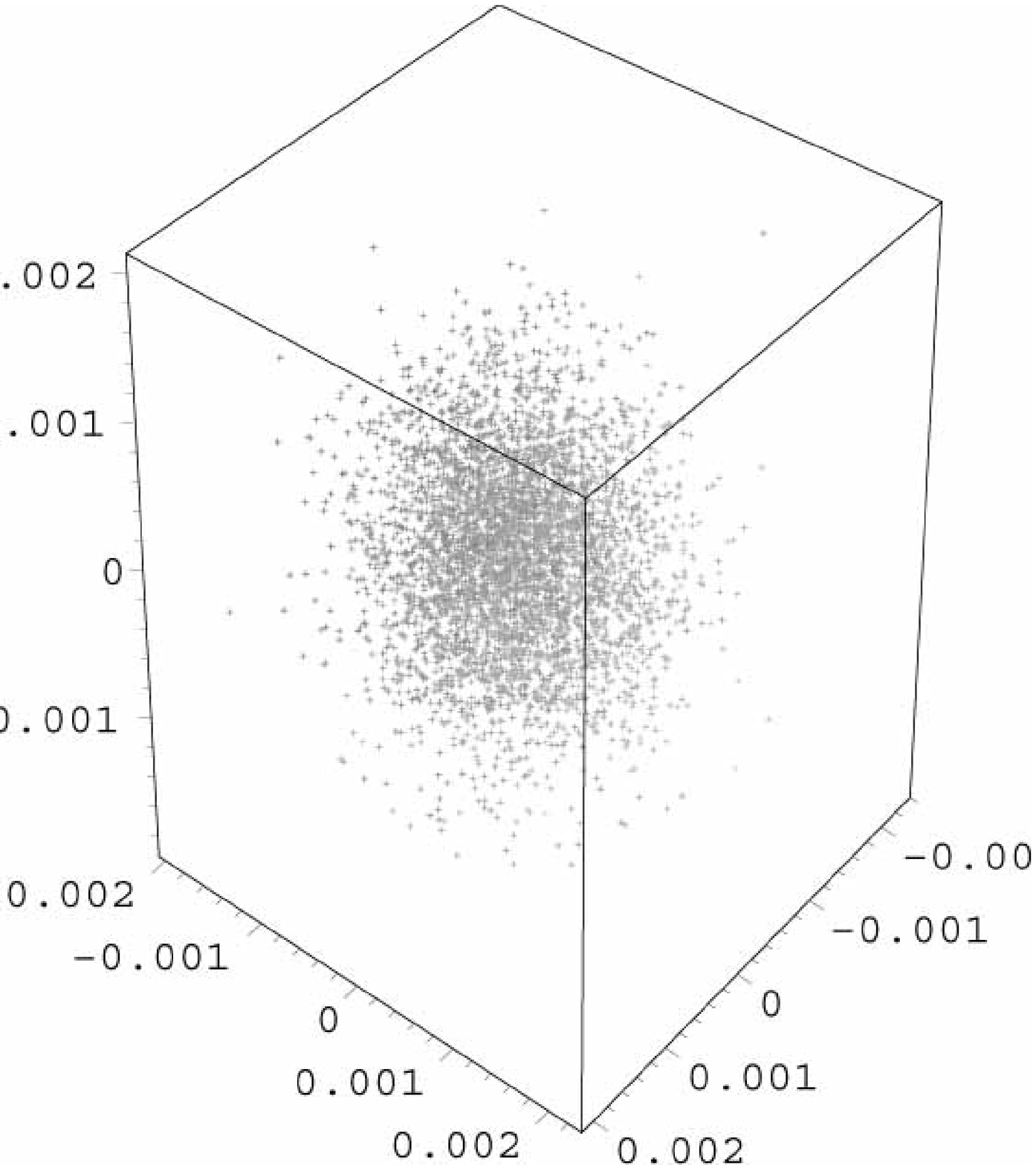}}
  \put(-40,200){\here{$\alpha=1$}}

 \put(20,0){\includegraphics[height=125\unitlength,width=135\unitlength]{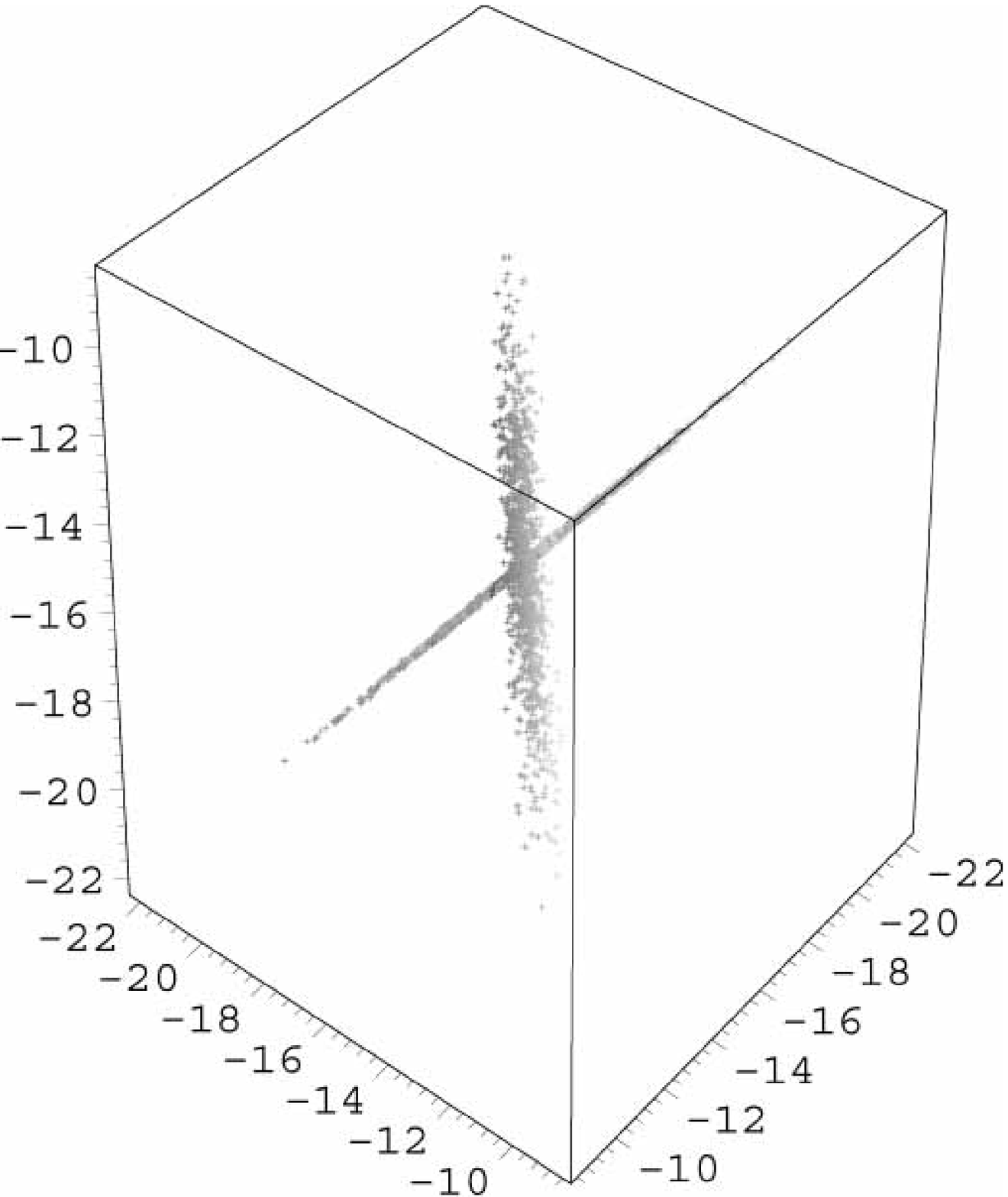}}
   \put(-40,50){\here{$\alpha=64$}}

   \put(190,300){\includegraphics[height=125\unitlength,width=135\unitlength]{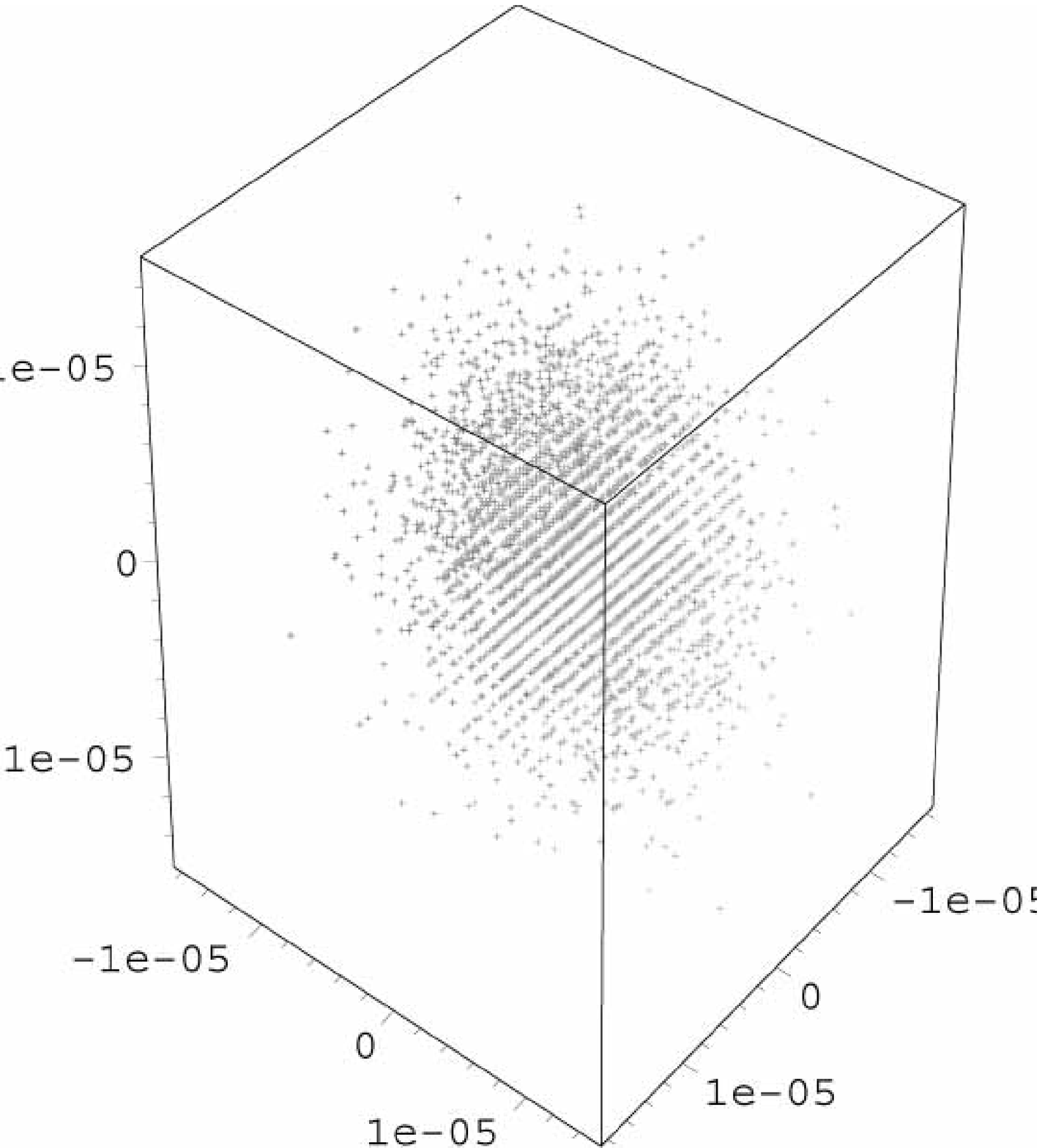}}

 \put(190,150){\includegraphics[height=125\unitlength,width=135\unitlength]{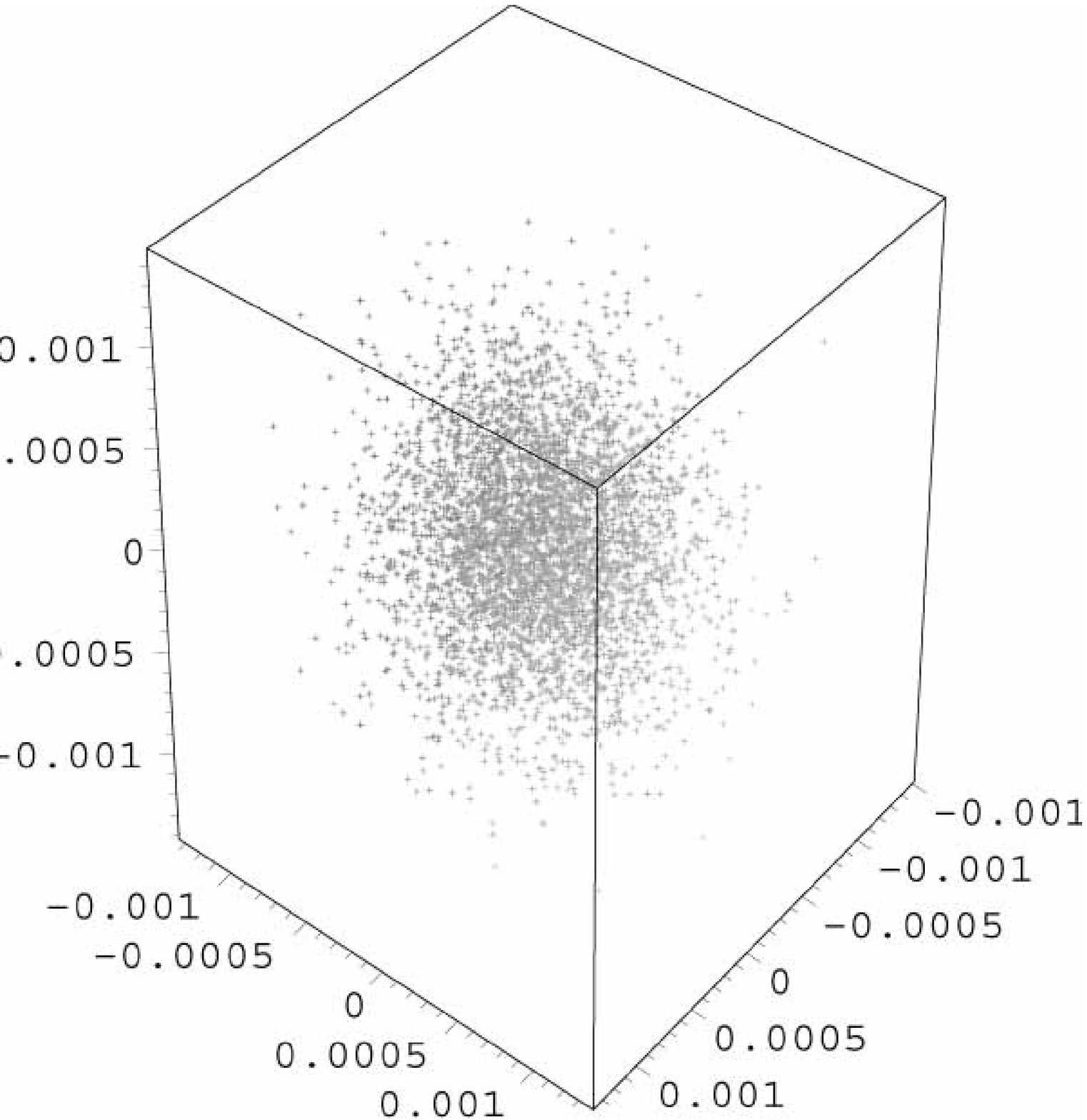}}

 \put(190,0){\includegraphics[height=125\unitlength,width=135\unitlength]{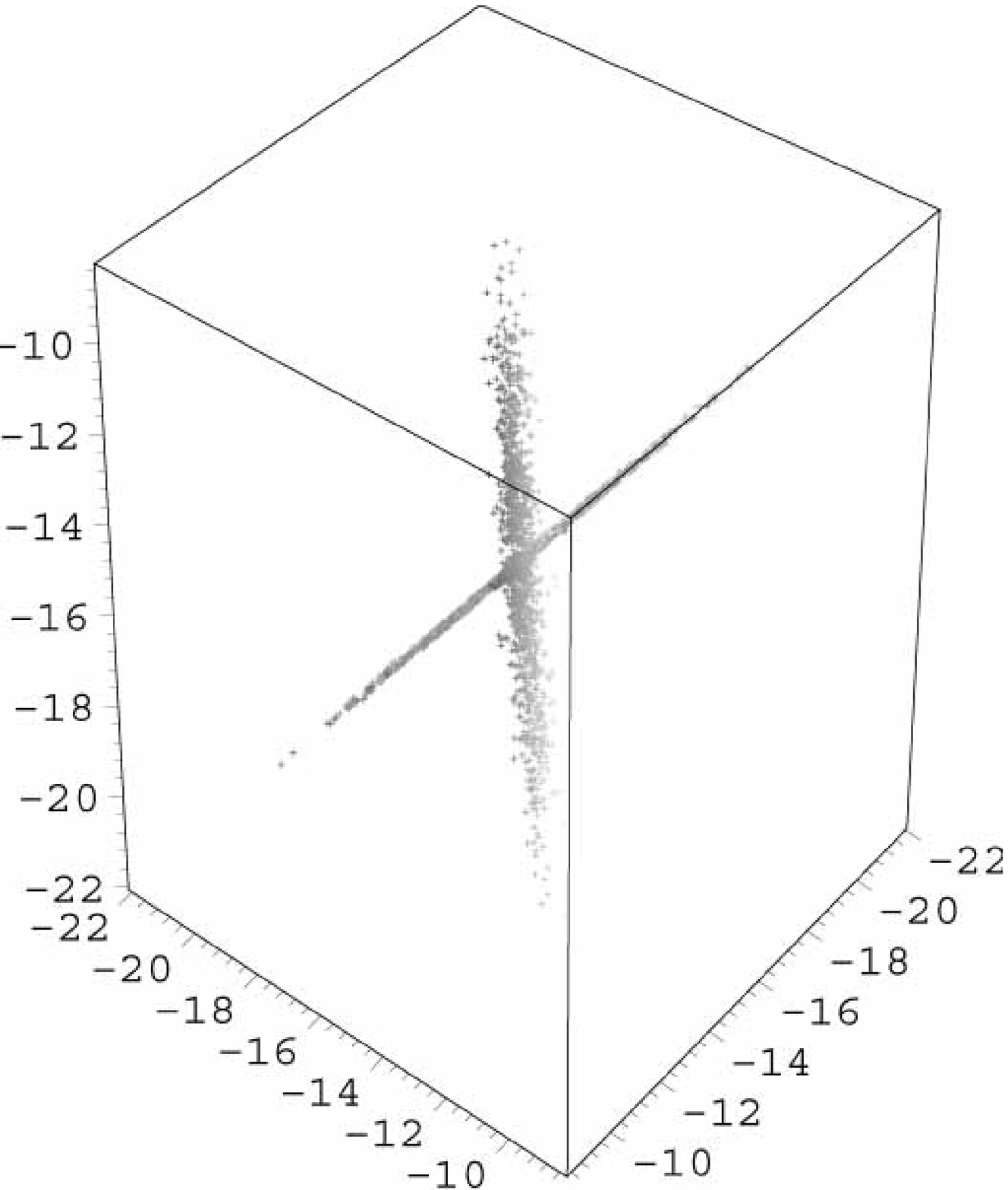}}

\end{picture}
 \vspace*{5mm}
\caption{ Each plot shows the first three components
$(\overline{v}_1^i,\overline{v}_2^i,\overline{v}_3^i)$ of the $N$
valuation velocity vectors as points in $\R^3$, obtained from
numerical simulations of an MG without decision noise, with
$N=4097$ and $S=4$. Left plots: unbiased initial conditions
(random initial strategy valuations drawn from
$[-10^{-4},10^{-4}]$). Right plots: biased initial conditions
(random initial strategy valuations drawn from $[-10,10]$). Note
the different scales in the three graphs (increasing from top to
bottom).} \label{fig:velo_3d_S4}
\end{figure}

For $S=4$ one  can obviously no longer plot the frequency vectors
${\bf f}_i=(f_1^i,f_2^i,f_3^i,f_4^i)$ as points in $[0,1]^3$, but
one has to resort to projection from 4D to 3D: we plot in figure
\ref{fig:freq_3d_S4} only the first three components
$(f_1^i,f_2^i,f_3^i)$. The corresponding strategy frequency
distributions  are shown in figure \ref{fig:S4freqhisto}. Once
more we observe the ergodicity/nonergodicity phase transition
(here occurring for a larger value of $\alpha$ than was the case
at $S=3$), and the tendency at small $\alpha$ for agents to play
only a specific subset of their four strategies. For unbiased
initial conditions the agents are seen to play in the nonergodic
regime always either one or two of their strategies; this curious
tendency, for which there is no immediate obvious explanation, is
found also for larger values of $S$ (we have confirmed this for
values up to $S=6$). Plotting the first three components of the
$N$ valuation velocity vectors
$(\overline{v}_1^i,\overline{v}_2^i,\overline{v}_3^i,\overline{v}_4^i)$
leads to the graphs shown in figure  \ref{fig:velo_3d_S4}. As with
$S=3$ we see the agents playing all $S$ strategies for large
$\alpha$, but selecting specific subsets for smaller $\alpha$,
generally with nontrivial frequencies. Here there are more options
than at $S=3$ for doing so: agents can and will go for either one,
two, three or four strategies.

\begin{figure}[t]
\vspace*{-9mm} \hspace*{15mm} \setlength{\unitlength}{0.40mm}
\begin{picture}(290,200)
  \put(0,0){\epsfxsize=160\unitlength\epsfbox{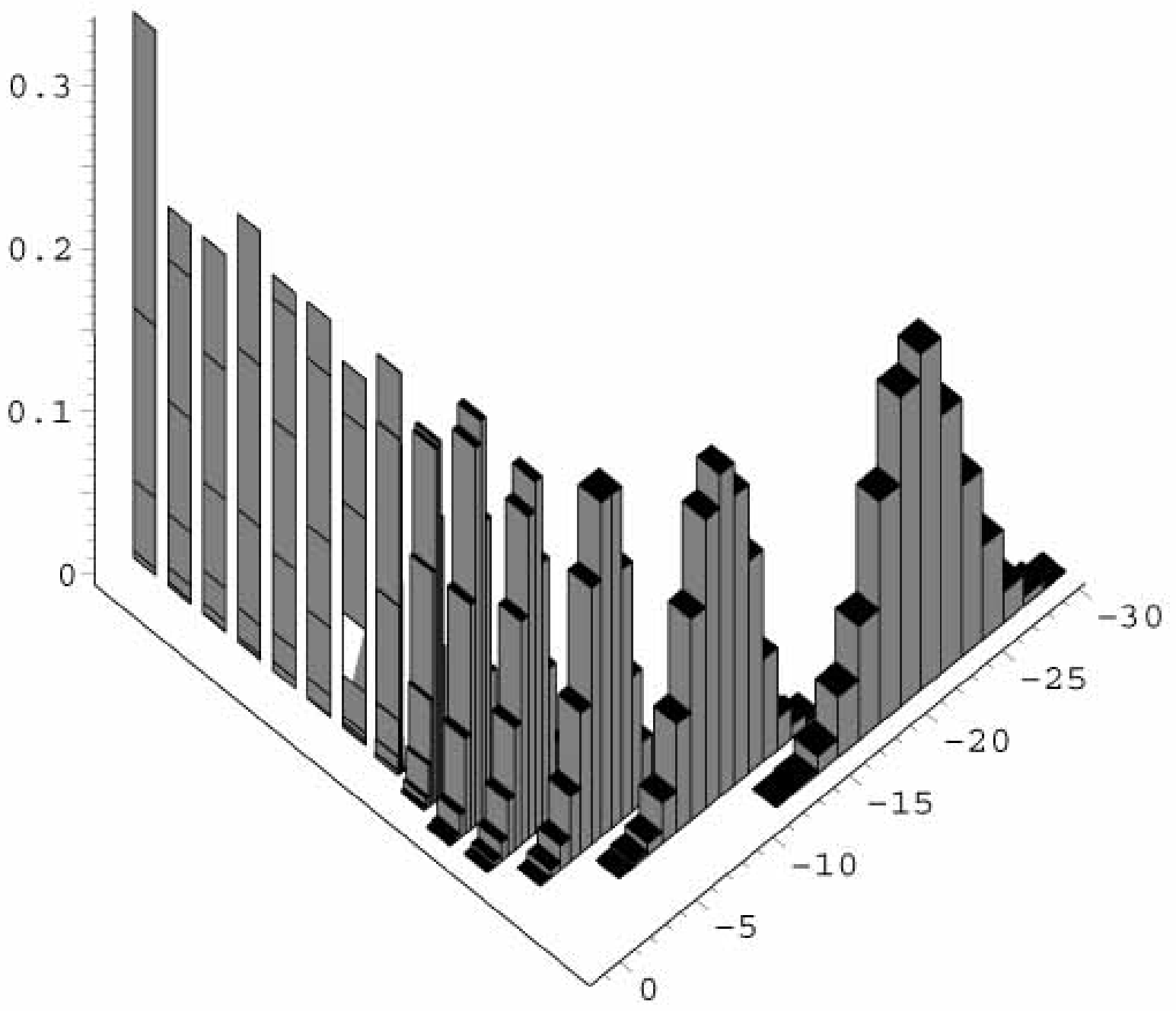}}
  \put(129,35){\here{$\overline{v}_1$}}\put(36,36){\here{$\alpha$}}
  \put(-20,110){$\varrho(\overline{v}_1)$}
    \put(200,0){\epsfxsize=160\unitlength\epsfbox{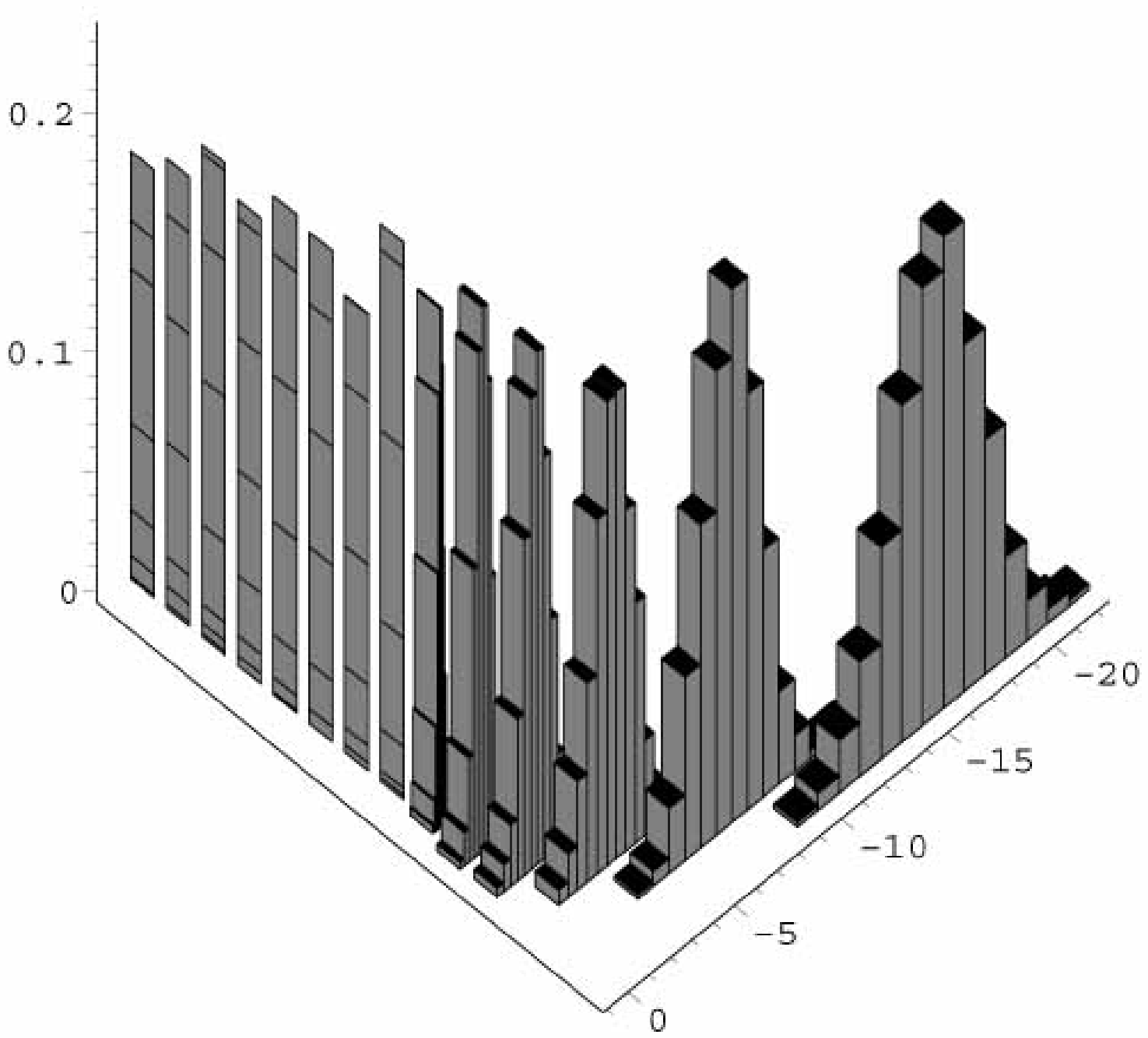}}
  \put(329,35){\here{$\overline{v}_1$}}\put(236,36){\here{$\alpha$}}
  \put(180,110){$\varrho(\overline{v}_1)$}
\end{picture}
\vspace*{-3mm} \caption{Histograms of the fraction
$\varrho(\overline{v}_1)$ of agents that have strategy velocity
$\overline{v}_1$ for strategy 1 in the stationary state (observed
in simulations with $N=4097$), for
$\alpha\in\{1/128,1/64,\ldots,32,64\}$ (increasing by a factor 2
at each step).
 Left: $S=3$. Right: $S=4$. All measurements were taken following unbiased
initial conditions (random initial strategy valuations drawn from
$[-10^{-4},10^{-4}]$).}
 \label{fig:velo_histograms}
\end{figure}

If we finally measure and plot the distribution
$\varrho(\overline{v}_a)=N^{-1}\sum_i
\delta(\overline{v}_a-\overline{v}_a^i)$ of the $N$ strategy
velocities for strategy $a$,  we  see clearly a consistent
tendency of the strategy velocities to shift towards negative
values as $\alpha$ increases (which indicates an increasing
inability of the agents to be successful in the game). Examples
are  shown in figure \ref{fig:velo_histograms} at different values
of $\alpha$, for $S=3$ and $S=4$, both following unbiased initial
conditions. The symmetry of the problem guarantees that for
$N\to\infty$ the distribution $\varrho(\overline{v}_a)$ will be
the same for all $a$ (although it will obviously depend on $S$).

It is clear from these figures that the complexities of MGs with
$S>2$ are in the nontrivial dependence on control parameters and
initial conditions of the frequencies with which the agents use
their available strategies. Mathematically one finds this
reflected in a nontrivial closure problem, as we will see below.

\section{Generating functional analysis for general $S$}

In solving the dynamics of the process
(\ref{eq:process1},\ref{eq:process2}) for general $S$ we can
follow the strategy of the $S=2$ case (see e.g.
\cite{HeimelCoolen01,CoolHeimSher01} or \cite{MGbook2}), although
we no longer benefit from simplifications such as a reduction to
equations and observables with valuation differences only. We will
therefore suppress details and give only relevant intermediate
stages of the calculation. For general $S$ the canonical
disorder-averaged\footnote{As always, the $pNS$ strategy entries
$R_{\mu}^{ia}$ are regarded as frozen disorder, and disorder
averages are written as $\overline{(\ldots)}$.} moment generating
functional will be
\begin{eqnarray}
\overline{Z[\bpsi]}&=& \overline{\bigbra \rme^{\rmi\sum_{i a
t}\psi_{ia}(t)\delta_{a,a_i(t)}}\bigket} \label{eq:Z}
\end{eqnarray}
where $a_i(t)=m(\bv^i(t),\bz^i(t))$ is the active strategy of
agent $i$ at time $t$. It generates dynamical observables such as
the correlation- and response functions
\begin{eqnarray}
C_{tt^\prime}&=& \frac{1}{N}\sum_{ia}\overline{\bra
\delta_{a,a_i(t)}\delta_{a,a_i(t^\prime)}\ket}=-
\lim_{\bpsi\to\bnull} \frac{1}{N}\sum_{ia}\frac{\partial^2
\overline{Z[\bpsi]}}{\partial
\psi_{ia}(t)\partial\psi_{ia}(t^\prime)} \label{eq:defineC}
\\
G_{tt^\prime}&=& \frac{1}{N}\sum_{ia}\frac{\partial \overline{\bra
\delta_{a,a_i(t)}\ket} }{\partial\theta_{ia}(t^\prime)}~~~~~~=-
\rmi\lim_{\bpsi\to\bnull} \frac{1}{N}\sum_{ia}\frac{\partial^2
\overline{Z[\bpsi]}}{\partial
\psi_{ia}(t)\partial\theta_{ia}(t^\prime)} \label{eq:defineG}
\end{eqnarray}
$C_{tt^\prime}$ gives the likelihood that the active strategies at
times $t$ and $t^\prime$ are identical; $G_{tt^\prime}$ measures
the increase in probability of a strategy being active at time $t$
following a perturbation of its valuation at time $t^\prime$ (both
averaged over all agents and over the disorder). Causality
guarantees that $G_{tt^\prime}=0$ for $t\leq t^\prime$. We note
that random strategy selection by the agents would give
$C_{tt^\prime}=S^{-1}+\delta_{tt^\prime}(1-S^{-1})$ and
$G_{tt^\prime}=0$. A fully frozen state would be characterized by
$C_{tt^\prime}=1$ and $G_{tt^\prime}=0$.

The functional (\ref{eq:Z}) is an average over all possible paths
of the combined $N$-agent state vector $\{\bv\}$ in time. The
probability density for each path is a product of the kernels
$W_t(\ldots)$ in (\ref{eq:process2}), so upon writing the
$\delta$-functions in (\ref{eq:process2}) in integral form one
finds
\begin{eqnarray}
\hspace*{-10mm} \overline{Z[\bpsi]}&=&
\int\!\left[\prod_{it}\frac{d\bv^i(t)d\hat{\bv}^i(t)}{(2\pi)^S}\right]P_0(\bv^1,\ldots,\bv^N)
~\rme^{\rmi
\sum_{iat}\hat{v}_a^i(t)[v_a^i(t+1)-v_a^i(t)-\theta_{ia}(t)]}
\nonumber \\ \hspace*{-10mm} && \times \bigbra \rme^{\rmi\sum_{i a
t}\psi_{ia}(t)\delta_{a,m(\bv^i(t),\bz^i(t))}}
\overline{\left[\rme^{\rmi (\tilde{\eta}/N)\sum_{ij a a^\prime
t\mu}R_\mu^{ia} R_{\mu}^{j
a^\prime}\hat{v}_a^i(t)\delta_{a^\prime,m(\bv^j(t),\bz^j(t))}}\right]}
\bigket_{\!\!\{\bz\}} ~~~~~
\end{eqnarray}
 The dependence on strategy entries in the exponent is linearized by introducing
auxiliary variables
$x_t^\mu=\tilde{\eta}N^{-1/2}\sum_{ia}\hat{v}_a^i(t)R_\mu^{ia}$
(via suitable $\delta$-functions), after which the disorder
average is carried out. In leading order in $N$, the result
depends on the $\{\bv\}$ only via
\begin{eqnarray}
C_{tt^\prime}(\{\bv,\bz\})&=&
\frac{1}{N}\sum_{ia}\delta_{a,m(\bv^i(t),\bz^i(t))}
\delta_{a,m(\bv^i(t^\prime),\bz^i(t^\prime))} \label{eq:Cf}
\\
K_{tt^\prime}(\{\bv,\bz\})&=&
\frac{1}{N}\sum_{ia}\delta_{a,m(\bv^i(t),\bz^i(t))}
\hat{v}_a^i(t^\prime) \label{eq:Kf}
\\
L_{tt^\prime}(\{\bv,\bz\})&=& \frac{1}{N}\sum_{ia}
\hat{v}_{a}^i(t) \hat{v}_{a}^i(t^\prime) \label{eq:Lf}
\end{eqnarray}
Upon isolating (\ref{eq:Cf},\ref{eq:Kf},\ref{eq:Lf}) via integrals
over suitable $\delta$-functions we can factorize and integrate
out the site-dependent variables, given site-factorized initial
conditions. We then find an expression for $\overline{Z[\bpsi]}$
which can for $N\to\infty$ be evaluated by steepest descent:
\begin{eqnarray}
\overline{Z[\bpsi]}&=&
\int\!\left[\prod_{tt^\prime}dC_{tt^\prime}d\hat{C}_{tt^\prime}dK_{tt^\prime}d\hat{K}_{tt^\prime}dL_{tt^\prime}d\hat{L}_{tt^\prime}
\right] \rme^{N(\Psi+\Phi+\Omega)+\order(\log N)}
\\
\Psi&=&
\rmi\sum_{tt^\prime}\left[C_{tt^\prime}\hat{C}_{tt^\prime}+K_{tt^\prime}\hat{K}_{tt^\prime}+L_{tt^\prime}\hat{L}_{tt^\prime}\right]
\label{eq:Psi}
\\
\Phi&=&\alpha \log \int\!\left[\prod_t\frac{dx_td\hat{x}_t}{2\pi}
\rme^{i\hat{x}_t x_t}\right]
\rme^{-\frac{1}{2}\sum_{tt^\prime}\left[x_tC_{tt^\prime}x_{t^\prime}+\tilde{\eta}^2\hat{x}_tL_{tt^\prime}\hat{x}_{t^\prime}
-2\hat{\eta}x_t K_{tt^\prime}\hat{x}_{t^\prime}\right]}
\label{eq:Phi}
\\
\Omega&=&\frac{1}{N}\sum_i\log
\bigbra\left[\prod_{at}\frac{dv_a(t)d\hat{v}_a(t)}{2\pi}\right]P_0(\bv(0))
\right.\nonumber \\ &&\left.\hspace*{-3mm}
\times~\rme^{-\rmi\sum_{a
tt^\prime}\left[\hat{C}_{tt^\prime}\delta_{a,m(\bv(t),\bz(t))}\delta_{a,m(\bv(t^\prime),\bz(t^\prime))}
+\hat{L}_{tt^\prime}
 \hat{v}_a(t)\hat{v}_a(t^\prime)+\hat{K}_{tt^\prime}\delta_{a,m(\bv(t),\bz(t))} \hat{v}_a(t^\prime)\right]}   \right. \nonumber \\ &&\left.
 \times~
\rme^{\rmi\sum_{at}\hat{v}_a(t)[v_a(t+1)-v_a(t)-\theta_{ia}(t)]+\rmi\sum_{at}\psi_{ia}(t)\delta_{a,m(\bv(t),\bz(t))}}
 \bigket
 \label{eq:Omega}
\end{eqnarray}
For $N\to\infty$ the order parameters
$\{C,\hat{C},K,\hat{K},L,\hat{L}\}$ are determined by the
saddle-point equations of the exponent $\Psi+\Phi+\Omega$. Working
out these equations is straightforward but lengthy, see
\ref{app:saddlepoint}. The resulting theory can be written solely
in terms of the kernels
 (\ref{eq:defineC},\ref{eq:defineG}), and,  upon choosing $\theta_{ia}(t)=\theta_a(t)$, formulated in terms of the
 following stochastic process for the strategy valuations $\bv=(v_1,\ldots,v_S)$ of a single `effective
 agent', with zero-average coloured Gaussian noise forces
 $\{\eta_a\}$ and a retarded self-interaction:
 \begin{eqnarray}
 &&
v_a(t+1)=v_a(t)+\theta_a(t)-\alpha \sum_{t^\prime\leq t}
R_{tt^\prime}
\delta_{a,m(\bv(t^\prime),\bz(t^\prime))}+\sqrt{\alpha}~\eta_a(t)
\label{eq:effective_agent1}
\\
&& R_{tt^\prime}=\tilde{\eta}(\one
+\tilde{\eta}G)^{-1}_{tt^\prime}~~~~~~~~
\bra\eta_a(t)\eta_b(t^\prime)\ket=\delta_{ab}
(RCR^\dag)_{tt^\prime} \label{eq:effective_agent2}
 \end{eqnarray}
 In terms of averages over the process (\ref{eq:effective_agent1},\ref{eq:effective_agent2}), the kernels
 $C$ and $G$ must be solved from
\be
C_{tt^\prime}=\sum_a \bra \delta_{a,m(\bv(t),\bz(t))}
\delta_{a,m(\bv(t^\prime),\bz(t^\prime))}\ket~~~~~~G_{tt^\prime}=\sum_a
\frac{\partial  \bra \delta_{a,m(\bv(t),\bz(t))}\ket}{\partial
\theta_a(t^\prime)} \label{eq:CandG}\ee\vsp

Finally, in a similar manner one derives exact expressions, in the
limit $N\to\infty$, for the disorder-averaged bid covariance
matrix $\overline{\Xi}_{tt^\prime}$ and also for the volatility
and predictability  $\sigma$ and $H$ (which both should be
self-averaging for $N\to\infty$). Here the appropriate generating
functional is
\begin{eqnarray}
\overline{Z[\bphi]}&=& \overline{\bigbra \rme^{\rmi\sum_{i \mu
t}\phi_{i\mu}(t)A_\mu(t)}\bigket} \label{eq:Zphi}
\end{eqnarray}
From (\ref{eq:Zphi}), which is calculated from the result of
evaluating the previous functional $\overline{Z[\bpsi]}$ upon
making simple substitutions (see \ref{app:vola_matrix} for
details), one obtains
\begin{eqnarray}
\lim_{N\to\infty}\overline{\bra A_\mu(t)\ket}&=&
-\rmi\lim_{N\to\infty} \lim_{\bphi\to\bnull} \frac{\partial
\overline{Z[\bphi]}}{\partial \phi_\mu(t)}= 0
\\
\lim_{N\to\infty}\frac{1}{p}\sum_\mu \overline{\bra
A_\mu(t)A_\mu(t^\prime)\ket}&=&-\lim_{N\to\infty}\frac{1}{p}\sum_\mu
\lim_{\bphi\to\bnull} \frac{\partial^2
\overline{Z[\bphi]}}{\partial
\phi_\mu(t)\partial\phi_\mu(t^\prime)}\nonumber
\\ &=&
\tilde{\eta}^{-2}(RCR^\dag)_{tt^\prime}
\end{eqnarray}
It follows that in the limit $N\to\infty$ we must have
$\overline{\Xi}_{tt^\prime}=\frac{1}{\tilde{\eta}^2}(RCR^\dag)_{tt^\prime}$,
and hence
\be
\sigma^2=\tilde{\eta}^{-2}\lim_{\tau\to\infty}\frac{1}{\tau}\sum_{t=1}^\tau(RCR^\dag)_{tt}
~~~~~~~~
H=\tilde{\eta}^{-2}\lim_{\tau\to\infty}\frac{1}{\tau^2}\sum_{tt^\prime=1}^\tau(RCR^\dag)_{tt^\prime}
\label{eq:sigma_and_H} \ee

\section{Stationary states: the strategy frequency problem}

To find time-translation invariant stationary solutions of our MG
without anomalous response and with weak long-term memory memory
we put $C_{tt^\prime}=C(t-t^\prime)$ and
$G_{tt^\prime}=G(t-t^\prime)$, and we define the usual static
order parameters, viz. the persistent correlations
$c=\lim_{t\to\infty}C(t)$ and the static susceptibility
$\chi=\sum_{t>0}G(t)$. Consequently also
$R_{tt^\prime}=R(t-t^\prime)$ and
\begin{eqnarray}
\chiR&=&\sum_{t\geq 0}R(t)=\frac{\tilde{\eta}}{1
+\tilde{\eta}\chi}
\end{eqnarray}
We extract from
(\ref{eq:effective_agent1},\ref{eq:effective_agent2}) an equation
for the valuation growth rates
$\overline{v}_a=\lim_{t\to\infty}v_a(t)/t$. This involves the
 `frozen' Gaussian fields
$\overline{\eta}_a=\lim_{t\to\infty}t^{-1}\sum_{t^\prime\leq t}
\eta_a(t^\prime)$, the persistent perturbations
$\overline{\theta}_a=\lim_{t\to\infty}t^{-1}\sum_{t^\prime\leq t}
\theta_a(t^\prime)$ and the strategy selection frequencies $f_a$:
\begin{eqnarray}
\overline{v}_a&=&\overline{\theta}_a+\sqrt{\alpha}~\overline{\eta}_a-\alpha\chiR
f_a ~~~~~~~~f_a= \lim_{t\to\infty}\frac{1}{t}\sum_{s\leq t} \bra
\delta_{a,m(\bv(s),\bz)}\ket_{\bz} \label{eq:freq1}
 \end{eqnarray}
 Since
$\bra\overline{\eta}_a\ket=0$ and $\bra
\overline{\eta}_a\overline{\eta}_b\ket=\delta_{ab}c\chiR^2$,  we
write $\overline{\eta}_a=\chiR\sqrt{c} x_a$, where
$\bx=(x_1,\ldots,x_S)$ is a vector of $S$ uncorrelated
zero-average and unit-variance frozen Gaussian variables. So
(\ref{eq:freq1}) are $2S$ relations for the $2S$ unknown variables
$(\overline{v}_1,\ldots,\overline{v}_S)$ and $(f_1,\ldots,f_S)$,
parametrized by $\bx$. Elimination of the valuation growth rates
would give a formula for $f_a(\bx)$, with $a=1\ldots S$. To
emphasize this structure of our problem we write (\ref{eq:freq1})
as
\begin{eqnarray}
f_a(\bx) &=& x_a\sqrt{\frac{c}{\alpha}} +
\frac{\overline{\theta}_a\!-\overline{v}_a(\bx)}{\alpha
\chiR}~~~~~~~f_a(\bx)= \lim_{t\to\infty}\frac{1}{t}\sum_{s\leq t}
\bra \delta_{a,m(\bv(s,\bx),\bz)}\ket_\bz~~~~~~ \label{eq:freq2}
 \end{eqnarray}
One easily derives from (\ref{eq:CandG}) closed equations for the
static order parameters, via
$c=\lim_{t\to\infty}t^{-2}\sum_{ss^\prime\leq t}C(s-s^\prime)$ and
$\chi=\lim_{t\to\infty}t^{-1}\sum_{s\leq
t}\partial\bra\delta_{a,m(\bv(s),\bz(s))}\ket/\partial\overline{\theta}_a$,
in terms of the solution $\{f_a(\bx)\}$ of (\ref{eq:freq2}). The
equation for $\chi$  can be simplified further upon noting that
$\partial/\partial\overline{\theta}_a=\alpha^{-1/2}\partial/\partial\overline{\eta}_a=
(\chiR\sqrt{\alpha c})^{-1}
\partial/\partial x_a$. The result is
\be
c=\int\!D\bx \sum_a
f_a^2(\bx)~~~~~~~~\chi=\frac{1}{\chiR\sqrt{\alpha c}} \int\!D\bx
\sum_a x_a f_a(\bx) \label{eq:c_and_chi}
 \ee (with the usual
short-hand $Dx=(2\pi)^{-\frac{1}{2}}e^{-\frac{1}{2}x^2}dx$). The
perturbation fields $\overline{\theta}_a$ are now no longer needed
and can be set to zero, which simplifies (\ref{eq:freq2}) to
\begin{eqnarray}
f_a(\bx) &=& x_a\sqrt{\frac{c}{\alpha}} -
\frac{\overline{v}_a(\bx)}{\alpha \chiR}~~~~~~~f_a(\bx)=
\lim_{t\to\infty}\frac{1}{t}\sum_{s\leq t} \bra
\delta_{a,m(\bv(s,\bx),\bz)}\ket_\bz \label{eq:freq3}
 \end{eqnarray}

 Solving the stationary state of the MG,
including finding the phase transition line marked by a divergence
of $\chi$, thus boils down to solving $\{f_a(\bx)\}$ from
(\ref{eq:freq3}). This is the strategy frequency problem. The
difficulty is in the second part of (\ref{eq:freq3}): even if two
strategies $(a,b)$ have $\overline{v}_a(\bx)=\overline{v}_b(\bx)$
it does not follow that $f_a(\bx)=f_b(\bx)$ (this is also clear in
simulations).  All valuation growth rates $\overline{v}_c(\bx)$,
including those with $c\notin\{a,b\}$, will influence $f_a(\bx)$
and $f_b(\bx)$.  The frequencies depend in a highly nontrivial way
on both the realization of the Gaussian vector $\bx$ (which
represents the diversity in the original $N$-agent population) and
the control parameter $\alpha$. Even the transients of the
valuations, i.e. the full $v_a(s,\bx)$ rather than just their
growth rates $\overline{v}_a(\bx)$, could in principle impact on
the long-term frequencies $\{f_a(\bx)\}$. All this appears to make
the problem practically insoluble. For $S=2$ the situation could
be saved upon translation of our equations into the language of
valuation differences; there was only one relevant quantity,
$\tilde{q}=\overline{v}_1-\overline{v}_2$, and what mattered was
only whether or not $\tilde{q}=0$. For $S>2$ this is no longer
true.

\section{Solution of the strategy frequency problem}

 We turn to the general solution of the strategy frequency
problem,  for arbitrary $S$ and
 additive\footnote{The solution in the case of multiplicative decision noise is not identical but very similar to the one discussed here;
 see \ref{app:multiplicative} for details.}    decision noise: $m(\bv,\bz)={\rm
 argmax}_{a}[v_a+Tz_a]$. This includes the deterministic case for $T=0$. We define
$\overline{v}^\star(\bx)=\max_b \overline{v}_b(\bx)$, and the set
$\Lambda(\bx)$ of all strategy indices for which
$\overline{v}_a(\bx)=\overline{v}^\star(\bx)$: \be
\Lambda(\bx)=\Big\{a|~\overline{v}_a(\bx)=\max_b
\overline{v}_b(\bx)\Big\}\subseteq
\{1,\ldots,S\}~~~~~~|\Lambda(\bx)|>0 \label{eq:defineLambda}\ee
The solution now proceeds in three stages:
\begin{itemize}
\item
 Since
$v_a(s,\bx)=s[\overline{v}_a(\bx)+\varepsilon_a(s,\bx)]$ with
$\lim_{s\to\infty}\varepsilon_a(s,\bx)=0$ we may write for the
second equation in (\ref{eq:freq2})
\begin{eqnarray*}\hspace*{-5mm}
f_a(\bx)&=& \lim_{s\to\infty} \Big\bra\prod_{b\neq
a}\Big\bra\theta\Big[\overline{v}_a(\bx)-\overline{v}_b(\bx)+\varepsilon_a(s,\bx)-\varepsilon_b(s,\bx)+\frac{T(z\!-\!z^\prime)}{s}\Big]\Big\ket_{\!z^\prime}
\Big\ket_{\!z}
\end{eqnarray*}
This quantity can be nonzero only for $a\in\Lambda(\bx)$. Hence,
once we know
 $\Lambda(\bx)$ and $\overline{v}^\star(\bx)$ the
problem is solved, since in combination with (\ref{eq:freq2}) we
may write
\begin{eqnarray}
a\notin\Lambda(\bx):&~~~ f_a(\bx)=0,~~~&\overline{v}_a(\bx)= x_a
\chiR \sqrt{\alpha c}
 \label{eq:bullet2a}
\\
a\in\Lambda(\bx):&~~~ f_a(\bx)= x_a\sqrt{\frac{c}{\alpha}} -
\frac{\overline{v}^\star(\bx)}{\alpha
\chiR},~~~&\overline{v}_a(\bx)=\overline{v}^\star(\bx)
\label{eq:bullet2b}
\end{eqnarray}
\item
Next we calculate $\overline{v}^\star(\bx)$. Probability
normalization guarantees that $\sum_{a}f_a(\bx)=1$ for any $\bx$,
but since $f_a(\bx)\neq 0$ only for $a\in\Lambda(\bx)$ we have in
fact $\sum_{a\in\Lambda(\bx)}f_a(\bx)=1$. Summing over the indices
in (\ref{eq:bullet2b}) therefore leads to
\begin{eqnarray}
\overline{v}^\star(\bx)&=&  \frac{\chiR\sqrt{\alpha
c}}{|\Lambda(\bx)|}\sum_{a\in\Lambda(\bx)}x_a  - \frac{\alpha
\chiR}{|\Lambda(\bx)|}
\end{eqnarray}
Upon abbreviating $|\Lambda(\bx)|^{-1}\sum_{b\in\Lambda(\bx)}U_b
=\bra U\ket_{\Lambda(\bx)}$ our equations
(\ref{eq:bullet2a},\ref{eq:bullet2b}) then become
\begin{eqnarray}
\hspace*{-15mm} a\!\notin\!\Lambda(\bx):&~~~ f_a(\bx)=0,~~~&
\overline{v}_a(\bx)= x_a \chiR \sqrt{\alpha c} \label{eq:bullet3a}
\\
\hspace*{-15mm}
 a\!\in\!\Lambda(\bx):&~~~ f_a(\bx)=
\frac{1}{|\Lambda(\bx)|} +
 \sqrt{\frac{c}{\alpha}}\left(x_a\! -\bra
 x\ket_{\Lambda(\bx)}\right),~~~&\overline{v}_a(\bx)=\overline{v}^\star(\bx)\label{eq:bullet3b}
\end{eqnarray}
\item
What remains is to determine the set $\Lambda(\bx)$. The
definition (\ref{eq:defineLambda}) of $\Lambda(\bx)$ demands that
$\overline{v}_a(\bx)<\overline{v}^\star(\bx)$ for all
$a\not\in\Lambda(\bx)$, i.e.
\begin{eqnarray}
a\not\in\Lambda(\bx):&~~~&  x_a  < \bra x\ket_{\Lambda(\bx)} -
\sqrt{\frac{\alpha}{c}} \frac{1}{|\Lambda(\bx)|}
\end{eqnarray}
However, we have similar inequalities for $a\in\Lambda(\bx)$, as
(\ref{eq:bullet3b}) must obey $f_a(\bx)\in[0,1]$:
\begin{eqnarray}
 a\in\Lambda(\bx):&~~~&
x_a \geq \bra
 x\ket_{\Lambda(\bx)}-\sqrt{\frac{\alpha}{c}} \frac{1}{|\Lambda(\bx)|}
 \\
 &&
 x_a \leq \bra
 x\ket_{\Lambda(\bx)}-\sqrt{\frac{\alpha}{c}} \frac{1}{|\Lambda(\bx)|} +\sqrt{\frac{\alpha}{c}}
 \end{eqnarray}
 These last three groups of inequalities turn out to determine the
 set $\Lambda(\bx)$ uniquely.
 Firstly, they tell us that $\Lambda(\bx)$ must contain the indices
 of the $\ell$  {\em largest} components of the vector $\bx$, where $\ell=|\Lambda(\bx)|$.
 With probability one we may assume all components of $\bx$ to be
 different, so for each $\bx$ there is a unique
 permutation $\pi_\bx:\{1,\ldots,S\}\to \{1,\ldots,S\}$ for which
 these components will be ordered according to $x_{\pi(1)}> x_{\pi(2)}> \ldots
 > x_{\pi(S)}$. We may now translate our key inequalities into the
 following three conditions, where $\ell=|\Lambda(\bx)|\in\{1,\ldots,S\}$ is the only quantity left to be solved, and with the
ordering permutation $\pi=\pi_\bx$:
 \begin{eqnarray}
x_{\pi(\ell)} & \geq \frac{1}{\ell\!-\!1}\sum_{m=1}^{\ell-1}
 x_{\pi(m)}-\frac{1}{\ell\!-\!1}\sqrt{\frac{\alpha}{c}}
 ~~~~~& {\rm
 if}~\ell>1
 \label{eq:ineq1}
 \\
 x_{\pi(\ell+1)}  & < \frac{1}{\ell}\sum_{m=1}^\ell
 x_{\pi(m)}-\frac{1}{\ell}\sqrt{\frac{\alpha}{c}}~~~~~& {\rm
 if}~\ell<S
 \label{eq:ineq2}
 \\
 x_{\pi(1)} &\leq \frac{1}{\ell}\sum_{m=1}^\ell
 x_{\pi(m)}-\frac{1}{\ell}\sqrt{\frac{\alpha}{c}} +\sqrt{\frac{\alpha}{c}}
 \label{eq:ineq3}
 \end{eqnarray}
The solution is the following: $\ell$ is the {\em smallest} number
in $\{1,\ldots,S\}$ for which (\ref{eq:ineq2}) holds (if any),
whereas if (\ref{eq:ineq2}) never holds then $\ell=S$. By
construction we thereby satisfy both (\ref{eq:ineq1}) and
(\ref{eq:ineq2}). What remains is to show that also
(\ref{eq:ineq3}) will be satisfied, and that the solution is
unique. Clearly (\ref{eq:ineq3}) holds when $\ell=1$.  To prove
(\ref{eq:ineq3}) for $\ell>1$ we first define
$X_k=k^{-1}\sum_{m\leq k}x_{\pi(m)}$ for $k\leq \ell$. We know
from (\ref{eq:ineq1}) that
\begin{eqnarray*}
\hspace*{-13mm}
X_{k}&=&\frac{1}{k}\sum_{m=1}^{k-1}x_{\pi(m)}+\frac{1}{k}x_{\pi(k)}
\\
\hspace*{-13mm} &\geq& \frac{1}{k}\sum_{m=1}^{k-1}x_{\pi(m)}+
 \frac{1}{k(k\!-\!1)}\sum_{m=1}^{k-1}
 x_{\pi(m)}-\frac{1}{k(k\!-\!1)}\sqrt{\frac{\alpha}{c}}
= X_{k-1}-\frac{1}{k(k\!-\!1)}\sqrt{\frac{\alpha}{c}}
\end{eqnarray*}
Hence
 $X_\ell \geq X_1-\sqrt{\frac{\alpha}{c}}\sum_{m=2}^\ell
\frac{1}{m(m-1)}=x_{\pi(1)}-\sqrt{\frac{\alpha}{c}}\sum_{m=2}^\ell
\frac{1}{m(m-1)}$, so that
\begin{eqnarray*}
x_{\pi(1)}&\leq& \frac{1}{\ell}\sum_{m=1}^\ell
 x_{\pi(m)}+\sqrt{\frac{\alpha}{c}}\sum_{m=1}^{\ell-1}
\frac{1}{m(m+1)} =\frac{1}{\ell}\sum_{m=1}^\ell
 x_{\pi(m)}+\sqrt{\frac{\alpha}{c}}\frac{\ell-1}{\ell}
\end{eqnarray*}
This is the inequality  (\ref{eq:ineq3}) that we set out to prove.
The corollary is that we have indeed defined a self-consistent
solution of our equations. The above solution must be unique: any
alternative choice of $\ell$ (rather than the smallest) that
satisfies  (\ref{eq:ineq2}) would always make the previous
(smallest) choice induce a violation of (\ref{eq:ineq1}).
\end{itemize}
\vsp

 \noindent We may now summarize the solution of the strategy
frequency problem for
 additive decision noise as follows:
\begin{eqnarray}
a\notin\Lambda(\bx):&~~~& f_a(\bx)=0 \label{eq:soln1}
\\
 a\in\Lambda(\bx):&~~~& f_a(\bx)=
\frac{1}{\ell(\bx)} +
 \sqrt{\frac{c}{\alpha}}\left(x_a\! -\bra
 x\ket_{\Lambda(\bx)}\right)
 \label{eq:soln2}
\\
\Lambda(\bx)&&= \{\pi_\bx(1),\ldots,\pi_\bx(\ell(\bx))\}
\\
 \pi_\bx:&~~~& {\rm permutation~such~that}~~x_{\pi(1)}> x_{\pi(2)}> \ldots
 > x_{\pi(S)}
 \\
 \ell(\bx):
 &~~~& {\rm defined~by~the~conditions~}\\
 && x_{\pi(\ell+1)}  < \frac{1}{\ell}\sum_{m=1}^\ell
 x_{\pi(m)}-\frac{1}{\ell}\sqrt{\frac{\alpha}{c}}~~~~~{\rm
 if}~\ell<S\\
 && x_{\pi(k+1)}  > \frac{1}{k}\sum_{m=1}^k
 x_{\pi(m)}-\frac{1}{k}\sqrt{\frac{\alpha}{c}}~~~~~{\rm
 for~all~}k<\ell
\end{eqnarray}
The meaning of this stationary state solution is as follows. The
randomness induced by the Gaussian variables $\bx$ represents the
variability in the original $N$-agent population. The set
$\Lambda(\bx)$ contains the strategies that will be played by the
effective agent, albeit with different frequencies. Strategies
$a\notin\Lambda(\bx)$ are never played. Apparently, for a strategy
$a$ to be in the `active' set $\Lambda(\bx)$, it  must have an
$x_a$ that is sufficiently large, and sufficiently close to the
components of the other strategies in the active set.

Since expressions (\ref{eq:soln1},\ref{eq:soln2}) depend only on
$\Lambda(\bx)$, i.e. on $\pi_\bx$ and $\ell$ (see above), we may
abbreviate these formulae as $f_a(\bx|\pi_\bx,\ell)$. Averages of
the form $\int\!D\bx ~\Phi(\{x_a,f_a(\bx))$ can now be written as
a sum over all permutations $\pi$ of $\{1,\ldots,S\}$, with a
function $C(\pi|\bx)=\delta_{\pi,\pi_\bx}$ that selects the right
component ordering permutation for each $\bx$:
\begin{eqnarray}
\hspace*{-10mm} \int\!D\bx~\Phi(\{x_a,f_a(\bx)\})&=& \sum_{\pi}
\sum_{\ell=1}^{S}\int\!D\bx ~C(\pi|\bx)
~\Phi(\{x_a,f_a(\bx|\pi,\ell)\}) \nonumber
\\
&&\times \prod_{a<\ell}\theta\Big[x_{\pi(a+1)}-
\frac{1}{a}\sum_{m=1}^a
 x_{\pi(m)}+\frac{1}{a}\sqrt{\frac{\alpha}{c}}\Big]
 \label{eq:general_averages}
 \\
 &&\times~\left\{\delta_{\ell S}+(1-\delta_{\ell S})~
\theta\Big[ \frac{1}{\ell}\sum_{m=1}^\ell
 x_{\pi(m)}-\frac{1}{\ell}\sqrt{\frac{\alpha}{c}}-x_{\pi(\ell+1)} \Big]
 \right\}\nonumber
 \end{eqnarray}
 where
 \begin{eqnarray}
 C(\pi|\bx)&=& \prod_{a=1}^{S-1}\theta[x_{\pi(a)}-x_{\pi(a+1)}]
\end{eqnarray}

\section{The static theory for arbitrary $S$}
\label{sec:finaltheory}

In those cases where one seeks to calculate the average of a
function $\Phi$ that is {\em invariant} under all permutations of
the index set $\{1,\ldots,S\}$, as in (\ref{eq:c_and_chi}), one
may use this invariance to simplify the calculation. Here the
average will equal the contribution from one particular ordering
of the components of $\bx$ (and its associated permutation
$\pi_\bx$, for which we may take the identity permutation) times
the number $S!$ of permutations:
\begin{eqnarray}
\hspace*{-10mm} \int\!D\bx~\Phi(\{x_a,f_a(\bx)\})&=& S!
\sum_{\ell=1}^{S}\int\!D\bx \prod_{a=1}^{S-1}\theta[x_{a}-x_{a+1}]
\Phi(\{x_a,f_a(\bx|\ell)\}) \nonumber
\\
&&\times \prod_{a<\ell}\theta\Big[x_{a+1}- \frac{1}{a}\sum_{m=1}^a
 x_{m}+\frac{1}{a}\sqrt{\frac{\alpha}{c}}\Big]\nonumber
 \\
 &&\times~\left\{\delta_{\ell S}+(1-\delta_{\ell S})~
\theta\Big[ \frac{1}{\ell}\sum_{m=1}^\ell
 x_{m}\!-\frac{1}{\ell}\sqrt{\frac{\alpha}{c}}-x_{\ell+1} \Big]
 \right\}
 \label{eq:permutation_invariant}
\end{eqnarray}
In particular, upon application to (\ref{eq:c_and_chi}) and using
(\ref{eq:soln1},\ref{eq:soln2}) we have now arrived at fully
explicit and closed equations for our static order parameters:
\begin{eqnarray}
\hspace*{-5mm} c&=& S!
\sum_{\ell=1}^{S}\sum_{a=1}^\ell\int\!D\bx~\prod_{a=1}^{S-1}\theta[x_{a}\!-\!x_{a+1}]
~\prod_{a<\ell}\theta\Big[x_{a+1}\!-\! \frac{1}{a}\sum_{m=1}^a
 x_{m}\!+\!\frac{1}{a}\sqrt{\frac{\alpha}{c}}\Big]\nonumber
 \nonumber
 \\
 \hspace*{-5mm}&&\times
\left\{\delta_{\ell S}+(1-\delta_{\ell S})~ \theta\Big[
\frac{1}{\ell}\sum_{m=1}^\ell
 x_{m}\!-\frac{1}{\ell}\sqrt{\frac{\alpha}{c}}-x_{\ell+1} \Big]
 \right\}
\nonumber
\\
 \hspace*{-5mm}
 &&\times~
 \Big[\frac{1}{\ell} \!+\!
 \sqrt{\frac{c}{\alpha}}\Big(x_a\! -\frac{1}{\ell}\sum_{m=1}^\ell x_m\Big) \Big]^2
 \label{eq:final_c_eq}
 \\
 \hspace*{-5mm}
\frac{\tilde{\eta}\chi
}{1\!+\!\tilde{\eta}\chi}&=&\frac{S!}{\sqrt{\alpha c}}
 \sum_{\ell=1}^{S}\sum_{a=1}^\ell
 \int\!D\bx~\prod_{a=1}^{S-1}\theta[x_{a}-x_{a+1}]~
\prod_{a<\ell}\theta\Big[x_{a+1}- \frac{1}{a}\sum_{m=1}^a
 x_{m}+\frac{1}{a}\sqrt{\frac{\alpha}{c}}\Big]
  \nonumber
\\
 \hspace*{-5mm}
 &&\times~\left\{\delta_{\ell S}+(1-\delta_{\ell S})~
\theta\Big[ \frac{1}{\ell}\sum_{m=1}^\ell
 x_{m}\!-\frac{1}{\ell}\sqrt{\frac{\alpha}{c}}-x_{\ell+1} \Big]
 \right\}\nonumber
\\
\hspace*{-5mm} &&\times~
 x_a\Big[\frac{1}{\ell} \!+\!
 \sqrt{\frac{c}{\alpha}}\Big(x_a\! -\frac{1}{\ell}\sum_{m=1}^\ell x_m\Big) \Big]
 \label{eq:final_chi_eq}
\end{eqnarray}
For any given value of $\alpha$ one first solves
(\ref{eq:final_c_eq}) for $c$, after which $\chi$ is calculated
via (\ref{eq:final_chi_eq}). The $\chi=\infty$ phase transition
occurs when the right-hand side of (\ref{eq:final_chi_eq}) equals
one, and defines the critical point $\alpha_c(S)$. Numerical
solution of (\ref{eq:final_c_eq},\ref{eq:final_chi_eq}) for
$S\in\{2,3,4,5\}$ gives the following values (accurate up to the
last digit given):\vsp

\begin{center}\vspace*{1mm}
\begin{tabular}{|c||c|c|c|c|}
 \hline\room $S$           & 2 & 3 & 4 & 5 \\
 \hline\room $\alpha_c(S)$ & 0.337 & 0.824 & 1.324 & 1.822\\
 \hline
\end{tabular}\end{center}
\vsp

\noindent These values are remarkably close to those one would
predict on the basis of  the heuristic relation $\alpha(S)\approx
\alpha_c(2)+\frac{1}{2}(S-2)$, as proposed in
\cite{MarsChalZecc00}, but not identical.

Below the critical point, in the nonergodic regime
$\alpha<\alpha_c(S)$ the above theory no longer applies, as its
assumption of finite integrated response $\chi$ is violated.  For
$S=2$ it was shown \cite{HeimelCoolen01,CoolHeimSher01} that upon
replacing (\ref{eq:final_c_eq}) by the equation $\chi^{-1}=0$,
where $\chi$ is calculated from  (\ref{eq:final_chi_eq}), the
agreement between ergodic theory and simulations with biased
initializations in the nonergodic regime could be improved. For
$S>2$ such heuristic improvements are again possible but more
awkward. Furthermore, they are entirely ad hoc and artificial, and
therefore mostly of cosmetic merit, so we have decided to restrict
ourselves here to the theory
(\ref{eq:final_c_eq},\ref{eq:final_chi_eq}).
 \vsp

 For arbitrary $S$ we must also generalize the concept of `frozen
agents' and the associated order parameter $\phi$: we define
$\phi_k$ as the fraction of agents that in the stationary state
play $k$ of their $S$ strategies. Since $k$ is simply the size of
the active set $\Lambda(\bx)$ in the language of
(\ref{eq:permutation_invariant}) the order parameter $\phi_k$ is
the average of the function
$\Phi(\{x_a,f_a(\bx|\ell)\})=\delta_{k\ell}$, so our solution
immediately tells us that
\begin{eqnarray}
 \phi_{\ell< S} &=& S! \int\!D\bx
\prod_{a=1}^{S-1}\theta[x_{a}-x_{a+1}]
\prod_{a=1}^{\ell}\theta\Big[x_{a}- \frac{1}{a}\sum_{m=1}^a
 x_{m}+\frac{1}{a}\sqrt{\frac{\alpha}{c}}\Big]\nonumber
 \\
 &&\hspace*{20mm}\times~
\theta\Big[ \frac{1}{\ell}\sum_{m=1}^\ell
 x_{m}\!-\frac{1}{\ell}\sqrt{\frac{\alpha}{c}}-x_{\ell+1} \Big]
 \label{eq:phi_l}
 \\
 \phi_S&=&  S! \int\!D\bx
\prod_{a=1}^{S-1}\theta[x_{a}-x_{a+1}]
\prod_{a=1}^{S}\theta\Big[x_{a}- \frac{1}{a}\sum_{m=1}^a
 x_{m}+\frac{1}{a}\sqrt{\frac{\alpha}{c}}\Big]
 \label{eq:phi_S}
\end{eqnarray}
One confirms easily that these expressions obey $\sum_{k=1}^S
\phi_k=1$, as they should. Our formulae
(\ref{eq:phi_l},\ref{eq:phi_S}) can be simplified, if needed, by
generating and exploiting permutation-invariant terms. For
instance, $\phi_1$ and $\phi_2$ can be rewritten as
\begin{eqnarray}
\phi_1&=&
S!\int\!D\bx~\theta[x_1-x_2-\sqrt{\frac{\alpha}{c}}]\prod_{m=3}^S\theta[x_{m-1}-x_m]
\nonumber
\\ &=& S!\int\!D\bx~\theta[x_1-\sqrt{\frac{\alpha}{c}}-\max_{a\geq
2}x_a ]\prod_{m=3}^S\theta[x_{m-1}-x_m]\nonumber
\\
&=&
\frac{S!}{(S-1)!}\int\!D\bx~\theta[x_1-\sqrt{\frac{\alpha}{c}}-\max_{a\geq
2}x_a ]\nonumber\\ &=& S \int\!Dx
\left\{\frac{1}{2}+\frac{1}{2}\erf\Big(\frac{x}{\sqrt{2}}-\sqrt{\frac{\alpha}{2c}}\Big)\right\}^{S-1}
\\
\phi_2&=&
S!\int\!D\bx~\theta[x_2\!-x_1\!+\sqrt{\frac{\alpha}{c}}]\theta[x_1\!+x_2\!-2x_3\!-\sqrt{\frac{\alpha}{c}}]\prod_{m=2}^S\theta[x_{m-1}\!-x_m]
\nonumber
\\ &=&
S!\int\!D\bx~\theta[x_1\!-x_2]\theta[x_2\!-x_1\!+\sqrt{\frac{\alpha}{c}}]\theta[x_2\!-\max_{a>
2}x_a] \nonumber
\\
&& \times ~\theta[x_1\!+x_2\!-2\max_{a>2}x_a\!
-\sqrt{\frac{\alpha}{c}}]\prod_{m=4}^S\theta[x_{m-1}\!-x_m]
\nonumber
\\ &=&
\frac{S!}{(S-2)!}\int\!D\bx~\theta[x_1\!-x_2]\theta[x_2\!-x_1\!+\sqrt{\frac{\alpha}{c}}]\theta[x_2\!-\max_{a>
2}x_a] \nonumber
\\
&& \times ~\theta[x_1\!+x_2\!-2\max_{a>2}x_a\!
-\sqrt{\frac{\alpha}{c}}] \nonumber
\\
&=& S(S\!-\!1) \int\!DxDy~
\theta[\sqrt{\frac{\alpha}{c}}\!+y-x]\theta[x-y]\nonumber\\
&&\times~\left\{\int\!Dz~\theta[y-z]\theta[x+y-2z-\sqrt{\frac{\alpha}{c}}]\right\}^{S-2}
\nonumber
\\
&=& S(S\!-\!1)\int\!Dx
\int_0^{\sqrt{\alpha/c}}\!\!\frac{du}{\sqrt{2\pi}}~e^{-\frac{1}{2}(x-u)^2}
\left\{\frac{1}{2}\!+\!\frac{1}{2}\erf\Big(\frac{2x\!-\!u}{2\sqrt{2}}\!-\!\frac{1}{2}\sqrt{\frac{\alpha}{2c}}\Big)\right\}^{S-2}
\nonumber
\\&&
\end{eqnarray}

We can now also calculate the disorder-averaged strategy frequency
distribution $\varrho(f)=\lim_{N\to\infty}N^{-1}\sum_i
\overline{\bra \delta[f-f_{a i}]\ket}$, which gives the fraction
of agents in the stationary state that use strategy $a$ with
frequency $f$. The problem is strategy permutation invariant, so
$\varrho(f)$ cannot depend on $a$. We may therefore write it in
the permutation-invariant form
$\varrho(f)=\lim_{N\to\infty}(SN)^{-1}\sum_{ia} \overline{\bra
\delta[f-f_{a i}]\ket}$, and  calculate it by applying our formula
(\ref{eq:permutation_invariant}) to the function
$\Phi(\{x_a,f_a(\bx|\ell)\})=S^{-1}\sum_{a=1}^S
\delta[f-f_a(\bx|\ell)]$. This gives, upon using our above
formulae for the $\phi_\ell$:
\begin{eqnarray}
\varrho(f)&=& (S\!-\!1)!
\sum_{\ell=1}^{S}\sum_{b=1}^S\int\!D\bx~\delta[f-f_b(\bx|\ell)]
\prod_{a=1}^{S-1}\theta[x_{a}-x_{a+1}]  \nonumber
\\
&&\times \prod_{a<\ell}\theta\Big[x_{a+1}- \frac{1}{a}\sum_{m=1}^a
 x_{m}+\frac{1}{a}\sqrt{\frac{\alpha}{c}}\Big]\nonumber
 \\
 &&\times~\left\{\delta_{\ell S}+(1-\delta_{\ell S})~
\theta\Big[ \frac{1}{\ell}\sum_{m=1}^\ell
 x_{m}\!-\frac{1}{\ell}\sqrt{\frac{\alpha}{c}}-x_{\ell+1} \Big]
 \right\}
 \nonumber
 \\
 &=&\delta(f)\sum_{\ell=1}^{S-1}(1\!-\!\frac{\ell}{S})\phi_\ell+
\delta(f-1) \frac{1}{S}\phi_1 \nonumber
\\
&&
 +(S\!-\!1)!
\sum_{\ell=2}^{S}\sum_{b=1}^\ell\int\!D\bx~\delta\Big[f\!-\!
\frac{1}{\ell} \!-\!
 \sqrt{\frac{c}{\alpha}}\Big(x_b\! -\!\frac{1}{\ell}\sum_{m=1}^\ell x_m\Big)\Big] \nonumber
\\
&&\times \prod_{a=1}^{S-1}\theta[x_{a}-x_{a+1}] ~
\prod_{a=1}^{\ell-1}\theta\Big[x_{a+1}- \frac{1}{a}\sum_{m=1}^a
 x_{m}+\frac{1}{a}\sqrt{\frac{\alpha}{c}}\Big]\nonumber
 \\
 &&\times\left\{\delta_{\ell S}+(1-\delta_{\ell S})~
\theta\Big[ \frac{1}{\ell}\sum_{m=1}^\ell
 x_{m}\!-\frac{1}{\ell}\sqrt{\frac{\alpha}{c}}-x_{\ell+1} \Big]
 \right\}
 \label{eq:rhof}
\end{eqnarray}
One can recover the full generating functional theory of the $S=2$
batch MG as in e.g. \cite{HeimelCoolen01,CoolHeimSher01} from the
above more general equations, as a test (taking into account
carefully the different definitions of correlation and response
functions that were made in earlier studies). We will not give
details of this in principle straightforward exercise here. \vsp

Finally, in the limit $\alpha\to\infty$ our theory predicts that
the MG will behave for any $S$ as if the agents were to select
strategies completely randomly, as one expects. Here, upon using
$S!\sum_{a=1}^S\int\!D\bx~\prod_{a=1}^{S-1}\theta[x_{a}\!-\!x_{a+1}]=1$
one finds that our equations simplify to
\begin{eqnarray}
& \lim_{\alpha\to\infty}c= S^{-1}~~~~~~~~&\lim_{\alpha\to\infty}
\chi=0
\\
& \lim_{\alpha\to\infty}\phi_\ell=\delta_{\ell S}~~~~~~~~
&\lim_{\alpha\to\infty} \varrho(f)
 =\delta\Big(f-\frac{1}{S}\Big)
\end{eqnarray}

\section{Application to MGs with $S=3$}

\begin{figure}[t]
\vspace*{-5mm} \hspace*{25mm} \setlength{\unitlength}{0.40mm}
\begin{picture}(290,160)
  \put(0,0){\epsfxsize=200\unitlength\epsfbox{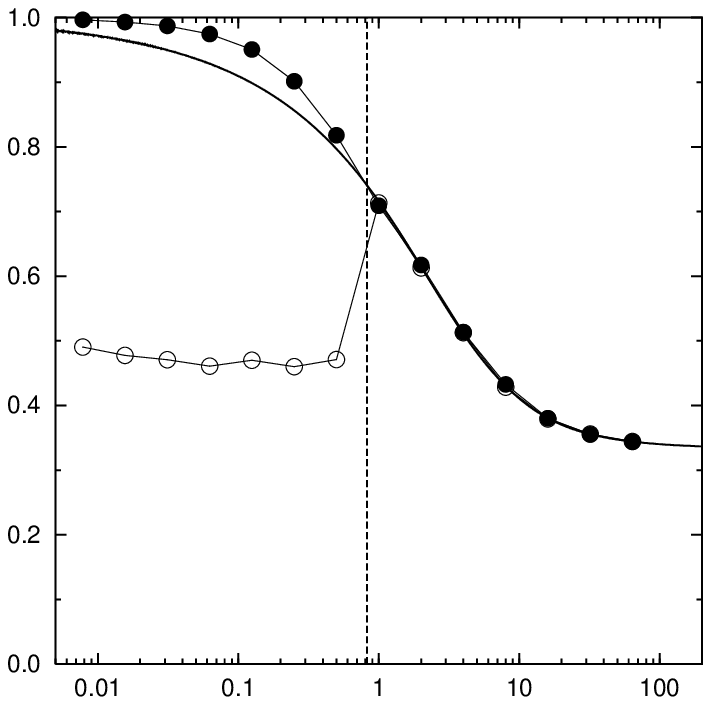}}
  \put(80,-10){\here{$\alpha$}}\put(-8,70){\large\here{$c$}}
    \put(150,0){\epsfxsize=200\unitlength\epsfbox{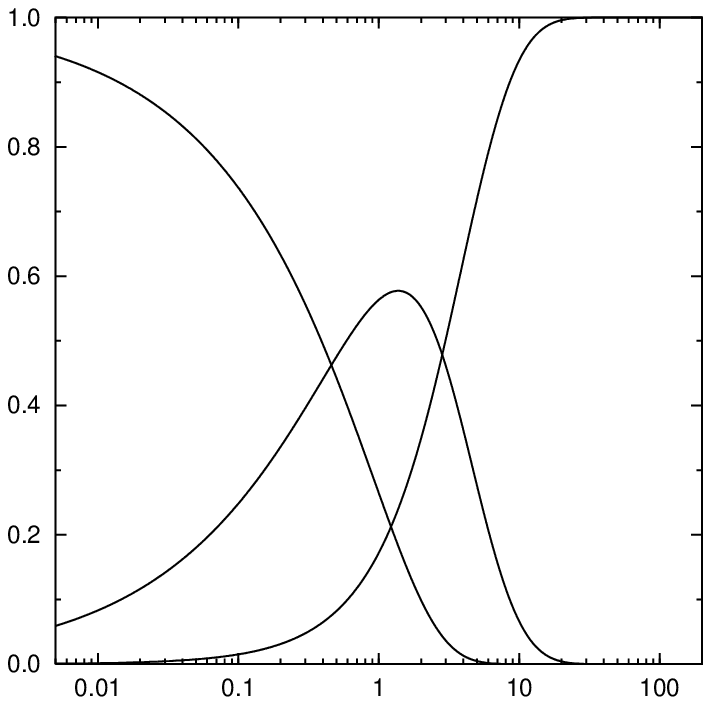}}
  \put(230,-10){\here{$\alpha$}}
  \put(173,107){$\phi_1$} \put(190,45){$\phi_2$} \put(260,115){$\phi_3$}
\end{picture}
\vspace*{6mm} \caption{Left: the predicted persistent correlations
$c=\lim_{\tau\to\infty}C(\tau)$ (solid curve) for $S=3$, as a
function of $\alpha=p/N$, calculated for ergodic stationary
states. It is shown together with corresponding data measured in
numerical simulations of MGs without decision noise, for both
biased initial conditions (random initial strategy valuations
drawn from $[-10,10]$, $\bullet$) and for unbiased initial
conditions (random initial strategy valuations drawn from
$[-10^{-4},10^{-4}]$, $\circ$). Simulation system size: $N=4097$.
Dashed: the phase transition point $\alpha_c(3)\approx  0.824$.
Right: the fractions $\phi_\ell$ of agents that play precisely
$\ell\in\{1,2,3\}$ of their $S=3$ strategies,
 as a
function of $\alpha=p/N$, calculated for ergodic stationary
states. Note that $\phi_1+\phi_2+\phi_3=1$.}
 \label{fig:c_S3}
\end{figure}

The case $S=3$ is the simplest situation where all the
complexities of having more than two strategies per agent can be
studied, so we will deal with this in detail. Here, with
persistence, one can do most of the nested integrations in
 (\ref{eq:final_c_eq},\ref{eq:final_chi_eq},\ref{eq:phi_l},\ref{eq:phi_S},\ref{eq:rhof})
 analytically. We give some of the basic identities that this involves in \ref{app:integrations}.
 The result is most easily expressed in parametric form in terms of the auxiliary variable $u=\sqrt{\alpha/c}$.
 The fundamental order parameter equation
 (\ref{eq:final_c_eq}) for the persistent correlations $c$ then becomes
 \begin{eqnarray}
 c&=& 1
 +(1-\frac{3c}{\alpha})I(u)-\frac{1}{4}\erf(\frac{u}{2})\Big[3+\erf(\frac{u}{2\sqrt{3}})\Big]+
 \frac{3}{2u^2}\Big[\erf(\frac{u}{2})-\frac{u}{\sqrt{\pi}}\rme^{-\frac{1}{2}u^2}\Big]
 \nonumber
 \\
 &&
 +\frac{3}{2u^2}\erf(\frac{u}{2})\Big[\erf(\frac{u}{2\sqrt{3}})-\frac{u}{\sqrt{3\pi}}
 \rme^{-\frac{1}{12}u^2}\Big]+\frac{3}{2u\sqrt{\pi}}\erf(\frac{u}{2\sqrt{3}})\rme^{-\frac{1}{4}u^2}
 \label{eq:c_eqn_S3}
 \end{eqnarray}
 with $I(u)=2\int_0^{u/\sqrt{2}}\!Dx~\erf(x/\sqrt{6})$ (this
latter integral we could unfortunately not do analytically, except
for the limit $I(\infty)=1/3$). Solving equation
(\ref{eq:c_eqn_S3}) for $c$ gives $c$ as a function of $\alpha$;
the result is shown and tested against simulation data in figure
\ref{fig:c_S3} (left panel).  Similarly, equation
(\ref{eq:final_chi_eq}) for the susceptibility $\chi$ takes the
explicit form
\begin{eqnarray}
\frac{\tilde{\eta}\chi}{1+\tilde{\eta}\chi}&=&
\frac{3}{2\alpha}\left\{\erf(\frac{u}{2})\Big[1+\erf(\frac{u}{2\sqrt{3}})\Big]-2I(u)\right\}
\end{eqnarray}
It follows that the $\chi=\infty$ phase transition occurs at a
value $\alpha_c$ that must be solved (numerically) from the
following two coupled equations
\begin{eqnarray}
\alpha&=&
\frac{3}{2}\left\{\erf(\frac{u}{2})\Big[1+\erf(\frac{u}{2\sqrt{3}})\Big]-2I(u)\right\}
\\
 \frac{\alpha}{u^2}&=& 1
 +(1-\frac{3c}{\alpha})I(u)-\frac{1}{4}\erf(\frac{u}{2})\Big[3\!+\!\erf(\frac{u}{2\sqrt{3}})\Big]+
 \frac{3}{2u^2}\Big[\erf(\frac{u}{2})\!-\!\frac{u}{\sqrt{\pi}}\rme^{-\frac{1}{2}u^2}\Big]
 \nonumber
 \\
 &&
 +\frac{3}{2u^2}\erf(\frac{u}{2})\Big[\erf(\frac{u}{2\sqrt{3}})-\frac{u}{\sqrt{3\pi}}
 \rme^{-\frac{1}{12}u^2}\Big]+\frac{3}{2u\sqrt{\pi}}\erf(\frac{u}{2\sqrt{3}})\rme^{-\frac{1}{4}u^2}
\end{eqnarray}
\begin{figure}[t]
\vspace*{-1mm} \hspace*{-2mm} \setlength{\unitlength}{0.53mm}
\begin{picture}(300,100)
 \put(0,0){\epsfxsize=133\unitlength\epsfbox{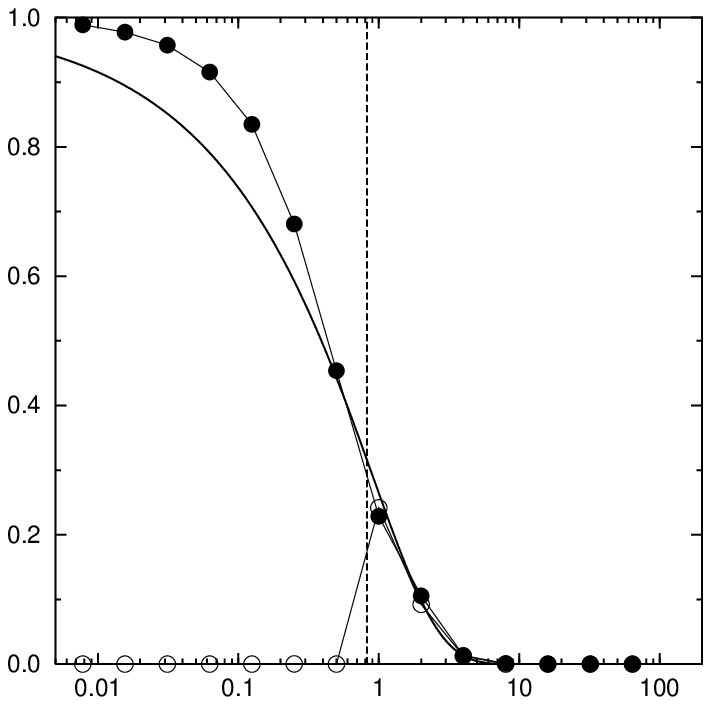}}
  \put(50,-10){\here{$\alpha$}} \put(80,78){$\phi_1$}
 \put(100,0){\epsfxsize=133\unitlength\epsfbox{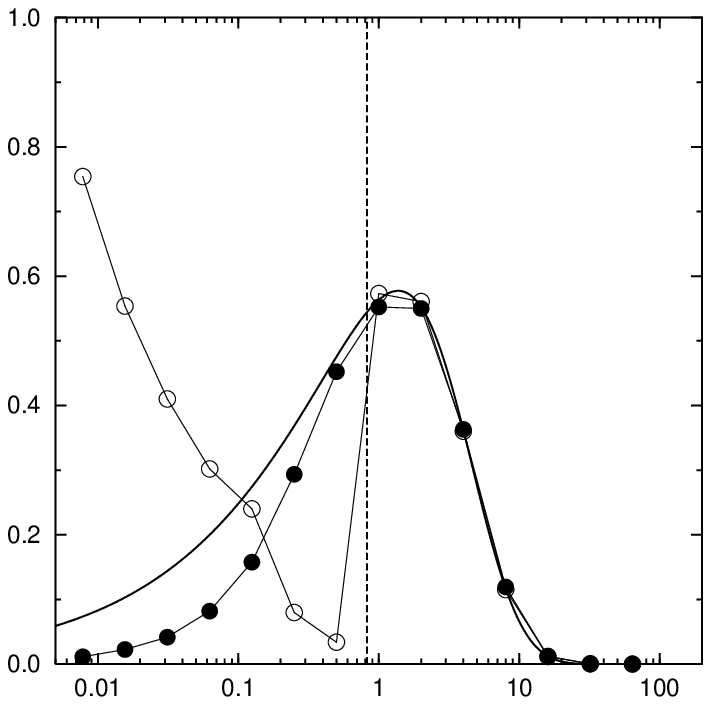}}
  \put(150,-10){\here{$\alpha$}} \put(180,78){$\phi_2$}
 \put(200,0){\epsfxsize=133\unitlength\epsfbox{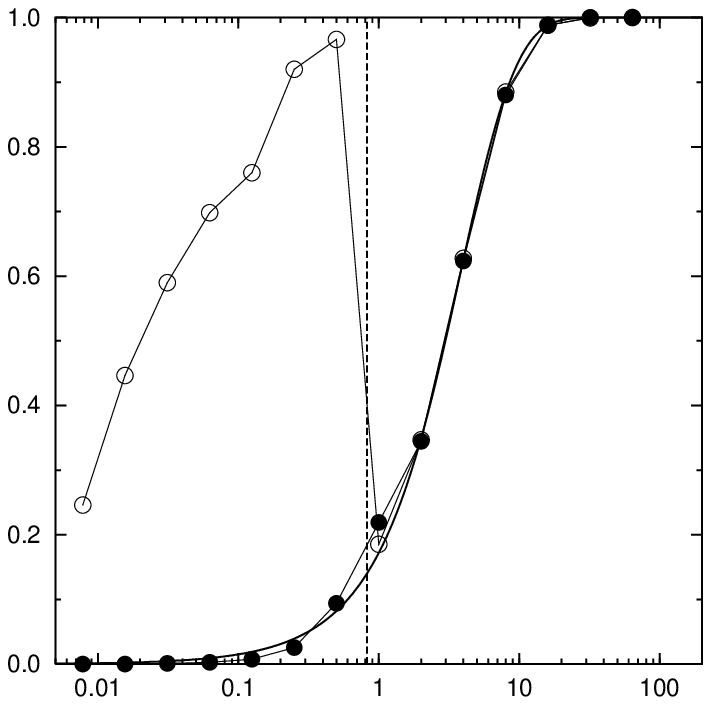}}
  \put(250,-10){\here{$\alpha$}}  \put(280,78){$\phi_3$}
\end{picture}
\vspace*{4mm} \caption{The fractions $\phi_\ell$ for $S=3$ are
shown together with corresponding data measured in numerical
simulations of MGs without decision noise, for both biased initial
conditions (random initial strategy valuations drawn from
$[-10,10]$, $\bullet$) and for unbiased initial conditions (random
initial strategy valuations drawn from $[-10^{-4},10^{-4}]$,
$\circ$). Simulation system size: $N=4097$. Dashed: the phase
transition point $\alpha_c(3)\approx 0.824$.
 } \label{fig:phi_S3}
\end{figure}
Upon solving these equations one finds that $\alpha_c\approx
0.824$. After doing the integrals in our expressions
 (\ref{eq:phi_l},\ref{eq:phi_S}) for the fractions
 $\phi_\ell$
 of agents that play $\ell$ of their three strategies one obtains
 the following explicit formulae (expressed once more in terms of $u=\sqrt{\alpha/c}$, i.e. in terms of the solution of our previous equation for $c$),
 which indeed obey
 $\phi_1+\phi_2+\phi_3=1$:
 \begin{eqnarray}
 \phi_1&=&1-\frac{3}{2}~\erf(\frac{u}{2})+\frac{3}{2}I(u)
 \\
 \phi_2&=&
 \frac{3}{2}~\erf(\frac{u}{2})\Big[1-\erf(\frac{u}{2\sqrt{3}})\Big]
 \\
 \phi_3&=&\frac{3}{2}~\erf(\frac{u}{2})~\erf(\frac{u}{2\sqrt{3}})-\frac{3}{2}I(u)
 \end{eqnarray}
These three expressions  are shown as functions of $\alpha$ in
figure \ref{fig:c_S3}, and tested against numerical simulations in
figure \ref{fig:phi_S3}. For large $\alpha$ the agents tend to use
all three strategies, upon reducing $\alpha$ one finds increasing
numbers switching to the use of only one or only two of their
strategies. Again, as in the previous figures involving $c$, we
observe a perfect agreement between theory and simulations in the
regime where our theory applies, i.e. for $\alpha>\alpha_c$.
Furthermore, the value found for $\alpha_c$ is consistent with the
simulation data in that non-ergodicity (a dependence on initial
conditions, i.e. on whether one chooses biased or unbiased
strategy valuations at time zero) is indeed seen to set in at the
predicted point.

Finally, perhaps the most sensitive test of our $S=3$ theory is to
work out and validate our formula (\ref{eq:rhof}) for the strategy
frequency distribution. Upon doing the relevant integrals we find
confirmed that $\varrho(f)=0$ for $f\notin[0,1]$ (as it should),
whereas for $f\in[0,1]$ we arrive at
\begin{eqnarray}
\varrho(f)&=&\frac{u}{\sqrt{\pi}}~\rme^{-u^2(f-\frac{1}{2})^2}\Big[1-\erf(\frac{u}{2\sqrt{3}})\Big]+\frac{u\sqrt{3}}{2\sqrt{\pi}}~\rme^{-\frac{3}{4}u^2(f-\frac{1}{3})^2}
\erf(\frac{u}{2}(1-f)) \nonumber
\\
&& +\delta(f)
(\frac{2}{3}\phi_1+\frac{1}{3}\phi_2)+\delta(f-1)\frac{1}{3}\phi_1
\label{eq:rhof_S3}
\end{eqnarray}
\begin{figure}[t]
\vspace*{-3mm} \hspace*{30mm} \setlength{\unitlength}{0.35mm}
\begin{picture}(400,450)
  \put(20,300){\epsfxsize=200\unitlength\epsfbox{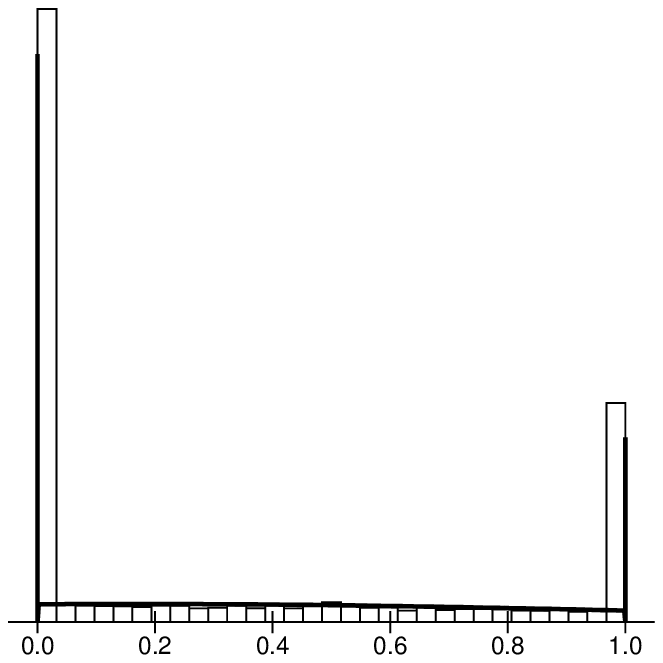}}
  \put(85,285){\here{$f$}}\put(-30,330){\here{$\alpha=0.5$}}

   \put(20,200){\epsfxsize=200\unitlength\epsfbox{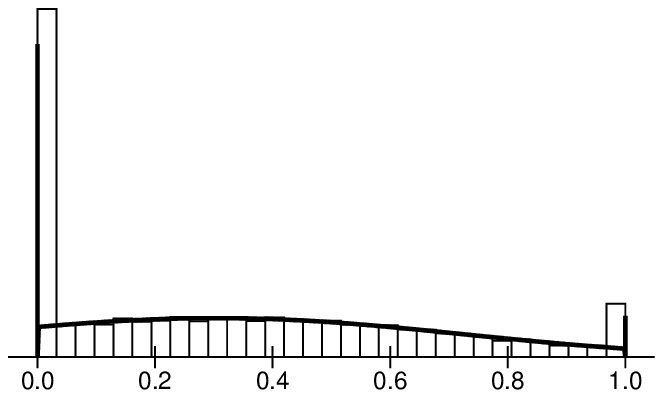}}
  \put(85,185){\here{$f$}}\put(-30,230){\here{$\alpha=2$}}

 \put(20,100){\epsfxsize=200\unitlength\epsfbox{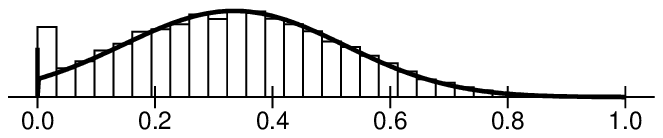}}
  \put(85,85){\here{$f$}}\put(-30,130){\here{$\alpha=8$}}

 \put(20,0){\epsfxsize=200\unitlength\epsfbox{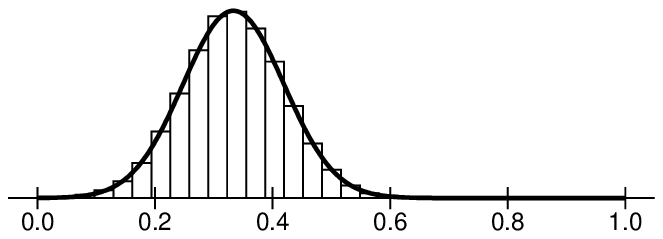}}
   \put(85,-15){\here{$f$}}\put(-30,30){\here{$\alpha=32$}}

  \put(190,300){\epsfxsize=200\unitlength\epsfbox{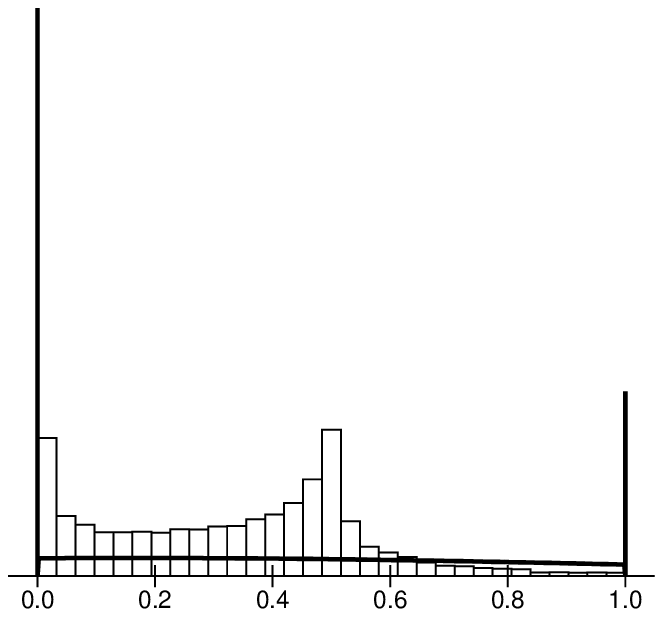}}
  \put(255,285){\here{$f$}}

   \put(190,200){\epsfxsize=200\unitlength\epsfbox{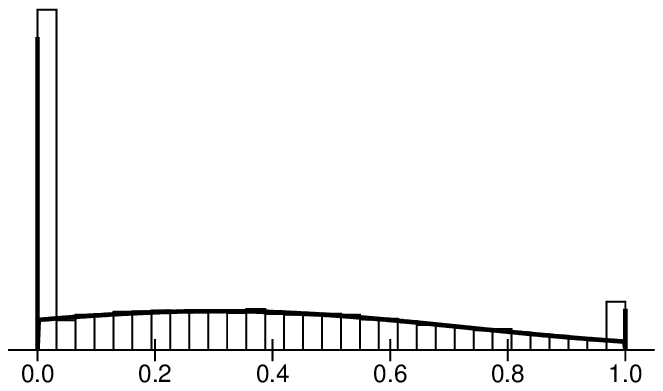}}
  \put(255,185){\here{$f$}}

 \put(190,100){\epsfxsize=200\unitlength\epsfbox{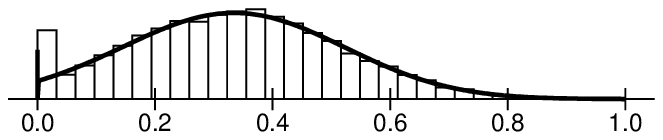}}
  \put(255,85){\here{$f$}}

 \put(190,0){\epsfxsize=200\unitlength\epsfbox{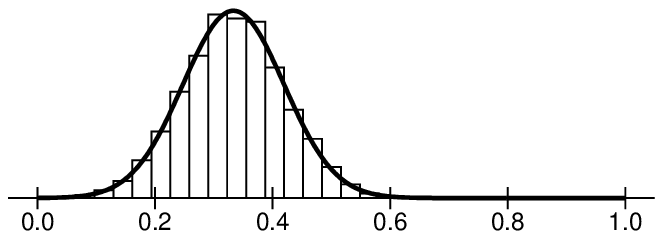}}
   \put(255,-15){\here{$f$}}

\end{picture}
 \vspace*{5mm}
\caption{Examples of the predicted strategy frequency distribution
$\varrho(f)$ (thick curves), for $S=3$ and different values of
$\alpha=p/N$, as calculated for ergodic stationary states. They
are shown together with strategy frequency  data measured in
numerical simulations of MGs without decision noise (shown as
histograms, averaged over three samples), for both biased initial
conditions (random initial strategy valuations drawn from
$[-10,10]$, left graphs) and for unbiased initial conditions
(random initial strategy valuations drawn from
$[-10^{-4},10^{-4}]$, right graphs). Simulation system size:
$N=4097$. Note that $\alpha_c(3)\approx  0.824$, so the top two
graphs refer to the nonergodic regime, where the present theory is
not supposed to apply.} \label{fig:rhof_S3} \vspace*{-2mm}
\end{figure}
One can understand this formula qualitatively. A strategy $a$ is
not played at all when either $\ell=1$ and it is among the two
non-selected strategies (this happens with probability
$\frac{2}{3}\phi_1$), or when $\ell=2$ and it is the one
non-selected strategy (this happens with probability
$\frac{1}{3}\phi_2$). A strategy $a$ is played permanently if
$\ell=1$ and it is the selected one (this happens with probability
$\frac{1}{3}\phi_1$). Hence the second line of (\ref{eq:rhof_S3}).
The first line of (\ref{eq:rhof_S3}) reflects $a$ being played now
and then (among other strategies), with non-trivial frequencies;
here one observes the expected local maxima at $f=\frac{1}{2}$
(corresponding to the `average' behaviour for $\ell=2$) and at
$f=\frac{1}{3}$ (corresponding to the `average' behaviour for
$\ell=3$). Also the prediction (\ref{eq:rhof_S3}) agrees perfectly
with numerical simulations, as shown in figure \ref{fig:rhof_S3},
in the regime $\alpha>\alpha_c$ for which is is supposed to be
correct.

In summary, we may conclude that for $S=3$ all our theoretical
predictions regarding stationary state order parameters, the
location of the phase transition, and even quantities such as the
strategy frequency distribution, make physical sense and find
perfect confirmation in numerical simulations.

\section{Application to MGs with $S=4,5$}

\begin{figure}[t]
\vspace*{-5mm} \hspace*{25mm} \setlength{\unitlength}{0.40mm}
\begin{picture}(290,160)
  \put(0,0){\epsfxsize=200\unitlength\epsfbox{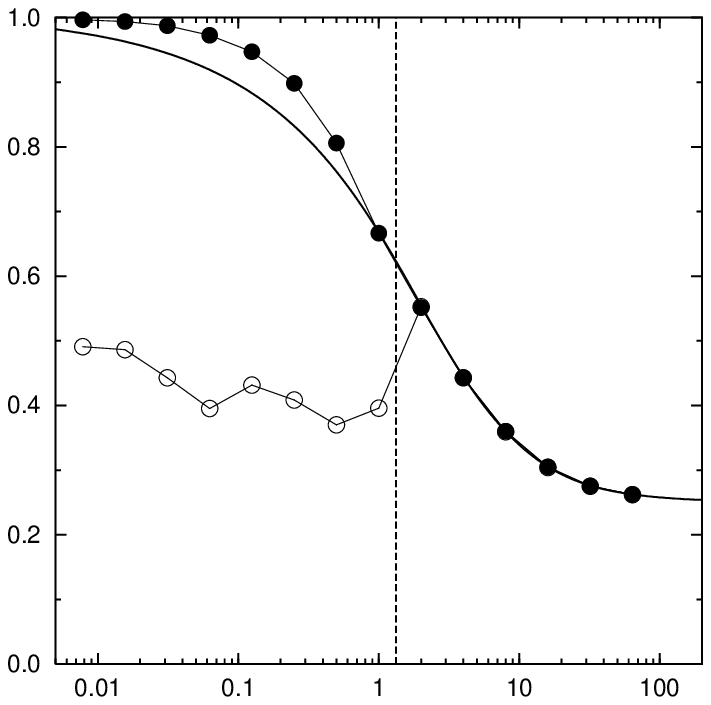}}
  \put(80,-10){\here{$\alpha$}}\put(-8,70){\large\here{$c$}}
    \put(150,0){\epsfxsize=200\unitlength\epsfbox{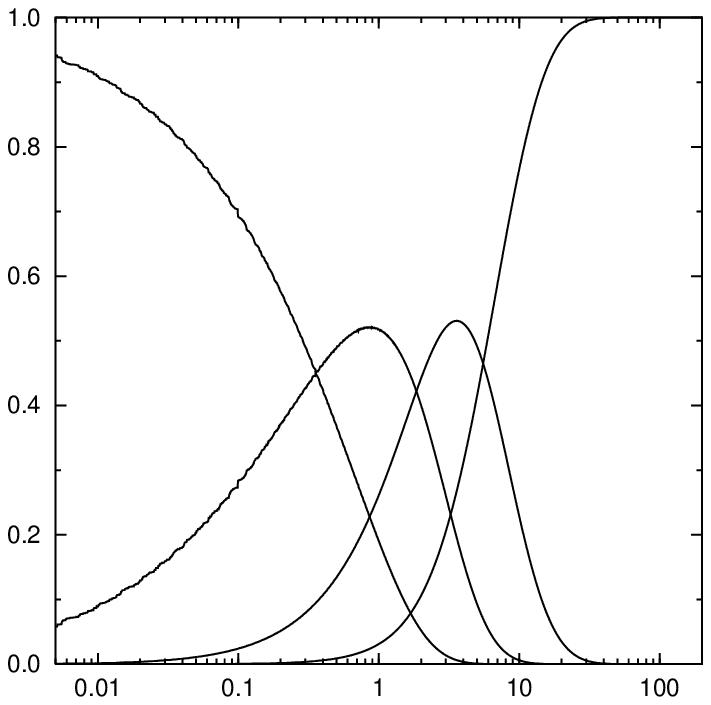}}
  \put(230,-10){\here{$\alpha$}}
  \put(173,107){$\phi_1$} \put(190,45){$\phi_2$} \put(265,25){$\phi_3$} \put(268,118){$\phi_4$}
\end{picture}
\vspace*{6mm} \caption{Left: the predicted persistent correlations
$c=\lim_{\tau\to\infty}C(\tau)$ (solid curve) for $S=4$, as a
function of $\alpha=p/N$, calculated for ergodic stationary
states. It is shown together with corresponding data measured in
numerical simulations of MGs without decision noise, for both
biased initial conditions (random initial strategy valuations
drawn from $[-10,10]$, $\bullet$) and for unbiased initial
conditions (random initial strategy valuations drawn from
$[-10^{-4},10^{-4}]$, $\circ$). Simulation system size: $N=4097$.
Dashed: the phase transition point $\alpha_c(4)\approx  1.324$.
Right: the fractions $\phi_\ell$ of agents that play precisely
$\ell\in\{1,2,3,4\}$ of their $S=4$ strategies,
 as a
function of $\alpha=p/N$, calculated for ergodic stationary
states. Note that $\phi_1+\phi_2+\phi_3+\phi_4=1$. }
 \label{fig:c_S4}
\end{figure}
\begin{figure}[t]
\vspace*{-1mm} \hspace*{32mm} \setlength{\unitlength}{0.53mm}
\begin{picture}(300,200)
 \put(0,100){\epsfxsize=133\unitlength\epsfbox{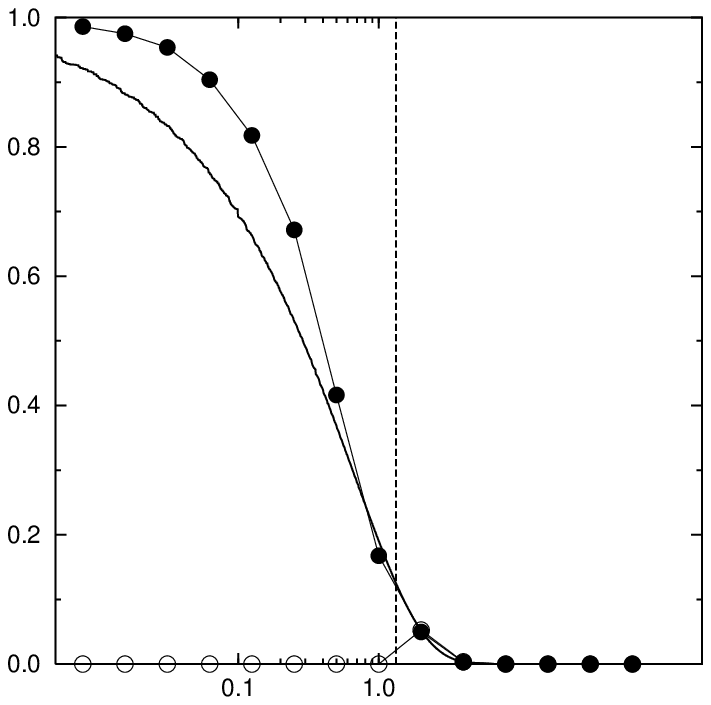}}
 \put(80,178){$\phi_1$}
 \put(100,100){\epsfxsize=133\unitlength\epsfbox{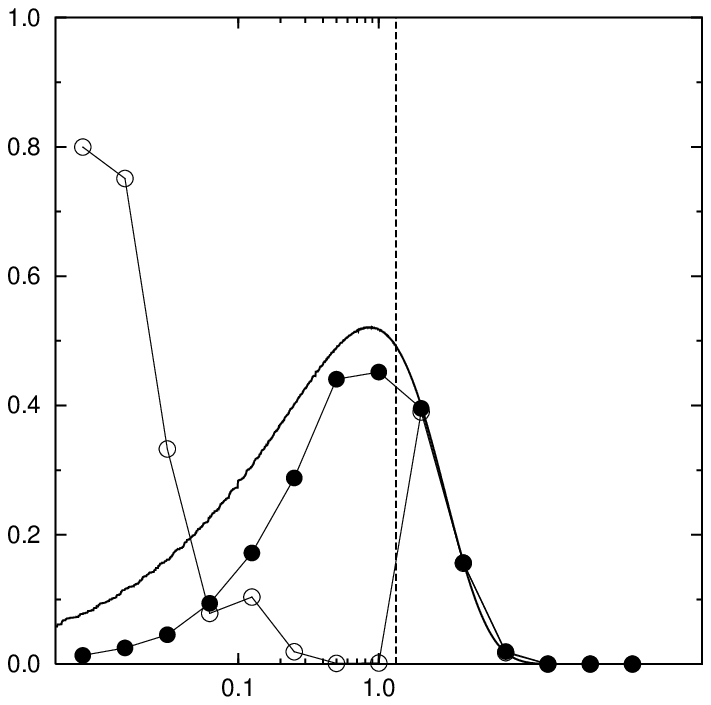}}
  \put(180,178){$\phi_2$}

   \put(0,0){\epsfxsize=133\unitlength\epsfbox{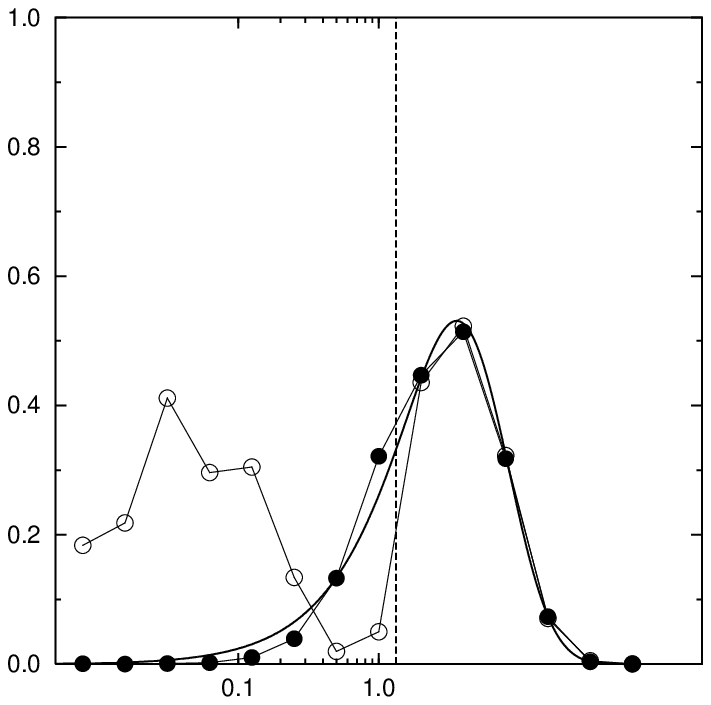}}
  \put(50,-10){\here{$\alpha$}} \put(80,78){$\phi_3$}
 \put(100,0){\epsfxsize=133\unitlength\epsfbox{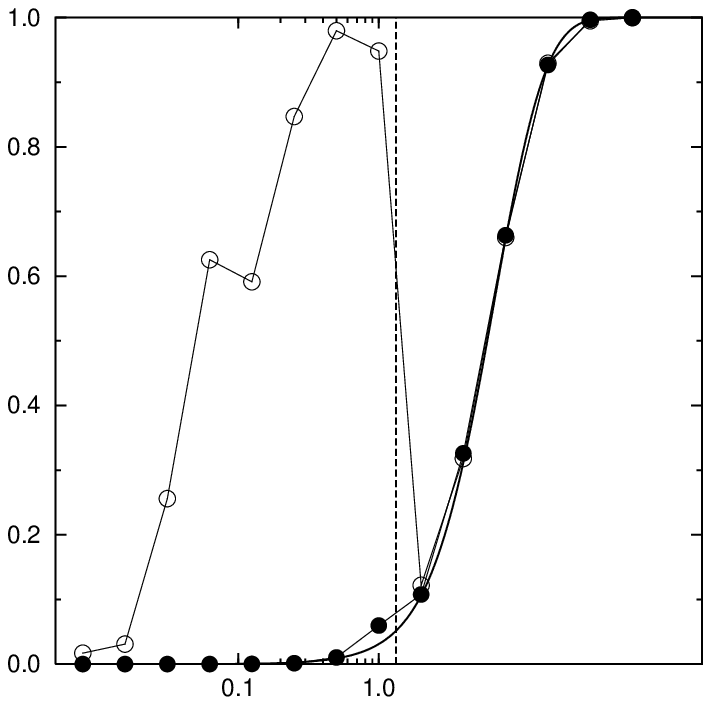}}
  \put(150,-10){\here{$\alpha$}} \put(180,78){$\phi_4$}
\end{picture}
\vspace*{4mm} \caption{The fractions $\phi_\ell$ for $S=4$ are
shown together with corresponding data measured in numerical
simulations of MGs without decision noise, for both biased initial
conditions (random initial strategy valuations drawn from
$[-10,10]$, $\bullet$) and for unbiased initial conditions (random
initial strategy valuations drawn from $[-10^{-4},10^{-4}]$,
$\circ$). Simulation system size: $N=4097$. Dashed: the phase
transition point $\alpha_c(4)\approx  1.324$.
 } \label{fig:phi_S4}
\end{figure}
We now turn to $S>3$. Although it is in principle still possible
to proceed with the various integrals in
(\ref{eq:final_c_eq},\ref{eq:final_chi_eq},\ref{eq:phi_l},\ref{eq:phi_S},\ref{eq:rhof})
analytically, for $S>3$ this would become a very tedious and
time-consuming exercise. We have here resorted to numerical
evaluation; one does not expect qualitatively different physics to
emerge compared to $S=3$, and life is short. For reasons of
brevity we have also restricted ourselves to the validation of the
static observables $c$ and $\phi_\ell$ only. It should be
mentioned that upon increasing the value of $S$, the accurate
numerical evaluation (based on the Gauss-Legendre method) of the
various nested integrals in our theoretical expressions becomes
nontrivial in terms of CPU quite quickly; especially for small
values of $\alpha$. Finding the expressions for $\alpha_c(S)$ as
presented in section \ref{sec:finaltheory} for $S>3$ already
involved a careful assessment of the scaling of these values with
the parameters that control the numerical integration accuracy.
One also finds that for increasing values of $S$ the finite size
effects in the non-ergodic regime $\alpha<\alpha_c(S)$ become more
prominent.

\begin{figure}[t]
\vspace*{-5mm} \hspace*{25mm} \setlength{\unitlength}{0.40mm}
\begin{picture}(290,160)
  \put(0,0){\epsfxsize=200\unitlength\epsfbox{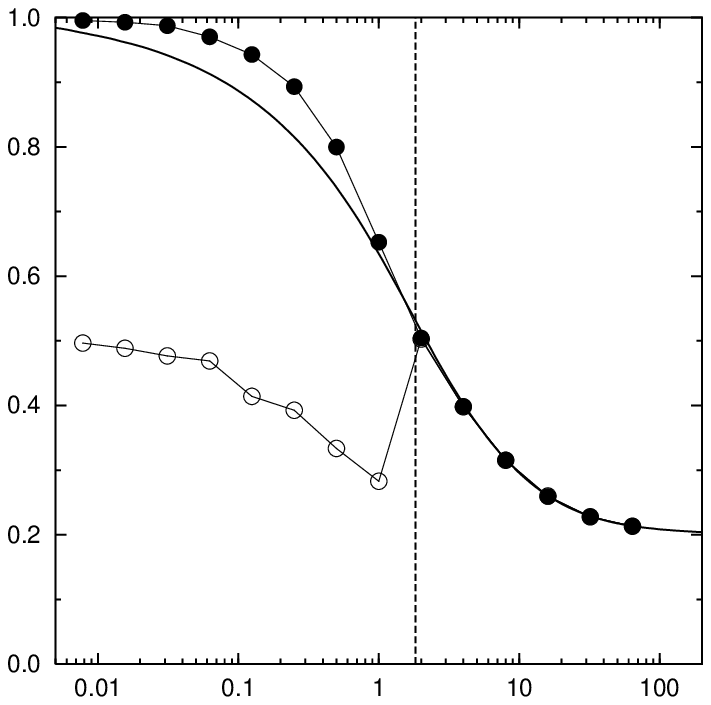}}
  \put(80,-10){\here{$\alpha$}}\put(-8,70){\large\here{$c$}}
    \put(150,0){\epsfxsize=200\unitlength\epsfbox{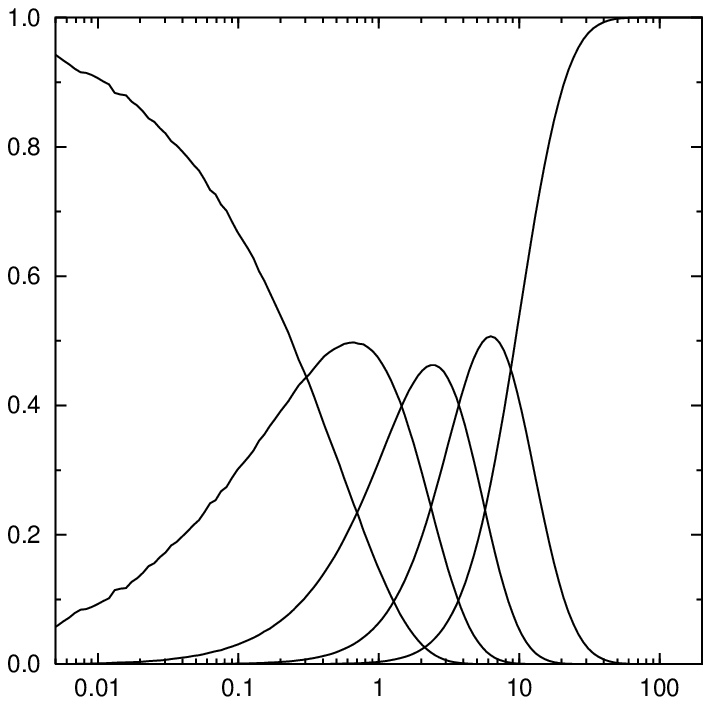}}
  \put(230,-10){\here{$\alpha$}}
   \put(173,107){$\phi_1$} \put(189,45){$\phi_2$} \put(205,23){$\phi_3$} \put(269,25){$\phi_4$}  \put(273,118){$\phi_5$}
\end{picture}
\vspace*{6mm} \caption{Left: the predicted persistent correlations
$c=\lim_{\tau\to\infty}C(\tau)$ (solid curve) for $S=5$, as a
function of $\alpha=p/N$, calculated for ergodic stationary
states. It is shown together with corresponding data measured in
numerical simulations of MGs without decision noise, for both
biased initial conditions (random initial strategy valuations
drawn from $[-10,10]$, $\bullet$) and for unbiased initial
conditions (random initial strategy valuations drawn from
$[-10^{-4},10^{-4}]$, $\circ$). Simulation system size: $N=4097$.
Dashed: the phase transition point $\alpha_c(5)\approx  1.822$.
Right: the fractions $\phi_\ell$ of agents that play precisely
$\ell\in\{1,2,3,4,5\}$ of their $S=5$ strategies,
 as a
function of $\alpha=p/N$, calculated for ergodic stationary
states.}
 \label{fig:c_S5}
\end{figure}

\begin{figure}[t]
\vspace*{-1mm} \hspace*{-2mm} \setlength{\unitlength}{0.53mm}
\begin{picture}(300,200)
 \put(0,100){\epsfxsize=133\unitlength\epsfbox{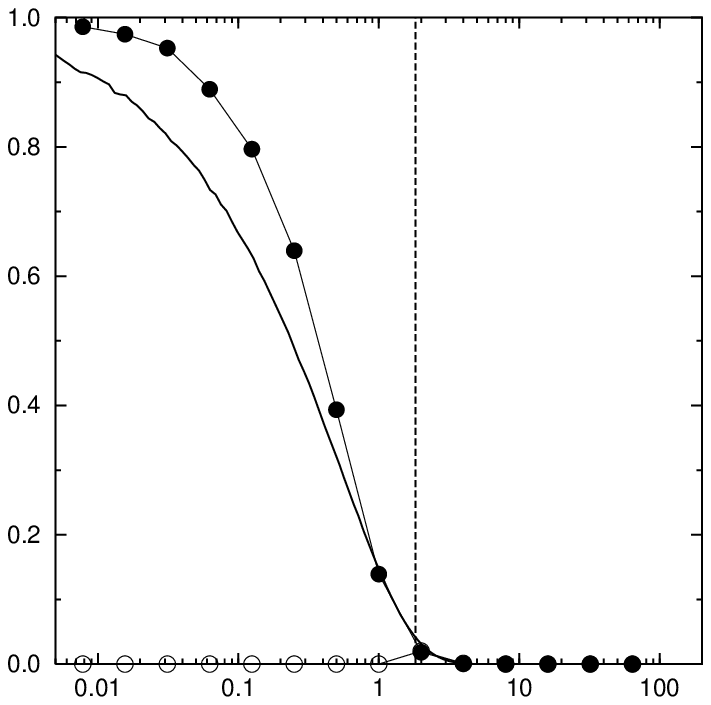}}
 \put(80,178){$\phi_1$}
 \put(100,100){\epsfxsize=133\unitlength\epsfbox{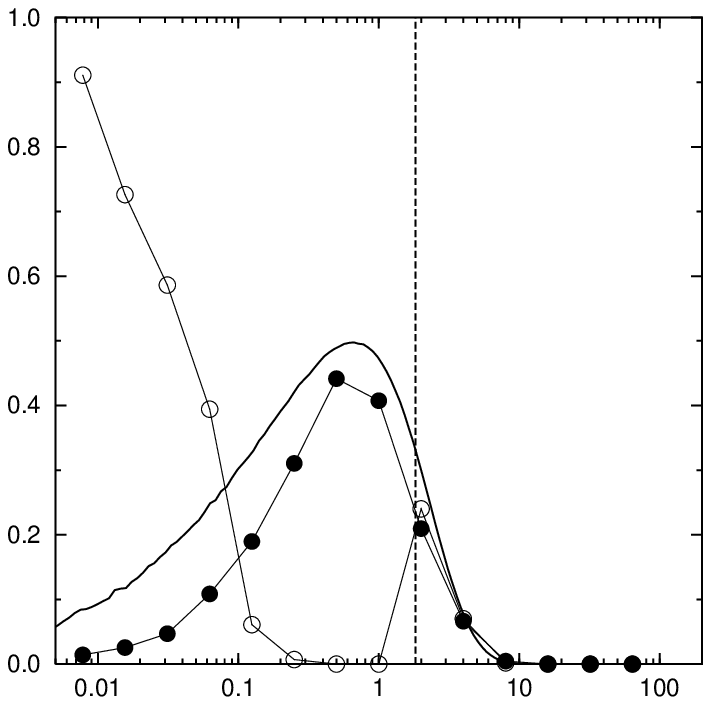}}
 \put(180,178){$\phi_2$}
 \put(200,100){\epsfxsize=133\unitlength\epsfbox{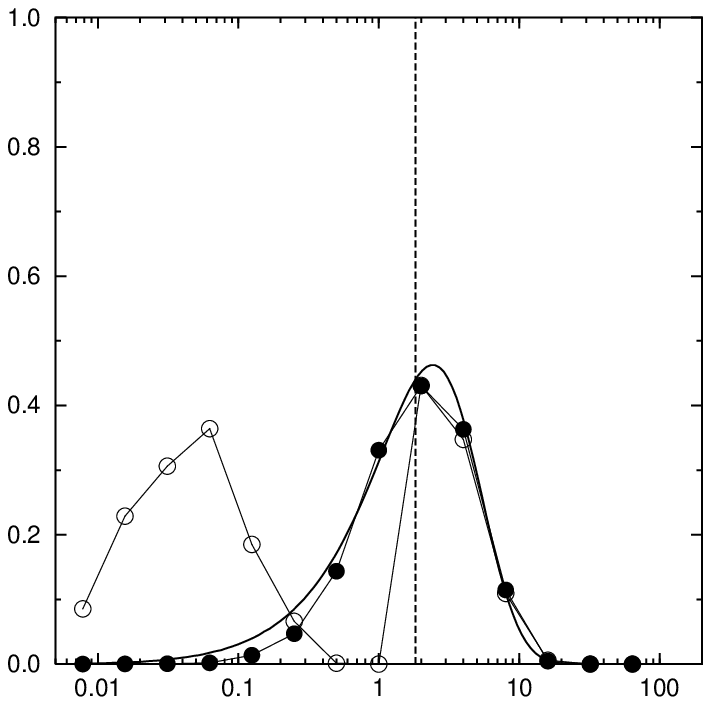}}
  \put(250,90){\here{$\alpha$}}  \put(280,178){$\phi_3$}
  \put(0,0){\epsfxsize=133\unitlength\epsfbox{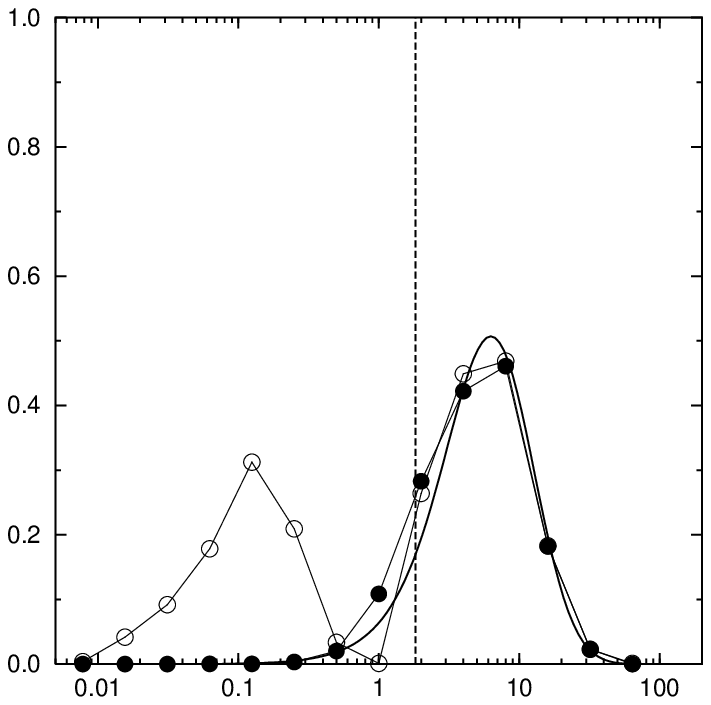}}
  \put(50,-10){\here{$\alpha$}} \put(80,78){$\phi_4$}
 \put(100,0){\epsfxsize=133\unitlength\epsfbox{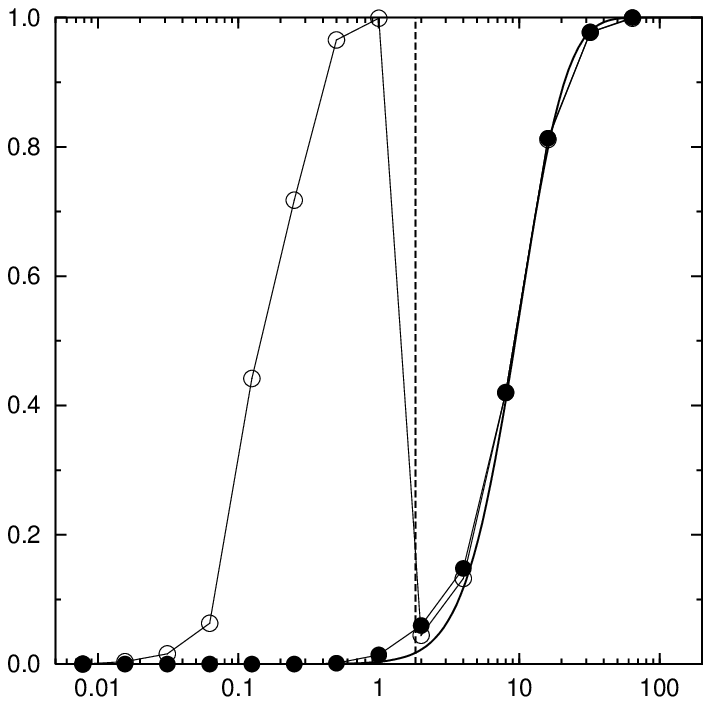}}
  \put(150,-10){\here{$\alpha$}} \put(180,78){$\phi_5$}
\end{picture}
\vspace*{4mm} \caption{The fractions $\phi_\ell$ for $S=5$ are
shown together with corresponding data measured in numerical
simulations of MGs without decision noise, for both biased initial
conditions (random initial strategy valuations drawn from
$[-10,10]$, $\bullet$) and for unbiased initial conditions (random
initial strategy valuations drawn from $[-10^{-4},10^{-4}]$,
$\circ$). Simulation system size: $N=4097$. Dashed: the phase
transition point $\alpha_c(5)\approx 1.822$.
 } \label{fig:phi_S5}
\end{figure}

For $S=4$ the results of evaluating numerically the theoretical
predictions
(\ref{eq:final_c_eq},\ref{eq:final_chi_eq},\ref{eq:phi_l},\ref{eq:phi_S})
are shown and tested against simulation data (obtained for MGs
without decision noise) in figures \ref{fig:c_S4} and
\ref{fig:phi_S4}. For $S=5$ they are shown and tested in figures
\ref{fig:c_S5} and \ref{fig:phi_S5}. It is very satisfactory that
once more we observe for $S\in\{4,5\}$ in all cases an excellent
agreement between theory and simulation data, both with regards to
the observables measured (viz. $c$ and the fractions $\phi_\ell$),
and in terms of the locations of the transition points
$\alpha_c(4)$ and $\alpha_c(5)$.

\section{Stationary state fluctuations: volatility and predictability}

\begin{figure}[t]
\vspace*{-5mm} \hspace*{45mm} \setlength{\unitlength}{0.40mm}
\begin{picture}(200,160)
  \put(0,0){\epsfxsize=200\unitlength\epsfbox{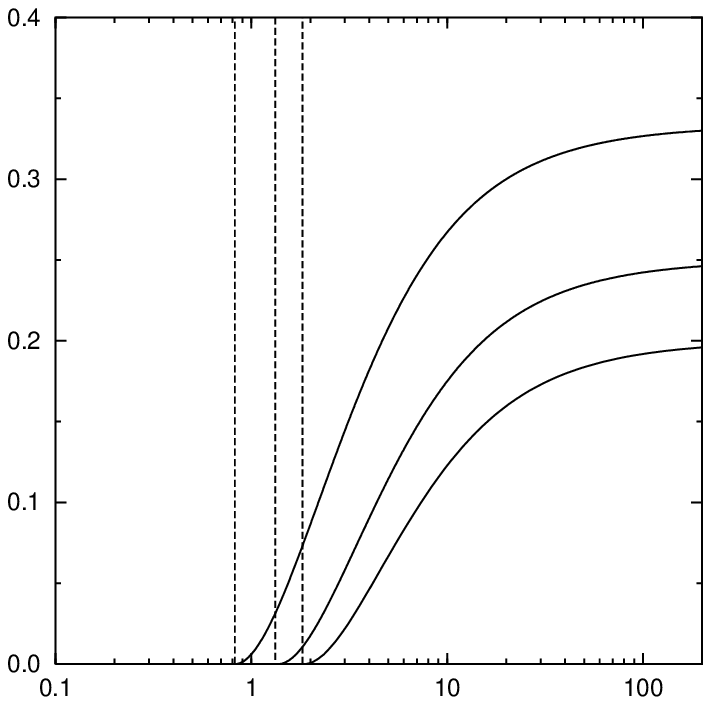}}
  \put(80,-10){\here{$\alpha$}}\put(-8,70){\here{$H$}}
  \put(150,109){$S=3$}   \put(150,82){$S=4$}   \put(150,65){$S=5$}
\end{picture}
\vspace*{6mm} \caption{The measure
$H=\lim_{N\to\infty}p^{-1}\sum_\mu
[\lim_{\tau\to\infty}\tau^{-1}\sum_{t=1}^\tau (A_\mu(t)-\bra
A(t)\ket)]^2$ of the overall market bid predictability, for
time-translation invariant states, as a function of the control
parameter  $\alpha=p/N$ for $S\in\{3,4,5\}$ (from top to bottom).
Dashed: the corresponding phase transition points $\alpha_c(S)$.}
 \label{fig:H}
\end{figure}

Once the static order parameters $c$ and $\chi$ are known, our
formulae (\ref{eq:sigma_and_H}) for the volatility $\sigma$ and
the predictability measure $H$ can be reduced to
\be
\sigma^2=\tilde{\eta}^{-2}\sum_{tt^\prime\geq 0}R(t)C(t-t^\prime)
R(t^\prime) ~~~~~~~~ H=\frac{c}{(1+\tilde{\eta}\chi)^2} \ee As for
$S=2$, $H$ can always be expressed fully in terms of persistent
order parameters and vanishes at the phase transition point. The
result is shown in figure \ref{fig:H} for $S\in\{3,4,5\}$. If in
the volatility formula
 we separate the correlation function into a
static and a non-persistent part,  $C(t)=c+\tilde{C}(t)$ with
$\lim_{t\to\pm\infty}\tilde{C}(t)=0$, we obtain
\begin{eqnarray}
\sigma^2&=&\frac{c}{(1+\tilde{\eta}\chi)^2}+\sum_{tt^\prime\geq
0}(\one+\tilde{\eta}G)^{-1}(t)\tilde{C}(t-t^\prime)
(\one+\tilde{\eta}G)^{-1}(t^\prime)\nonumber
\\
&=& \frac{c}{(1+\tilde{\eta}\chi)^2} +(1-c)
\sum_{t}[(\one+\tilde{\eta}G)^{-1}(t)]^2 \nonumber
\\
&& +\sum_{t\neq
t^\prime}(\one+\tilde{\eta}G)^{-1}(t)\tilde{C}(t-t^\prime)
(\one+\tilde{\eta}G)^{-1}(t^\prime)
 \label{eq:vola_worked_out}
\end{eqnarray}
Expression (\ref{eq:vola_worked_out}) is still exact. The first
term is recognized to be $H$. The other terms contain the
short-time fluctuations, involving non-persistent dynamic order
parameters. If one seeks a formula for $\sigma$ in terms static
order parameters only, one must pay the price of approximation. To
do this we generalize the two procedures described for the $S=2$
batch MG in e.g. \cite{HeimelCoolen01,CoolHeimSher01} and
\cite{MGbook2}. Both rely on the ansatz, motivated by observations
in simulations, that the response  function decays to zero very
slowly, subject to the constraint $\sum_t G(t)=\chi$. Upon putting
e.g. $G(t>0)=\chi(e^z-1)e^{-zt}$ with $z\to 0$ one finds for such
kernels that $\lim_{z\to 0} \sum_t[(\one+\tilde{\eta}G)(t)]^2=0$.
Consequently we may write
\begin{figure}[t]
\vspace*{2mm} \hspace*{-2mm} \setlength{\unitlength}{0.53mm}
\begin{picture}(300,100)
 \put(0,0){\epsfxsize=133\unitlength\epsfbox{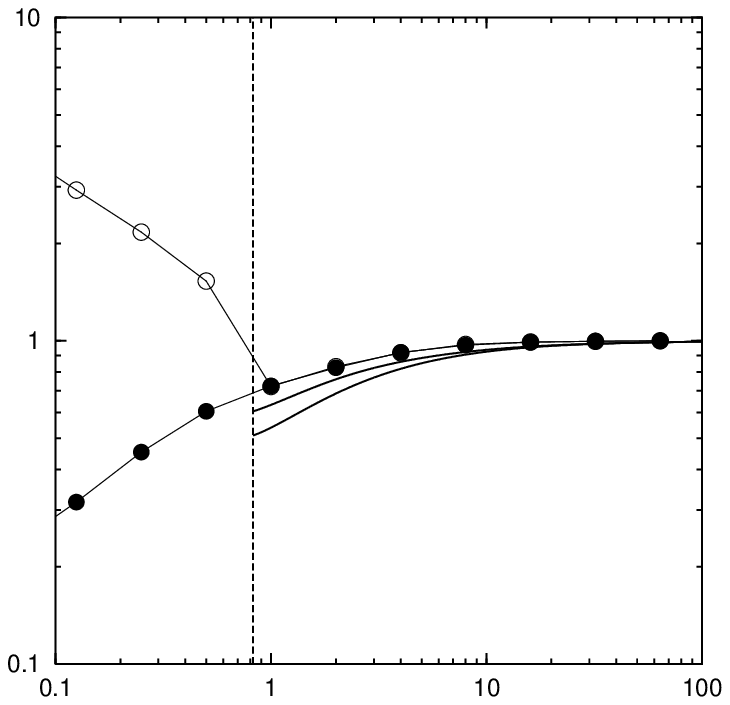}}
  \put(50,-10){\here{$\alpha$}}\put(-5,47){$\sigma$}
  \put(70,80){$S=3$}
 \put(100,0){\epsfxsize=133\unitlength\epsfbox{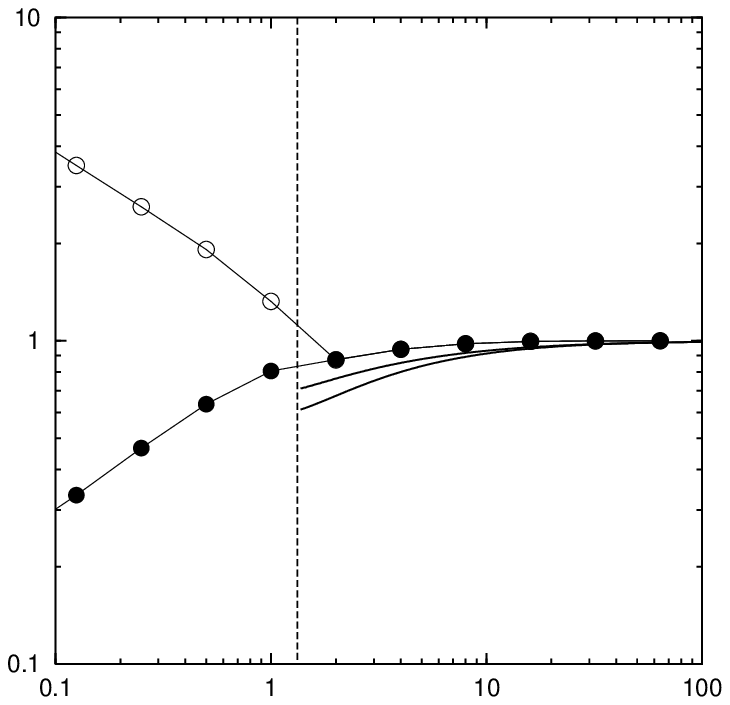}}
  \put(150,-10){\here{$\alpha$}}
  \put(170,80){$S=4$}
 \put(200,0){\epsfxsize=133\unitlength\epsfbox{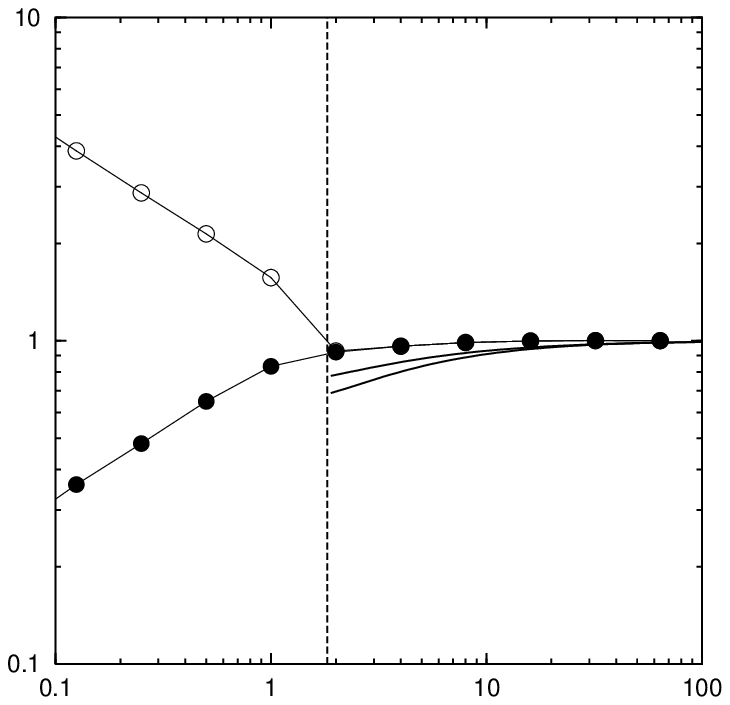}}
  \put(250,-10){\here{$\alpha$}}
  \put(270,80){$S=5$}
\end{picture}
\vspace*{4mm} \caption{The two approximate volatility formulas
$\sigma_A$ and $\sigma_B$  (lower and upper solid curves,
respectively) as functions of $\alpha=p/N$, for $S\in\{3,4,5\}$,
calculated in the ergodic regime $\alpha>\alpha_c$. They are shown
together with volatility data measured in numerical simulations of
MGs without decision noise, for both biased initial conditions
(random initial strategy valuations drawn from $[-10,10]$,
$\bullet$) and for unbiased initial conditions (random initial
strategy valuations drawn from $[-10^{-4},10^{-4}]$, $\circ$).
Simulation system size: $N=4097$. Dashed: the corresponding phase
transition points $\alpha_c(S)$.
 } \label{fig:volatility}
\end{figure}
\begin{eqnarray}
\sigma^2&\approx & \frac{c}{(1+\tilde{\eta}\chi)^2} +\sum_{t\neq
t^\prime}(\one+\tilde{\eta}G)^{-1}(t)\tilde{C}(t-t^\prime)
(\one+\tilde{\eta}G)^{-1}(t^\prime)
\end{eqnarray}
The most brutal approximation of $\tilde{C}$ is to assume the
correlations to decay to zero very fast, and simply put
$\tilde{C}(t)\to \tilde{C}(0)\delta_{t
0}=(1-c)\delta_{tt^\prime}$. This leads to the formula
\begin{eqnarray}
\sigma_{A}^2&=&\frac{c}{(1+\tilde{\eta}\chi)^2}+1-c
\label{eq:vola_A}
\end{eqnarray}
Alternatively, we may write the average in the definition
(\ref{eq:CandG}) of the correlation function as a sum of
contributions representing the possible sizes $\ell$ of the set of
active strategies:
\begin{eqnarray}
C(t-t^\prime)
&=&\delta_{tt^\prime}+(1-\delta_{tt^\prime})\sum_{\ell=1}^S
\phi_\ell  \sum_{a}\bra \delta_{a,m(\bv(t),\bz(t)}
\delta_{a,m(\bv(t^\prime),\bz(t^\prime)}\ket_{\ell~{\rm active}}
\end{eqnarray}
This is still exact, but now we approximate for $t\neq t^\prime$
\begin{eqnarray*}
\hspace*{-5mm} \sum_a \bra \delta_{a,m(\bv(t),\bz(t)}
\delta_{a,m(\bv(t^\prime),\bz(t^\prime)}\ket_{\ell~{\rm
active}}&\approx& \sum_a\bra
\delta_{a,m(\bv(t),\bz(t)}\ket_{\ell~{\rm active}}\bra
\delta_{a,m(\bv(t^\prime),\bz(t^\prime)}\ket_{\ell~{\rm active}}\\
&\approx&\ell^{-1}
\end{eqnarray*}
 This gives
 $\tilde{C}(t-t^\prime)\approx
 \delta_{tt^\prime}-c+(1-\delta_{tt^\prime})\sum_{\ell=1}^S\phi_\ell/\ell$
 and hence the approximation
\begin{eqnarray}
\sigma_B^2&=&
\frac{\sum_{\ell=1}^S\phi_\ell/\ell}{(1+\tilde{\eta}\chi)^2}+
1-\sum_{\ell=1}^S\phi_\ell/\ell \label{eq:vola_B}
\end{eqnarray}
Formula (\ref{eq:vola_A}) is identical to what was already
proposed for $S=2$ in \cite{MarsChalZecc00}; formula
(\ref{eq:vola_B}) generalizes to arbitary values of $S$ the $S=2$
volatility approximation first published in
\cite{HeimelCoolen01,CoolHeimSher01}. Our two approximation
formulae have more or less opposite deficits, since
(\ref{eq:vola_A}) fails to incorporate any transient contributions
to the correlations (here $\tilde{C}(t)=0$ for $t\neq 0$), whereas
(\ref{eq:vola_B}) incorporates too many (here
$\lim_{t\to\infty}\tilde{C}(t)=\sum_{\ell=1}^S\phi_\ell/\ell-c\neq
0$). For batch MGs one finds that (\ref{eq:vola_B}) is more
accurate than (\ref{eq:vola_A}), see e.g. figure
\ref{fig:volatility}.

\section{Dynamics for very small and very large $\alpha$}

 Solving our order parameters $C$ and $G$ from (\ref{eq:CandG})
for finite times is generally non-trivial due to the non-Markovian
nature of the effective process (\ref{eq:effective_agent1}). Only
for extreme values of $\alpha$, viz. for $\alpha\to 0$ and
$\alpha\to \infty$, and for short times it can to some extent be
done. The causality constraint $G_{tt'}=0$ for $t\geq t^\prime$
allows us to calculate with little difficulty the first few time
steps and gain a better understanding of the nature of the
dynamics, especially with regard to the role of the initial
conditions and the dependence on the control parameter $\alpha$.
 Causality implies that
$(G^n)_{tt^\prime}=0$ for $t-t^\prime<n$, so that for $t\geq
t^\prime\geq 0$:
\begin{eqnarray}
 R_{tt^\prime}&=&\tilde{\eta}(\one+\tilde{\eta}G)^{-1}_{tt^\prime}=
 \tilde{\eta}\sum_{n=0}^{t-t^\prime}(-\tilde{\eta})^{n}(G^n)_{tt^\prime}
 \\
 \Sigma_{tt^\prime}&=&
 \bra\eta_a(t)\eta_b(t^\prime)\ket=
 \delta_{ab}
(RCR^\dag)_{tt^\prime}\nonumber
\\
&=&\frac{\delta_{ab}}{\tilde{\eta}^2}\sum_{s=0}^{t}\sum_{s^\prime=0}^{t^\prime}
\sum_{n=0}^{t-s}\sum_{n^\prime=0}^{t^\prime-s^\prime}(-\tilde{\eta})^{n+n^\prime}(G^n)_{ts}(G^{n^\prime})_{t^\prime
s^\prime}C_{ss^\prime}
\end{eqnarray}
For short times all these expressions reduce to simple finite
sums; for instance:
\begin{eqnarray*}
&R_{00}=R_{11}=\tilde{\eta},~~~~~& \Sigma_{00}=\tilde{\eta}^2
\\
 & R_{10}=- \tilde{\eta}^2G_{10},~~ &
 \Sigma_{10}=\tilde{\eta}^2(C_{10}-G_{10})\\
 & R_{21}=-\tilde{\eta}^2 G_{21},~~~ & \Sigma_{11}= \tilde{\eta}^2
-2\tilde{\eta}^3G_{10}C_{10}+\tilde{\eta}^4 (G_{10})^2\\
 & R_{20}=\tilde{\eta}^3G_{21}G_{10}-\tilde{\eta}^2 G_{20}~~~~&
\end{eqnarray*}
One can now in the familiar manner solve the effective single
agent agent process for the first few time steps. The resulting
expressions would be fully exact (for any $\alpha$, including
those in the nonergodic regime $\alpha<\alpha_c(S)$), but
increasingly involved. \vsp

 For very large and very small $\alpha$ matters
become relatively simple, but the result is still quite
informative. We will discuss only the deterministic case; adding
decision noise will only introduce transparent stochastic
variations of the behaviour described below.
 For large $\alpha$
and finite times the single agent process tells us that
$G=\order(1/\tilde{\eta}\sqrt{\alpha})$, that the effective
Gaussian noise scales as $\eta_a(t)=\order(1)$,  and that
\begin{eqnarray}
t=0:&~~~& v_a(1)= -\alpha \delta_{a,{\rm argmax}_b [v_b(0)]}
+\order(\sqrt{\alpha}) \label{eq:la1}
\\
t>0:&~~~& v_a(t+1)= v_a(t) -\alpha \delta_{a,{\rm argmax}_b
[v_b(t)]}+\order(\sqrt{\alpha}) \label{eq:la2}
\end{eqnarray}
Let us assume that $v_a(0)=\order(\alpha^0)$ for all $\alpha$, and
exclude the pathological case where two or more initial valuations
are identical. According to (\ref{eq:la1}), the strategy with the
largest initial valuation is selected at time $t=1$, but will then
move to the back of the list of ordered valuations. At the next
step the second largest is selected, and then, following
(\ref{eq:la2}), also moved to the back of the ordered list, etc.
The net result is that all valuations $v_a(t)$ will become
increasingly negative (proportional to $\alpha$), growing on
average linearly with time, and that the effective agent will
continually alternate his $S$ strategies in a {\em fixed} order,
being the order in which the valuations are ranked initially. If,
for example, $v_1(0)>v_2(0)>\ldots
>v_S(0)$, then ${\rm argmax}_b[v_b(t)]= t+1~{\rm mod}~S$, and for $t>0$ the
effective agent equation gives
\begin{eqnarray}
v_a(t)&=& -\alpha \sum_{t^\prime=0}^{t-1}
\delta_{a,t^\prime+1~{\rm mod}~S}+\order(t\sqrt{\alpha})\nonumber
\\
&=& -\alpha ~{\rm int}\Big[\frac{S+t+1-a}{S}\Big]
+\order(t\sqrt{\alpha})
\end{eqnarray}
where ${\rm int}[z]$ denotes the largest integer $m$ such that
$z\geq m$.
 This
is the generalization to $S>2$ of the period-2 oscillations known
to occur in MGs with $S=2$. For arbitrary valuation
initializations $\bv(0)=\bv_0$ the above solution generalizes to
\begin{eqnarray}
v_a(t)&=&  -\alpha ~{\rm
int}\Big[\frac{S+t+1-\pi_{\bv_0}(a)}{S}\Big]
+\order(t\sqrt{\alpha})
\end{eqnarray}
with $\pi_{\bv}$ denoting that permutation of $\{1,\ldots,S\}$ for
which $v_{\pi(1)}> v_{\pi(2)}>\ldots >v_{\pi(S)}$. If $\bv_0$ is
drawn randomly from some finite-width distribution, corresponding
to the situation where the agents in the original $N$-agent system
are initialized non-identically, all agents would still
continually alternate their strategies in a fixed order, but the
orders would now generally be different for different agents. The
 dynamic order parameters
would in either case be $C_{tt^\prime}=\delta_{t,t^\prime{\rm
mod}~S}+\order(\alpha^{-1/2})$ and
$G_{tt^\prime}=\order(\alpha^{-1/2})$.

Let us finally turn to small $\alpha$. Here the effective process
(\ref{eq:effective_agent1}) describes only small valuation changes
at each time step, and one consequently finds that
$C_{tt^\prime}=1+\order(\sqrt{\alpha})$ and
$G_{tt^\prime}=\order(\sqrt{\alpha})$. The effective Gaussian
noise in (\ref{eq:effective_agent1}) is static in leading order in
$\alpha$, $\bra
\eta_a(t)\eta_b(t^\prime)\ket=\tilde{\eta}^2\delta_{ab}+\order(\sqrt{\alpha})$,
and in the absence of perturbation fields equation
(\ref{eq:effective_agent1}) gives simply
\begin{eqnarray}
v_a(t)&=& v_a(0) + t\tilde{\eta}\sqrt{\alpha}z_a +\order(\alpha)
\end{eqnarray}
where the $z_a$ are independent frozen random Gaussian variables,
with zero average and unit variance. The systems remains static
for a period of order $t\sim \Delta/\tilde{\eta}\sqrt{\alpha}$,
where $\Delta$ indicates the magnitude of the initial valuation
differences. A full analysis of the solution of our effective
agent equations following this transient stage is in the small
$\alpha$ regime a highly nontrivial exercise, which (to our
knowledge) even for $S=2$ has not yet been carried out, and would
merit a full and extensive study in itself.

\section{Discussion}

In this paper we have shown how the generating functional analysis
theory of minority games, developed in full initially only for
$S=2$, can be generalized to MGs with arbitrary values of $S$. The
key obstacle in this generalization turned out not to be the
derivation of closed equations for dynamic order parameters (via a
generalized effective single agent process) but rather the
solution of these equations in time-translation invariant
stationary states. In previous studies closure of persistent order
parameter equations could not yet be achieved analytically, and
equations had to be closed artificially with the help of
simulation data \cite{MarsChalZecc00,Bianconi_etal06}. At a
technical level the basic problem was the calculation of the
strategy selection frequencies of the effective agent. This
problem has now been solved, resulting in exact and explicit
closed equations for persistent order parameters and for phase
transition points, for any value of $S$.

In our applications of the resulting theory we have mainly
concentrated on the simplest nontrivial case $S=3$, complemented
by further applications to $S=4$ and $S=5$. In all cases, the
predictions of our theory in time translation invariant stationary
states without anomalous response  were shown to agree perfectly
with numerical simulation data, including sensitive measures such
as the strategy frequency distribution. We have not been able to
solve our order parameter equations in all possible situations,
however, those regimes where we could not proceed to full solution
(e.g. calculating stationary states in the regime
$\alpha<\alpha_c(S)$, and non-persistent order parameters at
arbitrary times) are the same as those which also for the simpler
case $S=2$ have so far resisted the efforts of statistical
mechanicists. Put differently, our objective and contribution here
has been to raise the solvability of MGs with arbitrary values of
$S$ to the same level as that of MGs with $S=2$.

It will clear that several further applications, developments and
generalizations of the theory could now be taken up. One could for
instance explore in more detail the effects of decision noise on
MGs with $S>2$, for which we have generated the required
mathematical tools but for which we have not worked out the full
consequences (such as the often counter-intuitive impact on the
volatility, or the phase diagrams for multiplicative noise in the
$(\alpha,T)$ plane). Alternatively, one could develop an $S>2$
generating functional analysis for the so-called fake history
on-line MGs \cite{onlineMG}, where valuation updates are made
after each randomly drawn sample of the global information.
Probably the most interesting and nontrivial next step, however,
would be to investigate the structure and the stationary state
solutions of an $S>2$ theory for MGs where the global information
is no longer drawn randomly but represents the actual global
history of the market, by generalization of \cite{historyMG}.
\\[10mm]
{\bf Note} \\ While finishing this paper we were made aware of
another study in progress, aiming also  to solve the strategy
frequency problem for minority games with more than two strategies
per agent, but in the context of multi-asset MGs \cite{Isaac}, and
using a somewhat different approach (which one must ultimately
expect to be mathematically equivalent). \vsp

\clearpage
\appendix
\section{Derivation of effective single agent equation}
\label{app:saddlepoint}

Extremization of the exponent $\Psi+\Phi+\Omega$, as defined by
(\ref{eq:Psi},\ref{eq:Phi},\ref{eq:Omega}), with respect to the
dynamic order parameters $\{C,\hC,K,\hK,L,\hL\}$ gives the
following saddle-point equations:
\begin{eqnarray}
C_{t\tp}=\sum_{a}\bra\delta_{a,\nt}\delta_{a,\ntp}\ket_{\star}
\\
K_{t\tp}=\sum_{a}\bra \delta_{a,\nt}\hv_{a}(\tp)\ket_{\star}
\\
L_{t\tp}=\sum_{a}\bra \hv_{a}(t)\hv_{a}(\tp)\ket_{\star}
\\
\hC_{t\tp}=\rmi\frac{\partial \Phi}{\partial
C_{t\tp}}~~~~~~~~\hK_{t\tp}=\rmi\frac{\partial \Phi}{\partial
K_{t\tp}}~~~~~~~~\hL_{t\tp}=\rmi\frac{\partial \Phi}{\partial
L_{t\tp}} \label{eq:hat_kernels}
\end{eqnarray}
 with the abbreviation $\bra f(\bv,\hbv,\bz)\ket_{\star}=
\lim_{N\to\infty}N^{-1}\sum_{i}\bra f(\bv,\hbv,\bz)\ket_{i}$,
where
\begin{eqnarray}
\hspace*{-15mm} \bra
f(\bv,\hbv,\bz)\ket_{i}&=&\frac{\int\!\left[\prod_{at}\frac{dv_a(t)d\hat{v}_a(t)}{2\pi}\right]\bra
f(\bv,\hbv,\bz)F_{i}(\bv,\hbv,\bz)\ket_{\bz}}{\int\!\left[\prod_{at}\frac{dv_a(t)d\hat{v}_a(t)}{2\pi}\right]\bra
F_{i}(\bv,\hbv,\bz)\ket_{\bz}} \label{eq:fi}
\\
\hspace*{-15mm} F_{i}(\bv,\hbv,\bz)&=& P_0(\bv(0))~
\rme^{\rmi\sum_{at}\hat{v}_a(t)[v_a(t+1)-v_a(t)-\theta_{ia}(t)]+\rmi\sum_{at}\psi_{ia}(t)\delta_{a,m(\bv(t),\bz(t))}}
\label{eq:Fi}
 \\ \hspace*{-15mm}  &&
 \times~
\rme^{-\rmi\sum_{a
tt^\prime}\left[\hat{C}_{tt^\prime}\delta_{a,m(\bv(t),\bz(t))}\delta_{a,m(\bv(t^\prime),\bz(t^\prime))}
+\hat{L}_{tt^\prime}
 \hat{v}_a(t)\hat{v}_a(t^\prime)+\hat{K}_{tt^\prime}\delta_{a,m(\bv(t),\bz(t))} \hat{v}_a(t^\prime)\right]}
 \nonumber
\end{eqnarray}
Via (\ref{eq:defineC},\ref{eq:defineG}) and the identity
$\overline{Z[\bnull]}=1$ one confirms as usual, for the physical
saddle-point, that the $C_{tt^\prime}$ are indeed the correlations
in (\ref{eq:defineC}), that $L_{tt^\prime}=0$, and that
$K_{tt^\prime}=iG_{tt^\prime}$. Putting $\bpsi\to\bnull$ (they are
no longer needed) and choosing $\theta_{ia}=\theta_a$
(site-independent perturbations) eliminates the dependence of
(\ref{eq:Fi}) on $i$: $F_{i}(\bv,\hbv,\bz)=F(\bv,\hbv,\bz)$. Next,
to evaluate equations (\ref{eq:hat_kernels}) we need to work out
the function $\Phi$ (\ref{eq:Phi}) for small $\{L_{tt^\prime}\}$.
Upon eliminating $K$ via $K=iG$, and with the short-hands $\one$
for the identity matrix and $(A^\dag)_{tt^\prime}=A_{t^\prime t}$,
we find
\begin{eqnarray}
\Phi&=&\alpha \log \int\!\left[\prod_{t}\frac{dx_t
d\hat{x}_t}{2\pi}\rme^{\rmi
x_t\hat{x}_t}\right]\rme^{-\frac{1}{2}\sum_{tt^\prime}x_tC_{tt^\prime}x_{t^\prime}+\rmi
\tilde{\eta}\sum_{tt^\prime}x_t G_{tt^\prime}\hat{x}_{t^\prime}}
\nonumber \\ &&\hspace*{20mm}\times
\Big[1-\frac{1}{2}\tilde{\eta}^2 \sum_{tt^\prime}\hat{x}_t
L_{tt^\prime}\hat{x}_{t^\prime}+\order(L^2)\Big] \nonumber
\\
&=&-\frac{1}{2}\alpha \log ~{\rm
det}\Big[(\one+\tilde{\eta}G^\dag) (\one+\tilde{\eta}G)\Big]
\nonumber \\ && -\frac{1}{2}\alpha \tilde{\eta}^2
\sum_{tt^\prime}L_{tt^\prime}\Big[(\one+\tilde{\eta}G)^{-1}C
(\one+\tilde{\eta}G^\dag)^{-1}\Big]_{tt^\prime}+\order(L^2)
\end{eqnarray}
For $L=0$ the three saddle-point equations (\ref{eq:hat_kernels})
now become
\begin{eqnarray}
\hC_{t\tp}&=&0~~~~~~~~\hK_{t\tp}=-\alpha
\tilde{\eta}(1+\tilde{\eta}G^\dag)^{-1}_{tt^\prime}
\\
\hL_{t\tp}&=& -\frac{1}{2}\rmi \alpha \tilde{\eta}^2
\Big[(\one+\tilde{\eta}G)^{-1}C
(\one+\tilde{\eta}G^\dag)^{-1}\Big]_{tt^\prime}
\end{eqnarray}
Upon inserting these expressions into (\ref{eq:Fi}), and  using
causality (viz. $G_{tt^\prime}=0$ for $t\leq t^\prime$) one can
now prove that the denominator of (\ref{eq:fi}) equals one. This,
in turn, implies that $\bra
g(\bv,\bz)\hat{v}_a(t)\ket_\star=\rmi\partial \bra
g(\bv,\bz)\ket_\star/\partial\theta_a$. We are then in a position
to integrate out all occurrences of the conjugate integration
variables $\{\hat{v}_a\}$, and end up with the remaining
saddle-point equations
\begin{eqnarray}
C_{t\tp}&=&\sum_{a}\bra\delta_{a,\nt}\delta_{a,\ntp}\ket_{\star}
\\
G_{t\tp}&=&\sum_{a}\frac{\partial}{\partial\theta_a(t^\prime)}\bra
\delta_{a,\nt}\ket_{\star}
 \end{eqnarray}
 where
\begin{eqnarray}
\hspace*{-23mm}
 \bra f(\bv,\bz)\ket_{\star}&=&
 \int\!\Big[\prod_{at}dv_a(t)\Big]\bra f(\bv,\bz)F(\bv,\bz)\ket_{\bz}
 \\
 \hspace*{-23mm}
F(\bv,\bz)&=& P(\bv(0))
\int\!\Big[\prod_{at}\frac{d\hat{v}_a(t)}{2\pi}\Big]\int\!\Big[\prod_{at}d\eta_a(t)\Big]~
\nonumber \\
 \hspace*{-23mm}
 && \times\prod_{at}
\delta\Big[\eta_a(t)\!-\!\frac{1}{\sqrt{\alpha}}\Big(
v_a(t\!+\!1)\!-\!v_a(t)\!-\!\theta_{a}(t)\!+\!\alpha\tilde{\eta}\sum_{t^\prime}(\one\!+\!\tilde{\eta}G)^{-1}_{tt^\prime}\delta_{a,m(\bv(t^\prime),\bz(t^\prime))}\Big)\Big]
\nonumber \\ \hspace*{-23mm}&& \times
 \rme^{\rmi\sqrt{\alpha}\sum_{at}\hat{v}_a(t)\eta_a(t)-\frac{1}{2}\alpha\tilde{\eta}^2\sum_a
 \sum_{tt^\prime}
 \hat{v}_a(t)[(\one+\tilde{\eta}G)^{-1}C(\one+\tilde{\eta}G^\dag)^{-1}]_{tt^\prime}\hat{v}_a(t^\prime)}
 \nonumber
 \\
 \hspace*{-23mm}&=& P(\bv(0))\int\!\Big[\prod_{at}\frac{d\eta_a(t)}{\sqrt{2\pi}}\Big]
 \prod_a\left\{\frac{\rme^{-\frac{1}{2}\tilde{\eta}^{-2}
 \sum_{tt^\prime}\eta_a(t)[(\one+\tilde{\eta}G^\dag)C^{-1}(\one+\tilde{\eta}G)]_{tt^\prime}\eta_a(t^\prime)}}{{\rm
 det}^{-\frac{1}{2}}[(\one\!+\tilde{\eta}G^\dag)C^{-1}(\one\!+\tilde{\eta}G)]}\right\}
 \nonumber
 \\
 \hspace*{-23mm}&&\times\prod_{at}\Big[v_a(t\!+\!1)\!-\!v_a(t)\!-\!\theta_{a}(t)\!
 +\!\alpha\tilde{\eta}\sum_{t^\prime}(\one\!+\!\tilde{\eta}G)^{-1}_{tt^\prime}\delta_{a,m(\bv(t^\prime),\bz(t^\prime))}\Big]
 \end{eqnarray}
We recognize the above measure to represent the statistics of an
effective single agent process, with dynamics defined as
\be
v_{a}(t+1)=v_{a}(t)+\theta_{a}(t)-\alpha\tilde{\eta}\sum_{\tp}R_{tt^\prime}
\delta_{a,\ntp}+ \sqrt{\alpha} ~\eta_{a}(t) \ee Here
$R_{tt^\prime}=\tilde{\eta}(\one+\tilde{\eta}G)^{-1}_{tt^\prime}$,
and  $\eta_a(t)$ is a Gaussian noise characterized by the moments
$\bra\eta_{a}(t)\ket=0$ and
$\bra\eta_{a}(t)\eta_{b}(\tp)\ket=\delta_{ab}(R CR^\dag)_{t\tp}$.

\section{The volatility matrix}
\label{app:vola_matrix}

Here we outline briefly how the generating functional
(\ref{eq:Zphi}) for overall bid fluctuations can be calculated via
simple modifications of the generating functional
$\overline{Z[\bpsi]}$ defined in (\ref{eq:Z}). Comparison with
(\ref{eq:Z}) shows that in the latter we should replace
 \bd
\rme^{\rmi\sum_{iat}\psi_{ia}(t)\delta_{a,m(\bv_i(t),\bz_i(t))}}~\to~
\rme^{\rmi\sum_{\mu a t}\phi_{\mu}(t)N^{-1/2}\sum_i
R_\mu^{ia}\delta_{a,m(\bv_i(t),\bz_i(t))}} \ed
 This is found to imply making the replacement $x_t^\mu\to
 x_t^\mu+\Phi_\mu(t)$ in the disorder average, and ultimately has
 an effect only on the exponent $\Phi$ of the saddle-point
 problem (leaving $\Psi$ and $\Omega$ unaffected). The latter will now become
 \begin{eqnarray}
\Phi&=&\frac{1}{N}\sum_\mu \log \int\!\left[\prod_{t}\frac{dx_t
d\hat{x}_t}{2\pi}\rme^{\rmi x_t\hat{x}_t}\right]
\Big[1-\frac{1}{2}\tilde{\eta}^2 \sum_{tt^\prime}\hat{x}_t
L_{tt^\prime}\hat{x}_{t^\prime}+\order(L^2)\Big] \nonumber
\\
&&\times
\rme^{-\frac{1}{2}\sum_{tt^\prime}(x_t+\phi_\mu(t))C_{tt^\prime}(x_{t^\prime}+\phi_\mu(t^\prime))+\rmi
\tilde{\eta}\sum_{tt^\prime}(x_t+\phi_\mu(t))
G_{tt^\prime}\hat{x}_{t^\prime}} \nonumber
\\
&=&-\frac{1}{2}\alpha \log ~{\rm
det}\Big[(\one+\tilde{\eta}G^\dag) (\one+\tilde{\eta}G)\Big]
\nonumber \\ && +\frac{1}{N}\!\sum_\mu\log\left[1
\!-\!\frac{1}{2}\sum_{tt^\prime}\Big(
L_{tt^\prime}\!+\!\frac{\phi_\mu(t)\phi_\mu(t^\prime)}{\tilde{\eta}^2}\Big)(RCR^\dag)_{tt^\prime}\!+\order(L^2\!,\phi^4)\right]~
~~~~~ \label{eq:Phiphi}
\end{eqnarray}
We can now calculate from $\overline{Z[\bphi]}$ the quantities of
interest. Upon emphasizing the dependence of (\ref{eq:Phiphi}) on
the fields $\bphi$, and using the normalization
$\overline{Z[\bnull]}=1$, we obtain:
\begin{eqnarray}
\hspace*{-19mm}
 \lim_{N\to\infty}\overline{\bra A_\mu(t)\ket}&=&
-\rmi\lim_{N\to\infty} \lim_{\bphi\to\bnull} \frac{\partial
\overline{Z[\bphi]}}{\partial \phi_\mu(t)}\nonumber
\\
\hspace*{-19mm} &=&-\rmi\lim_{N\to\infty} \lim_{\bphi\to\bnull}
\frac{\partial}{\partial \phi_\mu(t)}
~\rme^{N[\Phi(\bphi)-\Phi(\bnull)]}|_{\rm saddle}\nonumber
\\
\hspace*{-19mm} &=& -\rmi\lim_{\bphi\to\bnull}
\frac{\partial}{\partial \phi_\mu(t)} \prod_\lambda\left[1
\!-\!\frac{1}{2}\sum_{ss^\prime}\phi_\lambda(s)\frac{(RCR^\dag)_{ss^\prime}}{\tilde{\eta}^2}\phi_\lambda(s^\prime)\!+\ldots\right]
\nonumber \\
 \hspace*{-19mm}  &=& ~0
\\
\hspace*{-19mm}
 \lim_{N\to\infty}\overline{\bra
A_\mu(t)A_\nu(t^\prime)\ket}&=&-\lim_{N\to\infty}
\lim_{\bphi\to\bnull} \frac{\partial^2
\overline{Z[\bphi]}}{\partial
\phi_\mu(t)\partial\phi_\nu(t^\prime)}\nonumber
\\
\hspace*{-19mm} &=& -\lim_{N\to\infty} \lim_{\bphi\to\bnull}
\frac{\partial^2}{\partial \phi_\mu(t)\partial\phi_\nu(t^\prime)}
~\rme^{N[\Phi(\bphi)-\Phi(\bnull)]}|_{\rm saddle}\nonumber
\\
\hspace*{-19mm} &=& -\lim_{\bphi\to\bnull}
\frac{\partial^2}{\partial \phi_\mu(t)\phi_\nu(t^\prime)}
\prod_\lambda\left[1
\!-\!\frac{1}{2}\sum_{ss^\prime}\phi_\lambda(s)\frac{(RCR^\dag)_{ss^\prime}}{\tilde{\eta}^2}\phi_\lambda(s^\prime)\!+\ldots\right]
\nonumber \\ \hspace*{-19mm} &=&
~\tilde{\eta}^{-2}\delta_{\mu\nu}(RCR^\dag)_{tt^\prime}
\end{eqnarray}

\section{Integration identities}
\label{app:integrations}

Here we simply list (without proof) some of the basic identities
that one uses in doing the various integrals in the $S=3$ theory
analytically, for the benefit of the reader:
\begin{eqnarray}
\int_0^u\!Dx~x^2&=&
\frac{1}{2}\erf(\frac{u}{\sqrt{2}})-\frac{u}{\sqrt{2\pi}}~\rme^{-\frac{1}{2}u^2}
\\
\int\!Dx~\erf(A+Bx)&=& \erf\Big(\frac{A}{\sqrt{1+2B^2}}\Big)
\\
\int\!Dx~x~\erf(A+Bx)&=&
\frac{2B}{\sqrt{\pi(1\!+\!2B^2)}}~e^{-A^2/(1+2B^2)}
\\
\int\!Dx~\erf^2(Bx)&=&
\frac{4}{\pi}\arctan\Big(\sqrt{1+4B^2}\Big)-1
\\
\int\!Dx~\erf^2(A+Bx)&=&
\frac{4}{\pi}\arctan\Big(\sqrt{1+4B^2}\Big)-1\nonumber
\\
&&+\frac{4}{\sqrt{\pi}}\int_0^{A/\sqrt{1+2B^2}}\!dx~\rme^{-x^2}\erf\Big(\frac{x}{\sqrt{1+4B^2}}\Big)
\end{eqnarray}

\section{The strategy frequency problem for multiplicative noise}
\label{app:multiplicative}

Here
 we discuss briefly the solution of the strategy frequency
problem for
 multiplicative decision noise: $m(\bv,\bz)=r~ {\rm
 argmax}_{a}[v_a]+(1-r)b$, with random $r\in\{0,1\}$ and $b\in\{1,\ldots,S\}$.
 The deterministic case corresponds to $P(r)=\delta_{r,1}$. Again we
 define
$\overline{v}^\star(\bx)=\max_b \overline{v}_b(\bx)$ and the set
$\Lambda(\bx)$ via (\ref{eq:defineLambda}). Upon abbreviating
$\lambda=\sum_{r}P(r)r\in[0,1]$, the solution now proceeds as
follows
\begin{itemize}
\item
 Since
$v_a(s,\bx)=s[\overline{v}_a(\bx)+\varepsilon_a(s,\bx)]$ with
$\lim_{s\to\infty}\varepsilon_a(s,\bx)=0$ we may write the second
equation in (\ref{eq:freq2}) as
\begin{eqnarray*}
f_a(\bx)&=& \lim_{s\to\infty}\left\{ \lambda~ \delta_{a,{\rm
 argmax}_{b}[\overline{v}_b(\bx)+\varepsilon_b(s,\bx)]}+(1-\lambda)S^{-1}\right\}
\nonumber
\\
&=&\frac{1-\lambda}{S}+\lambda \lim_{s\to\infty}\prod_{b\neq
a}\Big\bra\theta\Big[\overline{v}_a(\bx)-\overline{v}_b(\bx)+\varepsilon_a(s,\bx)-\varepsilon_b(s,\bx)\Big]
\end{eqnarray*}
This second term can be nonzero only for $a\in\Lambda(\bx)$.
Hence, once we know
 $\Lambda(\bx)$ and $\overline{v}^\star(\bx)$ the
problem is solved:
\begin{eqnarray}
\hspace*{-10mm} a\notin\Lambda(\bx):&~~~
f_a(\bx)=(1\!-\!\lambda)/S,~~~&\overline{v}_a(\bx)= \alpha
\chiR(x_a\sqrt{\frac{c}{\alpha}}\!-\!\frac{1\!-\!\lambda}{S})
 \label{eq:bullet2a_mul}
\\
\hspace*{-10mm} a\in\Lambda(\bx):&~~~ f_a(\bx)=
x_a\sqrt{\frac{c}{\alpha}} - \frac{\overline{v}^\star(\bx)}{\alpha
\chiR},~~~&\overline{v}_a(\bx)=\overline{v}^\star(\bx)
\label{eq:bullet2b_mul}
\end{eqnarray}
\item
The value $\overline{v}^\star(\bx)$ is again calculated from
$\sum_{a}f_a(\bx)=1$, but now also strategies
$a\not\in\Lambda(\bx)$ are involved. We abbreviate
$|\Lambda(\bx)|^{-1}\sum_{b\in\Lambda(\bx)}U_b =\bra
U\ket_{\Lambda(\bx)}$, and sum over the indices in
 both (\ref{eq:bullet2a_mul})  and (\ref{eq:bullet2b_mul}) to get
\begin{eqnarray}
\overline{v}^\star(\bx)&=&  \chiR\sqrt{\alpha
c}\sum_{a\in\Lambda(\bx)}\bra x\ket_{\Lambda(\bx)} - \alpha
\chiR\Big(\frac{\lambda}{|\Lambda(\bx)|}+\frac{1\!-\!\lambda}{S}\Big)
\end{eqnarray}
equations (\ref{eq:bullet2a_mul},\ref{eq:bullet2b_mul}) thus
become
\begin{eqnarray}
\hspace*{-20mm} a\!\notin\!\Lambda(\bx):&~~~
f_a(\bx)=(1\!-\!\lambda)/S,~~~~~ \overline{v}_a(\bx)= \alpha
\chiR(x_a\sqrt{\frac{c}{\alpha}}\!-\!\frac{1\!-\!\lambda}{S})\label{eq:bullet3a_mul}
\\
\hspace*{-20mm}
 a\!\in\!\Lambda(\bx):&~~~ f_a(\bx)=
\frac{\lambda}{|\Lambda(\bx)|}+\frac{1\!-\!\lambda}{S} +
 \sqrt{\frac{c}{\alpha}}\left(x_a\! -\bra
 x\ket_{\Lambda(\bx)}\right),~~~&\overline{v}_a(\bx)=\overline{v}^\star(\bx)\label{eq:bullet3b_mul}
\end{eqnarray}
\item
What remains is to determine the set $\Lambda(\bx)$. The
definition (\ref{eq:defineLambda}) of $\Lambda(\bx)$ demands that
$\overline{v}_a(\bx)<\overline{v}^\star(\bx)$ for all
$a\not\in\Lambda(\bx)$, i.e.
\begin{eqnarray}
a\not\in\Lambda(\bx):&~~~&  x_a < \bra x\ket_{\Lambda(\bx)} -
\sqrt{\frac{\alpha}{c}} \frac{\lambda}{|\Lambda(\bx)|}
\end{eqnarray}
Demanding that expression (\ref{eq:bullet3b_mul}) obeys
$f_a(\bx)\in[0,1]$ gives, similarly
\begin{eqnarray}
 a\in\Lambda(\bx):&~~~&  x_a
\geq \bra
 x\ket_{\Lambda(\bx)}-\sqrt{\frac{\alpha}{c}}\Big(
\frac{\lambda}{|\Lambda(\bx)|}+\frac{1\!-\!\lambda}{S}\Big)
 \\
 &&
 x_a \leq \bra
 x\ket_{\Lambda(\bx)}-\sqrt{\frac{\alpha}{c}}\Big(
\frac{\lambda}{|\Lambda(\bx)|}+\frac{1\!-\!\lambda}{S}\Big)
+\sqrt{\frac{\alpha}{c}}
 \end{eqnarray}
So $\Lambda(\bx)$ again contains the indices
 of the $\ell$ largest components of the $\bx$, where $\ell=|\Lambda(\bx)|$.
 For each $\bx$ we define the
 permutation $\pi_\bx:\{1,\ldots,S\}\to \{1,\ldots,S\}$ for which
 these components will be ordered according to $x_{\pi(1)}> x_{\pi(2)}> \ldots
 > x_{\pi(S)}$, so that our three inequalities can be
 written as
\begin{eqnarray}
\hspace*{-5mm} x_{\pi(\ell)} &\geq
\frac{1}{\ell\!-\!1}\sum_{m=1}^{\ell-1}
 x_{\pi(m)}-\frac{1}{\ell\!-\!1}\sqrt{\frac{\alpha}{c}}\Big(
\lambda+(1\!-\!\lambda)\frac{\ell}{S}\Big) ~~~~~& {\rm if}~\ell>1
~~~~~ \label{eq:ineq1_mul}
 \\
 \hspace*{-5mm}
x_{\pi(\ell+1)} &< \frac{1}{\ell}\sum_{m=1}^\ell
 x_{\pi(m)} - \frac{1}{\ell}\sqrt{\frac{\alpha}{c}} ~\lambda
 ~~~~~&{\rm if}~\ell<S
  ~~~~~\label{eq:ineq2_mul}
\\
\hspace*{-5mm}
 x_{\pi(1)} &\leq \frac{1}{\ell}\sum_{m=1}^\ell
 x_{\pi(m)}-\frac{1}{\ell}\sqrt{\frac{\alpha}{c}}\Big(
\lambda+(1\!-\!\lambda)\frac{\ell}{S}\Big)
+\sqrt{\frac{\alpha}{c}} ~~~~~\label{eq:ineq3_mul}
 \end{eqnarray}
Similar to the case of additive noise we conclude that $\ell$ is
the {\em smallest} number in $\{1,\ldots,S\}$ for which
(\ref{eq:ineq2_mul}) holds (if any), whereas if
(\ref{eq:ineq2_mul}) never holds then $\ell=S$. We thereby satisfy
also (\ref{eq:ineq1_mul}) (compared to additive noise
 we here even satisfy a stronger condition).
The proof that also the third condition (\ref{eq:ineq3_mul}) will
be satisfied is virtually identical to that given for additive
noise, so need not be repeated here. Since (\ref{eq:ineq1_mul})
and (\ref{eq:ineq2_mul}) are no longer mutually exclusive for
$\lambda<1$, it is no longer immediately  obvious that the
solution is unique, but this can probably be proven.
\end{itemize}
From this point onwards one can translate the solution found into
expressions for averages of observables. This will once more
involve a sum over all possible permutations (from which the
component ordering permutation $\pi_{\bx}$ is selected via a
product of step functions) and further step functions to implement
the inequalities on the components of $\bx$ relative to the
average over the strategies in the set $\Lambda(\bx)$ that define
this active set.

\end{document}